\documentclass[twocolumn]{bmcart}

\usepackage[utf8]{inputenc}
\usepackage[T1]{fontenc}
\usepackage[english]{babel}
\usepackage{csquotes}
\usepackage{microtype}
\usepackage{amsmath}
\usepackage{amssymb}
\usepackage{mathtools}
\usepackage[group-digits=false,range-phrase={\text{~to~}}]{siunitx}
\usepackage{newfloat}
\usepackage{setspace}
\usepackage[noend]{algpseudocode}
\renewcommand{\algorithmicindent}{1em}
\usepackage{booktabs}
\usepackage{array}
\usepackage{hhline}
\usepackage{tabu}
\usepackage{colortbl}
\usepackage{multirow}
\usepackage[caption=false,font=footnotesize]{subfig}
\usepackage{tikz}
\usepackage{pgfplots}
\usepackage[switch]{lineno}
\usepackage{placeins}
\usepackage{ragged2e}
\usepackage{float}
\usepackage{varwidth}
\usepackage{version}
\usepackage[absolute,overlay]{textpos}
\usepackage[breaklinks=true,bookmarks=false]{hyperref}

\usetikzlibrary{matrix,positioning,fit,external,calc}
\usepgfplotslibrary{groupplots,statistics}
\pgfplotsset{compat=1.14}

\algdef{SE}[LOOP]{Loop}{EndLoop}[1]{\algorithmicloop\ #1}{\algorithmicend\ \algorithmicloop}
\algtext*{EndLoop}
\algrenewcommand\textproc{\textrm}

\setattribute{floatcaption}{size}{\footnotesize\sffamily\raggedright}

\startlocaldefs

\newcommand{\sabs}[1]{\lvert#1\rvert}

\newcommand{\snorm}[1]{\lVert#1\rVert}
\newcommand{\bignorm}[1]{\bigl\lVert#1\bigr\rVert}

\newcommand{\e}{e}
\newcommand{\dif}{\mathrm{d}}

\newcommand{\R}{\mathbb{R}}

\newcommand{\N}{\mathbb{N}}
\newcommand{\Z}{\mathbb{Z}}

\newcommand{\eps}{\varepsilon}

\newcommand*\patchAmsMathEnvironmentForLineno[1]{%
  \expandafter\let\csname old#1\expandafter\endcsname\csname #1\endcsname
  \expandafter\let\csname oldend#1\expandafter\endcsname\csname end#1\endcsname
  \renewenvironment{#1}%
     {\linenomath\csname old#1\endcsname}%
     {\csname oldend#1\endcsname\endlinenomath}}%
\newcommand*\patchBothAmsMathEnvironmentsForLineno[1]{%
  \patchAmsMathEnvironmentForLineno{#1}%
  \patchAmsMathEnvironmentForLineno{#1*}}%
\AtBeginDocument{%
\patchBothAmsMathEnvironmentsForLineno{equation}%
\patchBothAmsMathEnvironmentsForLineno{align}%
\patchBothAmsMathEnvironmentsForLineno{flalign}%
\patchBothAmsMathEnvironmentsForLineno{alignat}%
\patchBothAmsMathEnvironmentsForLineno{gather}%
\patchBothAmsMathEnvironmentsForLineno{multline}%
}
\DeclareFloatingEnvironment[name=Algorithm]{algoenv}

\endlocaldefs

\begin{document}

\begin{textblock*}{12cm}(8cm,1.1cm)
  \noindent
  \textcolor{red!70!black}{\bfseries\sffamily\large
  This is the authors' manuscript (post peer-review).\\
  In order to access the published article, please visit:\\
  \url{https://doi.org/10.1186/s13636-020-00190-4}}
\end{textblock*}

\begin{frontmatter}

\begin{fmbox}
\dochead{Research}


\title{Sparse Pursuit and Dictionary Learning for Blind Source
  Separation in Polyphonic Music Recordings}


\author[
   addressref={aff1},                   
   corref={aff1},                       
   email={sschulze@uni-bremen.de}   
]{\inits{SS}\fnm{Sören} \snm{Schulze}}
\author[
   addressref={aff2},
   email={emily.king@colostate.edu}
]{\inits{EJK}\fnm{Emily J.} \snm{King}}


\address[id=aff1]{
  \orgname{AG Computational Data Analysis, Faculty 3, University of Bremen}, 
  \street{Bibliothekstr.\ 5},                     %
  \postcode{28359}                                
  \city{Bremen},                              
  \cny{Germany}                                    
}
\address[id=aff2]{%
  \orgname{Mathematics Department, Colorado State University},
  \street{1874 Campus Delivery, 111 Weber Bldg},
  \postcode{CO 80523},
  \city{Fort Collins},
  \cny{USA}
}


\begin{artnotes}
\end{artnotes}



\begin{abstractbox}

\begin{abstract} 
We propose an algorithm for the blind separation of single-channel
audio signals.  It is based on a parametric model that describes the
spectral properties of the sounds of musical instruments independently
of pitch.  We develop a novel sparse pursuit algorithm that can
match the discrete {frequency}
spectra from the recorded signal with the continuous
spectra delivered by the model.  We first use this algorithm to
convert an STFT spectrogram from the recording into a novel form of
log-frequency spectrogram whose resolution exceeds that of the mel
spectrogram.  We then make use of the pitch-invariant properties of
that representation in order to identify the sounds of the instruments
via the same sparse pursuit method.  As the model parameters which
characterize the musical instruments are not known beforehand, we
train a dictionary that contains them, using a modified version of Adam.
Applying the algorithm on various audio samples, we
find that it is {capable of producing high-quality separation results
when the model assumptions are satisfied and the instruments are clearly
distinguishable, but combinations of instruments with similar spectral
characteristics pose a conceptual difficulty.  While a key feature of the
model is that it explicitly models inharmonicity, its presence can also still
impede performance of the sparse pursuit algorithm.  In general,
due to its pitch-invariance, our method is especially suitable for
dealing with spectra from acoustic instruments, requiring only a minimal
number of hyperparameters to be preset.}  Additionally, we demonstrate
that the dictionary that is constructed for one recording can be applied to a
different recording with similar instruments without additional
training.
\end{abstract}




\end{abstractbox}
\end{fmbox}

\end{frontmatter}



\section{Introduction}

\subsection{Problem Definition and Approach}

\begin{figure*}[btp]
  \tikzsetnextfilename{Figure_\the\numexpr(\thefigure+1)\relax}
  \footnotesize\centering\sffamily
  \begin{tikzpicture}[line width=0.55pt,scale=2]
    \tikzset{interface/.style={draw,rectangle,align=center},
             process/.style={draw,circle,align=center}}
    \matrix (m) [column sep=3mm,row sep=3mm]
    {
      \node[interface] (signal) {Time-domain\\signal};
      & \node[process] (stft) {STFT};
      & \node[interface] (spectrogram) {Spectrogram};
      & \node[process] (sp1) {Sparse\\Pursuit};
      & \node[interface] (logfreqspect) {Log-frequency\\spectrogram};
      &
      & \node[process] (sp2) {Sparse\\pursuit};
      & \node[above=0.5mm] (it) {\normalsize\itshape Iterate};
      \\
      \node[interface] (sepsignals) {Separated\\signals};
      & \node[process] (gl) {Griffin\\\& Lim};
      & \node[process] (spectmask) {Spectral\\masking};
      & \node[interface] (sepspects) {Separated\\spectrograms};
      & \node[process] (sp3) {Sparse\\pursuit};
      &
      & \node[interface] (finaldict) {Dictionary};
      & \node[process] (updatedict) {Adam};
      \\
    };
    \node[fit=(sp2) (updatedict), draw, inner sep=3.5mm,densely dashed] {};
    \draw [->] (signal) edge (stft)
               (stft) edge (spectrogram)
               (spectrogram) edge (sp1)
               (sp1) edge (logfreqspect)
               (logfreqspect) edge (sp2)
               (sp2) edge (updatedict)
               (updatedict) edge (finaldict)
               (finaldict) edge (sp2)
               (finaldict) edge (sp3)
               (logfreqspect) edge (sp3)
               (sp3) edge (sepspects)
               (sepspects) edge (spectmask)
               (spectrogram) edge (spectmask)
               (spectmask) edge (gl)
               (stft) edge (gl)
               (gl) edge (sepsignals);
  \end{tikzpicture}
  \caption{Data flow diagram for the proposed separation method.
    The sparse pursuit algorithm is used both for converting the STFT
    spectrogram into a log-frequency spectrogram and for identifying
    the instrument sounds in the log-frequency spectrogram.}
  \label{fig:overall}
\end{figure*}

{\emph{Source separation} concerns} the recovery of signals
$X_1,\dotsc,X_c$ from a mixture $X=X_1+\ldots+X_c$.  We speak of
\emph{blind separation} when no specific prior information to
characterize the sources of the signals is provided, especially not in
the form of labeled training data.  However, we do make
\emph{structural} assumptions about the signals;
in our case, we assume that they follow
the typical characterics of tones from wind and string instruments.

In order to {exploit this structure}, it is helpful
to regard a time-frequency representation (\emph{spectrogram}), which
subdivides the problem into smaller time frames and highlights the
frequency characteristics of the signal.  One simple spectrogram is
obtained via the modulus of the short-time Fourier transform (STFT)
(cf.\ \cite{Groechenig01}).  However, in the STFT spectrogram, 
different pitch of the instrument tones manifests in linear scaling of
the distances between the peaks on the frequency axis, which makes it
computationally hard to identify the tones in the spectrum.

Thus, we apply a novel \emph{sparse pursuit} algorithm that
represents the {time frames of the}
  STFT spectrogram via a limited number of peaks,
under the assumption that they originate from
sinusoidal signals in the recording.
We then place these {identified} peaks
in a new spectrogram {that has a logarithmic frequency
axis and is therefore \emph{pitch-invariant} (cf.\ Section \ref{sec:spectrogram})}.
On this, we apply {a \emph{dictionary learning} algorithm,
where the dictionary contains the learned relative amplitudes of the
harmonics for each instrument.  In an alternating loop, we 
identify the sounds of the instruments by now applying the sparse
pursuit algorithm on time frames of the log-frequency spectrogram
using the current value of the dictionary and then update the
dictionary based on that identification.  Both the problem of finding
the peaks in the STFT spectrogram and the problem of finding the
patterns representing the instrument sounds are generally
underdetermined (cf.\ Section \ref{sec:sparsity}), so
sparsity plays a crucial role in their regularization.}

{After training has
finished,} we apply the sparse pursuit algorithm on the entire
log-frequency spectrogram in order to obtain the separated
spectrograms, and after masking with the original mixture spectrogram,
we employ the algorithm by Griffin and Lim \cite{GriffinLim84} in
order to convert them back into time-domain signals, using the phase
of the original spectrogram as the initial value.
The overall procedure is displayed in Figure~\ref{fig:overall}.

The novelty of our approach lies in {the combination of
{pitch-invariant} representations with a
sparse pursuit algorithm during training:
Provided that the characteristics of the sounds of the
instruments are sufficiently stable, the relative amplitudes of the
harmonics saved in the dictionary represent}
the sounds of the instruments at any
arbitrary pitch, without making assumptions about their tuning or
range.  {At the same time, the use of a log-frequency axis
enables us to match the spectrogram with the modeled patterns of these
sounds in an efficient manner, and due to a non-linear
optimization step, the
parameters are locally optimal on a continuous scale.
As the outcome of
the training is sometimes sensitive with respect to the initial
dictionary, we typically use the method in an ensemble setting.
Apart from the sparsity condition, there is no need to hand-tune any
hyperparameters for a specific recording.}

The sparse pursuit algorithm that we propose is designed to match a
{generic}
sampled spectrum with shifted non-negative continuous patterns.  While
it was developed with audio frequency spectra in mind, it may be used
in other signal processing applications as well.

\subsection{Related Work}

During the past years, audio source separation has become a very wide
field, now incorporating a number of fundamentally different
applications and approaches.  A thorough overview can be found in
books on the subject that have recently appeared
\cite{Vincent18,Makino18,Chien18}.

The first instance of learning the harmonic structure of musical
instruments via non-negative matrix factorization (NMF)
\cite{LeeSeung99} on spectrograms was by Smaragdis and Brown
\cite{SmaradgisBrown03} for the purpose of polyphonic music
transcription.  This approach was then applied to audio source
separation by Wang and Plumbley
\cite{WangPlumbley05}.  The algorithm learns a dictionary where each
atom represents one instrument at a specific pitch.  By estimating the
tones of the instruments at specific points in time, it is thereby
possible to reconstruct the contributions of the individual
instruments.  An overview of single-channel NMF-based methods can be
found in \cite{Fevotte18}.

In many cases, a single musical instrument can generate different
sounds which are perceptually similar and only vary in the pitch of
the tones.  Using the \emph{constant-Q transform} (CQT) \cite{Brown91}
as a log-frequency spectrogram, Fitzgerald et al.\ \cite{Fitzgerald05a}
use non-negative tensor factorization to generate a dictionary
containing the frequency spectra of different
instruments, which can be shifted on a fixed grid of semitones
in order to apply them to different notes.  This approach was later
refined by Jaiswal et al.\ \cite{Jaiswal11a,Jaiswal11b,Jaiswal13}.

The advantage of this representation is that it can be applied to a
large variety of musical instruments, as long as pitch-invariance is
fulfilled.  The drawback is that it requires the instruments to be
tuned precisely to a known equal-temperament scale, which makes it
impractical for real-world recordings with acoustic instruments.

Alternatively, the source separation problem on spectrograms can be
formulated in probabilistic terms, which is done in the method of
probabilistic latent component analysis (PLCA)
\cite{Smaragdis06,Smaragdis07}. Here, the entire spectrogram is
regarded as a probability distribution, which is then decomposed via expectation
maximization (EM) into marginal distributions that depend on latent
variables.  In its original form, both the model and the
numerics are identical to NMF, but it can be argued that the
probabilistic notation is more powerful and especially beneficial when
incorporating priors.

The latent variables can be chosen so that separation via PLCA is also
pitch-invariant \cite{Smaragdis08,Fuentes11}, and it is also possible
to model the harmonics explicitly
\cite{Fuentes12a,Fuentes12b,Fuentes13}.  Those algorithms operate in
the discrete domain, so they effectively perform non-negative
tensor factorization.  {In this formulation, the approach was pioneered
by \cite{Vincent09} for application in multiple pitch estimation (MPE).}

Duan et al.\ \cite{Duan08} also follow a probabilistic approach, but
with a more explicit model of the spectral structure of the harmonics
of the instruments.  They first use a peak detection algorithm in
order to find the potential frequencies for the harmonics.  Using a
greedy maximum-likelihood model, the fundamental frequencies are
estimated, and the harmonic patterns are clustered in order to assign
them to certain instruments.  This approach is interesting because it
allows the representation of tones without a predetermined tuning.

In our algorithm, we apply a more advanced tone model that
{during optimization}
incorporates inharmonicity (cf.\ \cite{FletcherRossing12}) and also
deviations in the width of the peaks, which may occur in case of
volume changes.  While we also preselect peaks, we only do so in order
to generate a pitch-invariant log-frequency spectrogram that is
suitable for for wideband signals.

For {narrowband} signals, the CQT could be used instead.
Alternatively, one could employ the mel spectrogram (cf.\ \cite{Vincent18}) or the
method proposed in \cite{SchulzeKing19}, which combines favorable
properties from both time-frequency representations.  However, the
resolution of any spectrogram that was computed via classical means is
{ultimately} limited by the Heisenberg uncertainty principle
(cf.\ \cite{Groechenig01,Folland97}).

The pitch-invariance property of the representation is important since
it allows us to locate the sounds of the instruments via
cross-correlation, making the determination of the fundamental
frequencies much easier.  However, rather than \emph{explaining} the peaks
in the spectrogram via a parametric model of the harmonic
structure of the instruments via clustering, we use stochastic
optimization to train a dictionary containing the relative
amplitudes of the harmonics in order to \emph{reproduce} their
sounds.

In our model, we aim to be parsimonious in the number of parameters
and, following the the spirit of blind separation, also in the
assumptions on the data.  Therefore, we regard each time frame of the
spectrogram independently.  However, models that take the time axis
into account do exist.  Smaragdis \cite{Smaragdis04} introduced
\emph{NMFD} (non-negative matrix factor deconvolution), which is NMF with convolution in time (again, a form of
tensor factorization), and Schmidt and Mørup \cite{SchmidtMorup06}
combined time- and pitch-invariant approaches to \emph{NMF2D}
(non-negative matrix factor two-dimensional deconvolution).
Virtanen \cite{Virtanen04} added a temporal sparsity criterion, and
later, in \cite{Virtanen07}, a temporal continuity objective.
Blumensath and Davies \cite{BlumensathDavies05} operate completely in
the time domain, without any time-frequency representation.

The musical score that matches the piece in the recording is also a
valuable piece of information, as it resolves ambiguities about the
fundamental frequencies.  Hennequin et al.\ \cite{Hennequin10} first
proposed a pitch-invariant model that can accommodate local variation
from predetermined tuning via gradient descent, but the authors faced
the problem that this approach did not work on a global scale.
Therefore, in \cite{Hennequin11}, they use the score to give the
algorithm hints about the approximate frequencies and thereby reduce
the optimization problem to a local one.  One of the
main challenges in score-informed separation is the alignment of the
score with the audio recording.  For this, a combined approach has
recently been proposed by Munoz-Montoro et al.\ \cite{Munoz19}.

Due to the growing interest in deep learning among the machine
learning community, it is also applied to
audio source separation in a supervised manner.  However, this
approach requires labeled training data.  Huang et al.\ \cite{Huang15}
proposed a deep recurrent neural network architecture and achieved
respectable results.  In the SiSEC (Signal Separation Evaluation
Campaign) 2018 \cite{Stoeter18}, different state-of-the-art algorithms were compared,
and the reference implementation \emph{Open-Unmix}
\cite{Stoeter19} was determined as the overall winner.  The
network operates on magnitude spectrograms and combines different
kinds of layers, including long short-term-memory (LSTM) units.
Its performance was recently surpassed by Défossez et
al.\ \cite{Defossez19}, whose network is based on LSTMs and
convolutions, but operates directly in the time (i.e., waveform)
domain.

Due to their good performance, supervised deep learning methods
currently dominate the focus of many researchers.  They make only very
mild explicit prior structural assumptions on the data and instead rely on
training to sort out the separation process.  Thus, whenever
appropriate training data is available, they make a very powerful and
versatile tool.

{Naturally, using more prior information in a machine learning problem
typically improves the quality of the results.  Conversely, purely blind
approaches can only work under very controlled conditions, and they have
therefore received relatively little attention in recent years.}
We aim to show that progress on
this problem is {nevertheless}
still possible, and that even blind separation can
profit from the modern machine learning techniques that have been
developed.

{Our sparse pursuit algorithm is a greedy approximation to
$\ell_0$ sparsity, based on concepts from orthogonal matching pursuit
(OMP) \cite{TroppGilbert07} and subspace pursuit
\cite{DaiMilenkovic09} while making use of the pitch-invariance of the
time-frequency representation.  However, a similar problem has been
formulated in an $\ell_1$ setting as \emph{convolutional sparse coding}
for image processing \cite{Bristow13}.  While it is relatively fast,
the drawback of this method is that it is still limited to discrete
convolutions.  In \emph{continuous basis pursuit} \cite{Ekanadham11},
this problem is approached by either Taylor or polar interpolation.
\emph{Beurling LASSO} \cite{Castro12,Catala17} first solves the sparse
representation
problem in the \emph{dual} space, but finding the corresponding primal
solution generally remains a challenge.  Whereas the general advantage
of $\ell_1$-based formulations lies in their convexity, greedy methods
allow for a more flexible optimization step while keeping the
dimensionality low.}

\subsection{The Musical Role of Sparsity}
\label{sec:sparsity}

The representation of the time frames of a spectrogram of a music
recording with a pitch-invariant dictionary is in general not unique.
If we consider
wind and string instruments, their sound is dominated by a linear
combination of sinusoids, which show up as horizontal lines in the
spectrogram.  Thus, there exists a trivial solution that assumes a
single sinusoidal instrument which plays a large number of
simultaneous tones.  While this solution is valid, it is undesirable,
as no separation is performed at all.

A similarly trivial solution is to construct different instruments for
each time frame of the spectrogram.  This, however, leaves us with the
problem of matching the constructed instruments with the actual
instruments.  This process would need to be done either manually or
via an appropriate clustering algorithm, such as the one used in
\cite{Duan08}.  Also, instruments which play harmonically
related notes may be mistaken for a single instrument, and this case
would need special consideration.

In order to attain meaningful solutions, we thus decide to limit both
the total number of instruments and the number of tones that are
assumed to be played at the same time.  The former is controlled by
the layout of the dictionary, while the latter is a sparsity
condition that requires the use of appropriate algorithms.

The constraints imposed by these numbers are supposed to encourage
solutions that will appear meaningful to a human listener.  Good
results can be achieved if both numbers are known and sufficiently
low, but blind separation meets its conceptual limits in case of very
polyphonic works such as orchestral symphonies.  One particularly
difficult instrument would be the pipe organ, where the combination of
organs stops blurs the borders of what should be considered a single
instrument (cf.\ \cite{FletcherRossing12,Barnes64}).

\subsection{Structure of this Paper}

In Section \ref{sec:pursuit}, we propose a novel general-purpose sparse
pursuit algorithm that matches a sampled spectrum with
non-negative continuous patterns.  The algorithm is a modified version
of orthogonal matching pursuit (OMP) \cite{TroppGilbert07} with a
non-linear optimization step for refinement.

In Section \ref{sec:spectrogram}, we use this algorithm in order to
convert an STFT magnitude spectrogram into a wideband pitch-invariant
log-frequency spectrogram.  In Section \ref{sec:dictionary}, we
explain how we use the same algorithm (with slightly different
parameter choices) and a dictionary representation of the harmonics in
order to identify patterns of peaks related to the sounds of musical
instruments in time frames of the spectrogram.  Due to the non-linear
optimization, we can identify the fundamental frequency, the width of
the Gaussian, and the inharmonicity individually for each tone on a
continuous scale.

In Section \ref{sec:algorithm}, we expound the learning algorithm:
For the dictionary update, we employ a modified version of Adam
\cite{KingmaBa14}, which is a popular
stochastic gradient descent algorithm that was initially developed for
the training of deep neural networks.  Our modifications adapt this
algorithm to dictionary learning, preserving the relative scaling of
certain components of the gradient and periodically resetting parts of
the dictionary as needed.  In Section \ref{sec:resynthesis}, we
explain how we use the trained dictionary in order to perform the
separation and obtain the audio signals for the separated
instruments.

In Section \ref{sec:evaluation}, we apply our algorithm on
mixtures that we recorded using acoustic instruments as well as on
samples from the literature.  We evaluate the performance of the
overall algorithm via standard measures and discuss the results.  We
also provide spectrograms of the separation result.

{A pseudo-code implementation of the algorithm
as well as an additional elaboration about the choice of the
time-frequency representation can be found in the appendix (Section
\ref{sec:appendix}).}

\section{Sparse Pursuit Algorithm for Shifted Continuous Patterns}
\label{sec:pursuit}

For both the transformation of the spectrogram and the identification
of instruments inside the spectrogram, we need an algorithm to
approximate a non-negative
discrete mixture spectrum $Y[s]\geq0$, $s\in\Z$, via shifted versions of
continuous patterns $y_{\eta,\theta}(s)\geq0$, $s\in\R$.
{The exact meaning of the variables depends on the specific application,
but in general,}
$\eta\in\{0,\dotsc,N_{\mathrm{pat}}-1\}$ is a discrete index, and
$\theta\in\R^{N_{\mathrm{par}}}$ is a set of continuous, real-valued
parameters. {The fixed values
$N_{\mathrm{pat}},N_{\mathrm{par}}\in\N$ specify the number of patterns and the
number of parameters in $\theta$.}

{Mathematically speaking}, we aim to identify amplitudes $a_j>0$, shifts
$\mu_j$, indices $\eta_j$, and parameter sets $\theta_j$ such that:
\begin{equation}\label{eq:approx}
Y[s]\approx\sum_ja_jy_{\eta_j,\theta_j}(s-\mu_j),
\end{equation}
for $s\in\Z$.

{For a preliminary intuition}, $y_{\eta_j,\theta_j}$ can
be {understood} as the spectrum of the instrument with the number
$\eta_j$, and $\theta_j$ can contain additional parameters
that influence the exact shape of the pattern, like the width of the
peaks and the inharmonicity.

In order to formalize the approximation, we define a
\emph{loss function} to be minimized.  The first natural choice for
such a loss function is the $\ell_2$
distance, but it is not ideal for use in magnitude frequency spectra,
as it focuses very much on the high-volume parts of the spectrum, and
the same applies to other $\ell_p$ (quasi-)distances for $p>0$.

This problem is often approached by use of the $\beta$-divergence {(cf.\ \cite{Fevotte11,Fevotte18})},
which puts a high penalty on “unexplained” peaks in the spectrum.
However, it is asymmetric, and while it is natural in NMF-based
methods, it is difficult to integrate in the algorithm that we
propose.

Instead, we remain with $\ell_2$, but we \emph{lift} low-volume parts
of the spectrum via a concave power function:
\begin{equation}\label{eq:loss}
\begin{gathered}
L\bigl(Y,(a_j),(\mu_j),(\eta_j),(\theta_j)\bigr)=
  \sum_s\Biggl(\bigl(Y[s]+\delta\bigr)^q\\\qquad-\biggl(\delta+\sum_{j}a_jy_{\eta_j,\theta_j}(s-\mu_j)\biggr)^q\Biggr)^2,
\end{gathered}\end{equation}
with $q\in(0,1]$, where $\delta>0$ is a small number merely used to
ensure differentiability.

Furthermore, we impose the \emph{sparsity condition} that every
value of $\eta_j$ may only occur at most {$N_\mathrm{spr}\in\N$}
times in the linear combination.

Minimizing $L$ is a highly non-convex and partly combinatorial
problem, so we cannot hope to reach the perfect solution.
Instead, we follow a \emph{greedy} approach, using ideas from
orthogonal matching pursuit (OMP) \cite{TroppGilbert07} and subspace
pursuit \cite{DaiMilenkovic09}.

\begin{figure*}[t]
  \tikzsetnextfilename{Figure_\the\numexpr(\thefigure+1)\relax}
  \centering
  \footnotesize
  \begin{tikzpicture}
    \begin{groupplot}[group style={group size=2 by 3,
          vertical sep=14mm,horizontal sep=14mm},
      enlargelimits=false,xmin=0,xmax=20,ymin=0,ymax=1,
      width=8.5cm,height=2.7cm,ytick={0,0.5,1},
      title style={yshift=-1.5ex},xlabel={$s$}]
  \nextgroupplot[title={Discrete spectrum $Y[s]$}]
  \addplot[thick,only marks,mark size=1.8pt] table
          {test_pursuit_sample.dat};
  \nextgroupplot[title={Discrete spectrum $Y[s]$ with model reconstruction}]
  \addplot[thick,only marks,mark size=1.8pt] table
          {test_pursuit_sample.dat};
  \addplot[thick] table {test_pursuit_real.dat};
  \nextgroupplot[title={Pattern $y_{0,\theta}(s)$}]
  \addplot[thick] table[y index=1] {test_pursuit_patterns.dat};
  \nextgroupplot[title={Pattern $y_{0,\theta}(s)$, shifted and scaled in amplitude}]
  \addplot[thick] table[y index=1] {test_pursuit_patterns_appl.dat};
  \nextgroupplot[title={Pattern $y_{1,\theta}(s)$}]
  \addplot[thick] table[y index=2] {test_pursuit_patterns.dat};
  \nextgroupplot[title={Pattern $y_{1,\theta}(s)$, shifted and scaled in amplitude}]
  \addplot[thick] table[y index=2] {test_pursuit_patterns_appl.dat};
  \end{groupplot}
  \node[below=1mm] at (group c1r3.outer south) {\sffamily (a) Input};
  \node[below=1mm] at (group c2r3.outer south) {\sffamily (b) Model representation};
  \end{tikzpicture}
  \caption{{Example of the pursuit algorithm applied on a
    spectrum which is a sampled superposition of two shifted patterns.
    The algorithm finds appropriate shifts and amplitudes such that
    the linear combination of the shifted patterns reconstructs the
    spectrum.}}
  \label{fig:pursuit}
\end{figure*}

We start with an empty index set $\mathcal{J}$ and then run the
following steps in a loop:
\begin{enumerate}
\item Compute the (discrete) cross-correlation between the residual
  \begin{equation}\label{eq:residual}
  r[s]=Y[s]^q-\biggl(\sum_{j}a_jy_{\eta_j,\theta_j}(s-\mu_j)\biggr)^q
  \end{equation}
  (i.e., the lifted difference between the raw spectrum and the
  current reconstruction) and the sampled patterns.
  Assume a default parameter set $\theta_{\mathrm{nil}}$,
  {and with}
  \begin{equation}\label{eq:xcorr}
  \rho[\mu,\eta]=\sum_{i}
  \frac{r[i]\,\bigl(y_{\eta,\theta_{\mathrm{nil}}}[i-\mu]\bigr)^q}
       {\bignorm{y_{\eta,\theta_{\mathrm{nil}}}[\cdot]^q}_2},
  \end{equation}
  preselect the {$N_{\textrm{pre}}\in\N$} combinations
  $(\mu,\eta)\in\Z\times\{0,\dotsc,N_{\mathrm{pat}}-1\}$
  with the greatest $\rho[\mu,\eta]$, {equip them with indices}, and add {those} to the index set
  $\mathcal{J}$.  For each preselected pair $(\mu_j,\eta_j)$,
  initialize
  $a_j=(\rho[\mu_j,\eta_j]/\snorm{y_{\eta_j,\theta_{\mathrm{nil}}}[\cdot]^q}_2)^{1/q}$.
  Skip the combinations for which $a_j$ is non-positive.  If none are
  left, terminate.
  
\item Do non-linear optimization on
  $a_j$, $\mu_j$, and $\theta_j$, $j\in\mathcal{J}$, in order to
  minimize $L$, where $a_j\geq0$ and $\theta_j\in\Omega_\theta$
  with $\Omega_\theta\subseteq\R^{N_{\mathrm{par}}}$.

\item For each $\eta=0,\dotsc,N_{\mathrm{pat}}-1$, find
  the indices $j\in\mathcal{J}$ where $\eta_j=\eta$, and remove all
  but those with the $N_{\mathrm{spr}}$ highest amplitudes
  $a_j$ such that, in the end, each pattern $\eta$ is represented at
  most $N_{\mathrm{spr}}$ times in the index set $\mathcal{J}$.

  Re-run the non-linear optimization procedure on the now smaller
  index set $\mathcal{J}$.

\item If the loss $L$ has decreased by less than the factor of
  $1-\lambda$ compared to the previous iteration, with
  $\lambda\in(0,1]$, restore all the values from the
  previous iteration and return them as the result.

  Otherwise, if the count of iterations has reached
  {$N_{\mathrm{itr}}\in\N$}, return the current parameters.  If this is not
  the case, do another iteration.
\end{enumerate}

The hyperparameters $N_{\mathrm{pat}}$ and $N_{\mathrm{spr}}$
{determine the number of given patterns and the maximum number of
times that any pattern can be selected for the representation of a
spectrum.  Both are assumed to be known from}
the application.  For the $q$ exponent, we
usually pick $q=1/2$, as this is the lowest one to keep $L$ convex in
$a_j$, which is beneficial to the optimization procedure.  In some
cases, better results can be achieved by choosing the value of $q$
even lower, but this also increases the chance of divergence.

{Further, the} hyperparameters $\lambda$ and $N_{\mathrm{itr}}$ are
safeguards to limit the runtime of the algorithm, such that the loop
is not run indefinitely with marginal improvement in the non-linear
optimization step.  They also mitigate the problem of overfitting.
The value of $\lambda$ should be chosen slightly below $1$; in
practice, we find that $\lambda=0.9$ yields good results.  We limit
the number of iterations to
$N_{\mathrm{itr}}=2N_{\mathrm{spr}}N_{\mathrm{pat}}$, which is twice
the overall sparsity level.  The loop typically terminates due to
insufficient decrease in $L$, not by exceeding $N_{\mathrm{itr}}$.

{The value for $\theta_{\mathrm{nil}}$ should be determined so
that the point-wise difference
$y_{\eta_j,\theta_j}-y_{\eta_j,\theta_{\mathrm{nil}}}$
is as close to $0$ as possible over a reasonable range of $\theta_j$.
This is because the cross-correlation in \eqref{eq:xcorr} is always
computed using $y_{\eta_j,\theta_{\mathrm{nil}}}$ while the value of
the loss function \eqref{eq:loss} depends on $y_{\eta_j,\theta_j}$.
Thus, if the difference is too large, a suboptimal $\eta_j$ may be
selected.  This especially becomes a problem when inharmonicity is
considered.}

As continuous functions are highly correlated with slightly shifted
versions of themselves, we typically choose $N_{\mathrm{pre}}=1$ in
order to avoid the preselection of the same pattern multiple times for
one feature in the spectrum.

The choice of the non-linear optimization algorithm is not critical,
as long as it supports box
bounds.  We decided to employ the L-BFGS-B algorithm
\cite{Byrd95,Zhu97,Morales11}, which is fast even for high-dimensional
problems.

{Figure~\ref{fig:pursuit} provides an illustrative example of the
sparse pursuit algorithm.  The input is displayed in
Figure~\ref{fig:pursuit}a:  It consists of a discrete spectrum
$Y$ and two continuous patterns $y_{0,\theta},y_{1,\theta}$.
For simplicity, we assume that these patterns are perfectly constant, so
they do not depend on any additional parameters (therefore, $\theta\in\R^0$),
and we set the
exponent {to} $q=1$
(cf.\ \eqref{eq:loss},\eqref{eq:residual},\eqref{eq:xcorr}).
The algorithm selects $\eta_0=0$ and $\eta_1=1$
one after another
and finds appropriate amplitudes $a_0,a_1>0$ and shifts
$\mu_0,\mu_1\in\R$ such that the superposition of these patterns
matches the discrete spectrum $Y$ within numerical precision
($L(Y,a_0,a_1,\mu_0,\mu_1,\eta_0,\eta_1)=0$), as is displayed in Figure~\ref{fig:pursuit}b.}

{The patterns used for this example are purely synthetic,
but similar patterns will in appear
both in the computation of the
pitch-invariant spectrogram and in the separation of the instrument
sounds, and they} could also originate from
other physical phenomena.

\section{Computation of the Pitch-Invariant Spectrogram}
\label{sec:spectrogram}

A spectrogram is a function defined on the time-frequency plane that
is supposed to indicate to what extent a certain frequency is present
in the recording at a given point in time.

The “canonical” time-frequency representation is the spectrogram
obtained from the modulus of the STFT
(cf.\ \cite{Groechenig01}), which is defined via:
\begin{equation}
\mathcal{V}_wX(t,f)=\int_{-\infty}^\infty
X(\tau)\,w(\tau-t)\,\e^{-i2\pi f\tau}\;\dif\tau.
\end{equation}
One particularly popular window with very favorable properties is the
Gaussian:
\begin{equation}
w(t)={\frac{1}{\sqrt{2\pi\zeta^2}}}\exp\bigl(-t^2/(2\zeta^2)\bigr),\qquad\zeta>\R.
\end{equation}

For a sinusoidal signal $X(t)={a}\,\exp(i2\pi\nu t)$
{with amplitude $a\geq0$}, this results in a horizontal
line in the spectrogram:
\begin{equation}\label{eq:stftsin}
\mathcal{V}_wX(t,f)={a}\,\mathcal{F}w(f-\nu)\,\e^{-i2\pi(f-\nu)t},
\end{equation}
and
\begin{equation}\label{eq:fouriergauss}
\mathcal{F}w(f-\nu)=\exp(-(f-\nu)^2/(2\sigma^2))
\end{equation}
with standard deviation $\sigma=1/(2\pi\zeta)$, where $\mathcal{F}$ is the unitary Fourier
transform.  In practice, we use an FFT-computed sampled
version $Z[f,t]=\sabs{\mathcal{V}_wX(t/T,f/F)}$, where $T,F>0$ are
time and frequency units.  While $X$ has a sampling frequency of
$f_{\mathrm{s}}=\SI{48}{kHz}$, we want the time resolution of $Z$ to
be lower by a factor of $256$; thus,
$1/T=256/f_{\mathrm{s}}=5.\overline{3}\,\si{ms}$.
Further, we set $\zeta=1024/f_{\mathrm{s}}$ and cut $w$ at
$\pm6\zeta$, yielding $1/F=f_{\mathrm{s}}/(12\cdot1024)=\SI{3.90625}{Hz}$.

Note that contrary to the definition of the
  spectrogram in \cite{Groechenig01}, we do \emph{not} square the
  magnitude of the STFT, as we
require positive homogeneity:  If the signal $X$ is multiplied by a
positive factor, then we need $Z[f,t]$ to be multiplied by the same
factor.

The problem is that the STFT spectrogram is not
\emph{pitch-invariant}:  We would like a representation where varying
the pitch of the tone of an instrument \emph{shifts} the pattern,
but, for instance, changing the pitch of a tone by an octave
\emph{scales} it by a factor of $2$ on a linear frequency axis,
which is a different distance depending on the original pitch of the
tone.\footnote{When we speak of \emph{pitch}, we refer to the ability
  of one musical instrument to generate tones whose harmonics have the
  same relative amplitudes but different locations on the frequency
  axis, whether that be linear of logarithmic.}

In order to achieve pitch-invariance, one needs a representation with
a logarithmic frequency axis.  However, a naive transform
of the modulus of the STFT would not only influence the position of
the horizontal lines, but also their width.  In order to overcome this
problem, there exist two classical approaches:
\begin{itemize}
\item The mel spectrogram (cf.\ \cite{Vincent18}) performs a
  logarithmic transform on the frequency axis of the STFT spectrogram
  and then applies smoothing along that axis in order to keep the
  widths consistent.  The frequency range that can be represented by
  this approach is limited by the Heisenberg
  uncertainty principle, which states that one cannot have arbitrarily
  good time and frequency resolution at the same time.
\item The constant-Q transform \cite{Brown91} is a discrete wavelet
  transform and can thus be understood as an STFT with differently
  dilated windows for each frequency.  While it keeps the width of the
  horizontal lines constant on a logarithmic frequency axis, the time
  resolution will vary for different frequencies.  This is
  problematic, as it results in simultaneously starting sinusoids
  first appearing in different time frames of the spectrogram.
\end{itemize}

As was shown by \cite{Doerfler17}, the constant-Q transform can be
turned into a mel spectrogram by applying additional smoothing along
the time axis, but it is not possible to overcome the limitations of
the Heisenberg uncertainty principle by classical means.

For {narrowband} signals, this is not a problem; the above methods
can and have been used in order to provide a time-frequency
representation for audio source separation.  However,
as we experimentally show in Section
\ref{sec:mel}, the time-frequency resolution of the mel spectrogram
is too low for the data that we consider, leading to significantly
inferior quality of the separation.  Instead, as we already
have the algorithm from Section~\ref{sec:pursuit} at hand, we can use it
{as an another way to transform the linear-frequency STFT spectrogram
into a pitch-invariant log-frequency spectrogram.  Since this method
gives us sharp frequency values, we are no longer constrained by the
Heisenberg uncertainty principle.  On wideband signals, this
\enquote{super-resolution} gives us an advantage in the subsequent
separation procedure.}

We set $Y=Z[\cdot,t]$ and assume a single Gaussian pattern
\begin{equation}
y_{0,\theta}(s)=\exp\bigl(-s^2/(2F^2\sigma^2)\bigr),\qquad
\theta=(\sigma),
\end{equation}
with $N_{\mathrm{pat}}=1$ and $\theta_{\mathrm{nil}}=(1/(2\pi\zeta))$.

Since the number of Gaussian peaks in a spectrum can be high, we set
$N_{\mathrm{spr}}=\num{1000}$ to make sure they can all be
represented.  This makes the algorithm rather slow, so we choose $q=1$
in order to {bring $L$ closer to a quadratic objective};
as we aim to represent the
spectrum with very low overall error, there is no need to lift certain
features of the spectrum.

To reduce the number of iterations, we also set $N_{\mathrm{pre}}=1000$.
However, this comes with the aforementioned problem that the algorithm
would select a lot of neighboring shifts.  Thus, instead of computing
the cross-correlation, we simply select the $1000$ largest local
maxima of the residual that satisfy $r[i]\geq r[i+k]$ for
$\sabs{k}\leq3$ and assume their heights as initial values for the
amplitudes.

To allow for high-detail representation, we set $\lambda=1$.  The
maximum number of iterations is $N_{\mathrm{itr}}=20$, but the
algorithm often terminates before that. 

After having identified the Gaussian peaks in the sampled STFT
magnitude spectrogram $Z[f,t]$, we resynthesize them in another
magnitude spectrogram $U[\alpha,t]$, applying a logarithmic frequency
transform $\alpha(f)=\alpha_0\,\log_2(f/f_0)$ to the mean frequencies
$\mu_j$, $j\in\mathcal{J}$.  With
$f_0=\SI{20}{Hz}/f_{\mathrm{s}}\cdot12\cdot1024=5.12$ and
$\alpha_0=1024/10=102.4$,
we can, assuming a sampling frequency of $f_{\mathrm{s}}=\SI{48}{kHz}$ and
$\alpha=\{0,\dotsc,1023\}$, represent $10$ octaves from $\SI{20}{Hz}$ to
$\SI{20.48}{kHz}$.

The algorithm can also be used without modification for compact disc
(CD) recordings with a sampling frequency of
$f_{\mathrm{s}}=\SI{44.1}{kHz}$.  In this case, the represented audio
frequency range {consists of the $10$ octaves} from
$\SI{18.375}{Hz}$ to $\SI{18.816}{kHz}$.

\begin{figure}[htbp]
  \tikzsetnextfilename{Figure_\the\numexpr(\thefigure+1)\relax}
  \centering
  \footnotesize
  \begin{tikzpicture}
  \begin{groupplot}[group style={group size=1 by 3,
                                 vertical sep=20mm},
                    enlargelimits=false,
                    axis on top,
                    scale only axis,
                    width=6.50cm,
                    ymode=log,
                    xlabel={Time [$\si{s}$]},
                    ylabel={Frequency [$\si{kHz}$]},
                    ylabel absolute,
                    ylabel style={yshift=-4mm},
                    log ticks with fixed point]
    \nextgroupplot[height=2.6261cm]
    \addplot graphics[xmin=0,xmax=10,ymin=0.53019,ymax=18.816] {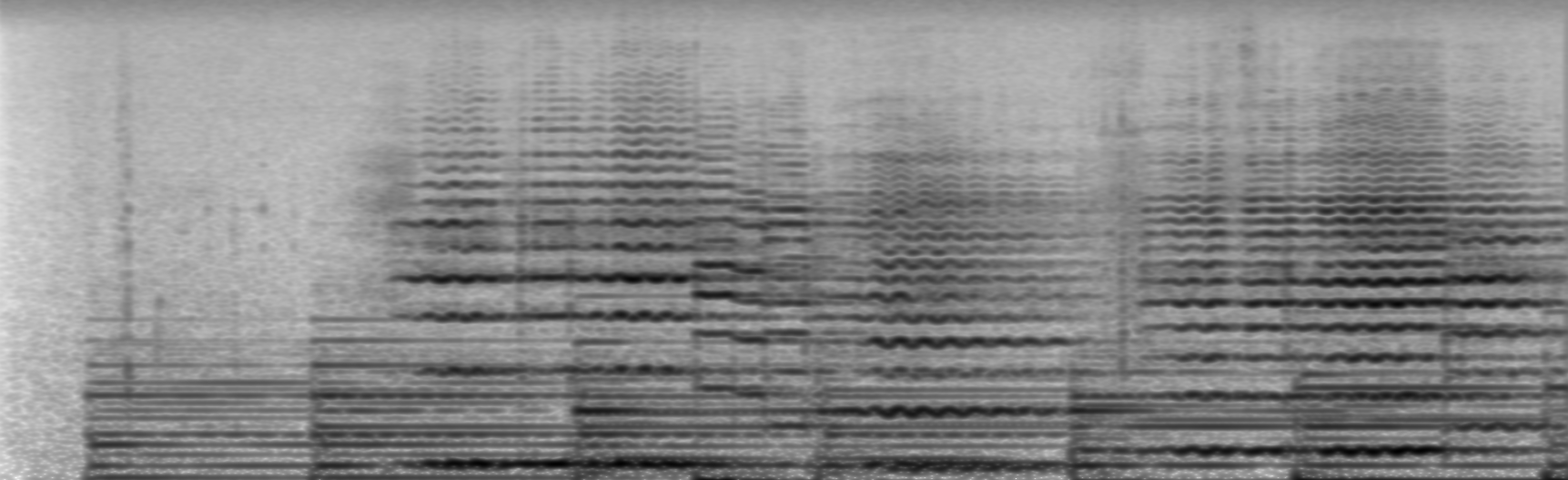};
    \nextgroupplot[height=5.1cm]
    \addplot graphics[xmin=0,xmax=10,ymin=0.018375,ymax=18.816] {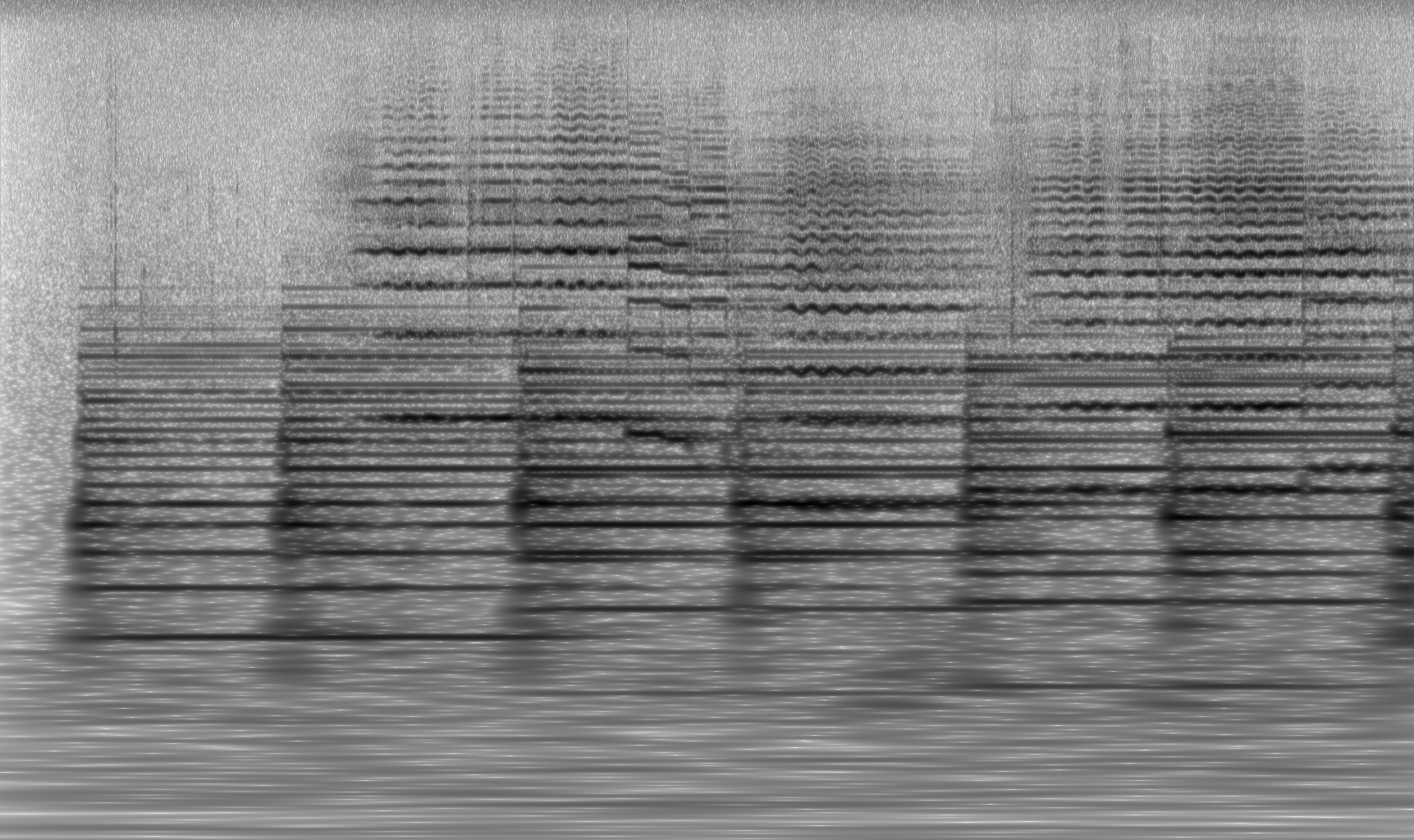};
    \nextgroupplot[height=5.1cm]
    \addplot graphics[xmin=0,xmax=10,ymin=0.018375,ymax=18.816] {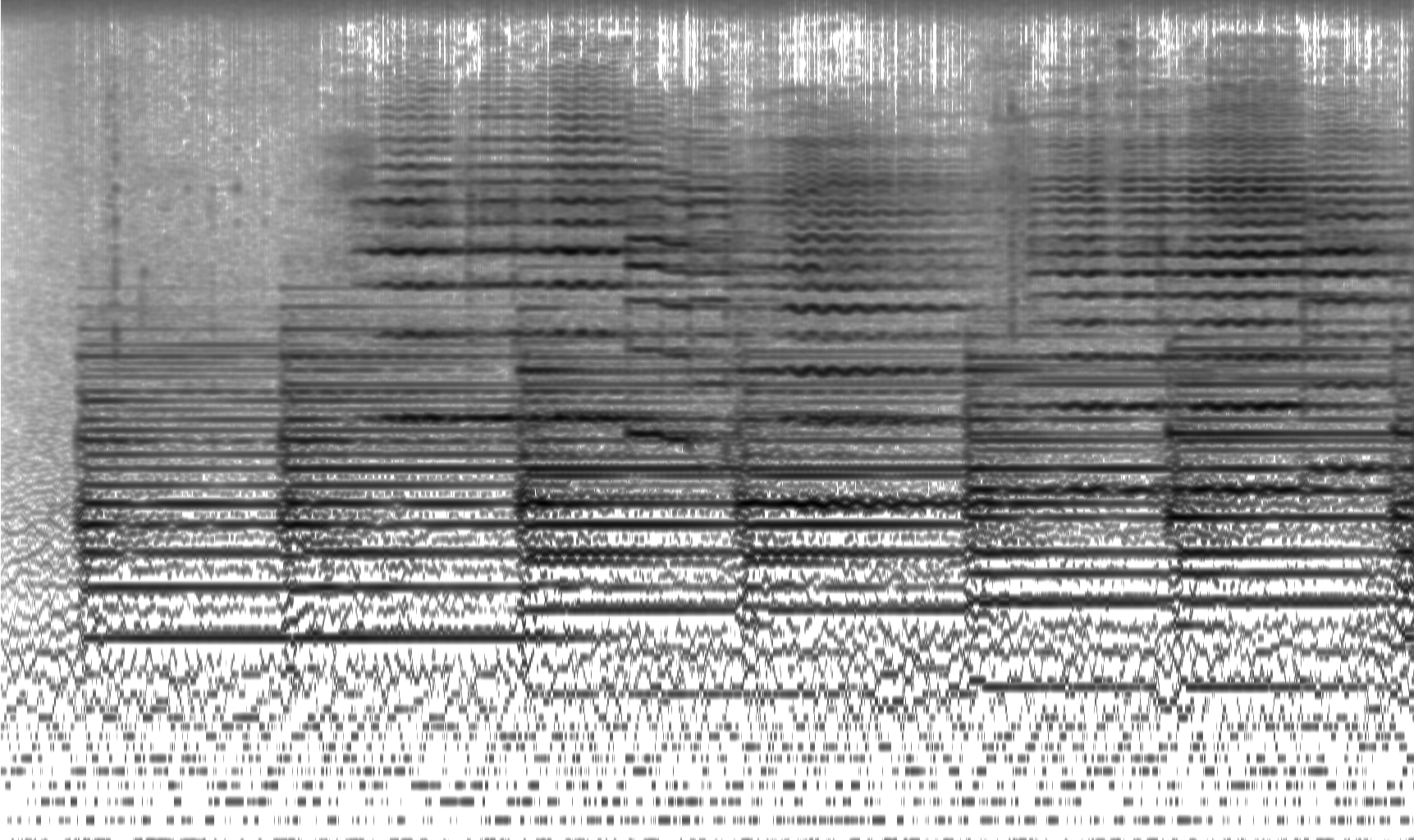};
  \end{groupplot}
  \node[below] at (group c1r1.outer south) {\sffamily (a) Mel spectrogram of the recording};
  \node[below] at (group c1r2.outer south) {\sffamily (b) Constant-Q transform of the recording};
  \node[below] at (group c1r3.outer south) {\sffamily (c) Sparsity-based representation of the recording};
  \end{tikzpicture}
  \caption{Log-frequency spectrograms of the beginning of the 1st
    mvt.\ of the sonata no.\ 1 for violin and piano by Johannes
    Brahms (op.\ 100).  The grayscale axis is logarithmic and
    normalized to a dynamic range of $\SI{100}{dB}$ for each plot.
    Performance by Itzhak Perlman and Vladimir
    Ashkenazy.  Remastered CD recording by EMI Classics, 1999.}
  \label{fig:brahms}
\end{figure}

For Figure~\ref{fig:brahms}, we performed different transforms on an
excerpt of a commercial recording of a piece for violin and piano.
The mel spectrogram in Figure~\ref{fig:brahms}a had to be cut off
at $\SI{530}{Hz}$ in order to
maintain a constant time-log-frequency resolution.  The constant-Q
transform in Figure~\ref{fig:brahms}b can represent lower
frequencies, but its time-log-frequency resolution varies with
frequency:  Clearly, the tones with lower frequencies have a wider time
spread in the representation than those with higher frequencies,
giving an inconsistent image in the individual time frames.

Our proposed sparsity-based transform in
Figure~\ref{fig:brahms}c does not have this problem:  It aligns
the tones properly along the time axis like the mel spectrogram, but
it can represent much lower frequencies.

As our proposed representation is specifically designed for sinusoids,
it largely fails to represent other sounds; in this case, however,
this is even beneficial, as it removes portions of the spectrogram
that do not correspond to the tones that we aim to represent (creating
the white regions in Figure~\ref{fig:brahms}c).  From
this perspective, we can say that it \emph{denoises} the
spectrogram.\footnote{To our separation algorithm, anything
  non-sinusoidal is noise.  This does not imply, however, that these
  parts of the signal are undesirable for a human listener.}

However, it should be kept in mind that the uncertainty principle
cannot be \enquote{tricked} arbitrarily; if two sinusoids have very
low and very similar frequencies, their representations in the STFT
spectrogram will overlap greatly, and our algorithm may fail to tell
them apart.  On the other hand, if a peak is slightly perturbed, it
may also occur that the algorithm will identify one single sinusoid as
two.

Some parts of the noise do get mistaken for sinusoids and are thus
carried over to the log-frequency spectrogram.  In the low
frequencies, this creates the illusion of sparsity in the
log-frequency spectrogram, causing horizontal lines that do not belong
to the music to appear in Figure~\ref{fig:brahms}c.  Their
vertical positions correspond to the transformed frequencies of the
pixels in the linear-frequency spectrogram.  However, we do not
consider these artifacts as a problem from the algorithm, as the noise
was already present in the STFT spectrogram.  Our algorithm merely
creates the white space between the lines.

\section{Model Representation of the Spectrogram}
\label{sec:dictionary}

{In the previous section, we have described how to obtain
a discrete log-frequency spectrogram $U[\alpha,t]$, $\alpha,t\in\Z$
from an audio signal that} contains the superposed sound
of the musical instruments.  {Now, the goal is to} represent
{$U[\alpha,t]$} via a parametric model of the sounds of
the individual instruments, while the parameter values that
characterize the instruments are not known {beforehand}.

A simple model for {the tone production of} many musical instruments
(particularly string and
wind instruments) is the wave equation,
which has sinusoidal solutions (the \emph{harmonics}) at frequencies 
$f_h=hf_1^\circ$, $h=1,\dotsc,N_{\mathrm{har}}$, where $f_1^\circ>0$
is the \emph{fundamental frequency} {and $N_{\mathrm{har}}\in\N$
is the number of harmonics to be considered}.  However, many string instruments
(especially the piano in its high notes) have
non-negligible stiffness in their strings, leading to a fourth-order
equation which has solutions $f_h=(1+bh^2)^{1/2}hf_1^\circ$,
{$h=1,\dotsc,N_{\mathrm{har}}$}, with the
inharmonicity parameter $b\geq0$ (cf.\ \cite{FletcherRossing12}).

Neglecting any negative frequencies, we model
our time-domain signal for the $j$th tone as a linear combination of
complex exponentials:
\begin{equation}
x_j(t)=\sum_{h=1}^{N_{\mathrm{har}}} a_{j,h}
\cdot \e^{i2\pi (f_{j,h}t+\varphi_{j,h})},
\end{equation}
with {amplitudes $a_{j,h}\geq0$ and} phase values $\varphi_{j,h}\in[0,2\pi)$.
{This could locally be
interpreted as an extension of the McAulay-Quatieri model
\cite{McAulayQuatieri86}.}

We assume that the images of these sinusoids
superpose linearly in the spectrograms.  In reality, this is not the
case in the presence of non-constructive interference (\emph{beats}),
but if we accept the error introduced by this common simplification,
we can {set $\varphi_{j,h}=0$,}
apply {\eqref{eq:stftsin} and} \eqref{eq:fouriergauss}{,} and approximate $Z[f,t]$ via:
\begin{equation}\label{eq:fmodel}\begin{aligned}
z[f,t]\coloneqq\sum_{j,h}
a_{j,h,t}\cdot\exp\Biggl(-\frac{\bigl(f-f_{j,h,t}\bigr)^2}{2F^2\sigma_{j,t}^2}\Biggr),
\end{aligned}\end{equation}
where $a_{j,h,t}$ is the amplitude of the $h$th harmonic of the $j$th
tone in the $t$th time frame, and $f_{j,h,t}$ is the respective
frequency.  For the log-frequency spectrogram $U[\alpha,t]$, this
transforms to the following approximation:
\begin{equation}\label{eq:umodel}
\begin{aligned}
u[\alpha,t]\coloneqq\sum_{j,h}
a_{j,h,t}\cdot\exp\Biggl(-\frac{\bigl(\alpha-\alpha_{j,h,t}\bigr)^2}{2F^2\sigma_{j,t}^2}\Biggr),
\end{aligned}\end{equation}
with
$\alpha_{j,h,t}=\alpha(f_{j,h,t})=\alpha((1+b_{j,t}h^2)^{1/2}h)+\alpha(f_{j,1,t}^\circ)$.

We further make the simplifying assumption that the sound of a musical
instrument is constant over the duration of a tone and that the
relation of the amplitudes of the harmonics is constant with respect to pitch
and volume.
We thus save the relative amplitudes of the instruments in a
\emph{dictionary}, which is a matrix
$D\in[0,1]^{N_{\mathrm{har}}\times N_{\mathrm{pat}}}$.
Introducing an overall amplitude $a_{j,t}$ for each tone, we can express
$a_{j,h,t}=D[h,\eta_{j,t}]\,a_{j,t}$, where $\eta_{j,t}$ is the instrument by which
the tone is played.
{For practical acoustic instruments, this assumption is never fully
satisfied, so the deviation between the modeled amplitudes and the
true amplitudes introduces a certain error.  However, we will later
apply a \emph{spectral masking} step (Section \ref{sec:masking}) that
restores the amplitudes of each harmonic directly from the recording
in order to mitigate this error in the final output.}

Our pursuit algorithm can now be applied to \eqref{eq:umodel} by
setting the patterns as:
\begin{equation}\begin{aligned}
&y_{\eta_{j,t},\theta_{j,t}}(\alpha)=\sum_{h}D[h,\eta_{j,t}]\cdot\\
&\quad\exp\Biggl(-\frac{\bigl(\alpha-\alpha((1+b_{j,t}h^2)^{1/2}h)\bigr)^2}{2F^2\sigma_{j,t}^2}\Biggr),
\end{aligned}\end{equation}
with $\theta_{j,t}=(\sigma_{j,t},b_{j,t})$ and
$\mu_{j,t}=\alpha(f_{j,1,t}^\circ)$, according to the notation from
\eqref{eq:approx} with time dependency added.  The
initial value is $\theta_{\mathrm{nil}}=(1/(2\pi\zeta),0)$.

As the patterns now depend on the dictionary, this dependency is
carried over to the loss function \eqref{eq:loss}, which we thus
denote as $L_D$.

\section{Dictionary Learning}
\label{sec:algorithm}

\subsection{Scheme}
\label{sec:scheme}

In order to train the dictionary, we pursue a stochastic
alternating-optimization approach.  First the dictionary is
initialized; for each $\eta=0,\dotsc,N_{\mathrm{pat}}-1$, we generate a
uniformly distributed random vector $d\in[0,1)^{N_{\mathrm{har}}}$ and an
exponent $e$ that is Pareto-distributed with a scale
parameter of $1/2$ (to make sure that $e\geq1$, guaranteeing a minimum
decay rate), and we set $D[h,\eta]=d[h]/h^e$.

Given an initial dictionary, a random time frame $U[\cdot,t]$ of the
log-frequency spectrogram of the recording is chosen, and the sparse
pursuit algorithm is applied on it.  Afterwards, the gradient
$\nabla_DL_D$ of the dictionary-dependent loss function is computed
with the parameters from the sparse pursuit algorithm, and this is
used to update the dictionary in order to reduce the loss.  The
process is repeated {$N_{\mathrm{trn}}\in\N$} times, which is the number of
training iterations as specified by the user.

We set the number of patterns to be generated from the dictionary to
twice the expected number of instruments in the recording
($N_{\mathrm{pat}}=2N_{\mathrm{ins}}$, {$N_{\mathrm{ins}}\in\N$}), allowing for some redundancy
during the training.

\subsection{Dictionary Update}
\label{sec:dictupdate}

Classically, dictionary learning is performed via techniques like NMF
\cite{LeeSeung99,LeeSeung01}, K-SVD \cite{Aharon06}, or tensor
factorization (cf.\ \cite{Chien18}).  However, the first two methods
do not account for
the pitch-invariant structure of
our data.  Tensor factorization
does, but only
for a fixed number of frequency shifts. Moreover, all of these methods
become slow when the amount of data is large.

While the use of stochastic gradient descent for dictionary learning
has been common for many years (cf., e.g., \cite{AharonElad08}), new
methods have been arising very recently due to their applications in
deep learning.  One of the most popular methods for this purpose is
Adam \cite{KingmaBa14}.  Its underlying idea is to treat the gradient
as a random variable and, for each component, compute unbiased
estimates $\hat{v}_1,\hat{v}_2$ for the first and second moments, and
choose the step
size proportional to $\hat{v}_1/\sqrt{\hat{v}_2}$.  If the derivative
of the $i$th
component is constant, then $\hat{v}_1[i]/\sqrt{\hat{v}_2[i]}=\pm1$,
in which case
a large step size can be used.  If the derivative oscillates a lot,
however, then $\hat{v}_1[i]/\sqrt{\hat{v}_2[i]}$ will also be small
and thereby dampen the oscillation in that direction.

The standard formulation of Adam is completely independent of the
scale of the derivatives.  This makes it easy to control the absolute
step size of the components, but it destroys the Landweber
regularization property of gradient descent, which
automatically decreases the step size for components whose partial
derivative is small, taking into account the scaling of different
harmonics.

Our first modification to Adam is that while we still estimate the
first moments for each dictionary entry (i.e., for each instrument
and for each harmonic), we only compute one second moment estimate for
each instrument, which is the arithmetic mean over the all the
estimates for the harmonics.  With this, we restore the regularization
property and prevent excessive change of the components that have small
values. 

Furthermore, we require all entries in the dictionary to be
non-negative, since negative harmonic amplitudes would violate our model assumptions.
For consistency, we also require that no entries be larger than $1$,
so we end up with the box constraint that $D[h,\eta]\in[0,1]$ for
$h=1,\dotsc,N_{\mathrm{har}}$, $\eta=0,\dotsc,N_{\mathrm{pat}}-1$.  To
enforce this, we project each component to $[0,1]$ after the end of
a step.

Finally, we have to tackle the problem that due to the stochastic
nature of the optimization procedure, dictionary entries for a
particular supposed instrument may diverge to a point where they will
not be used by the identification algorithm anymore and thus not
contribute to the separation.  For this purpose, we track the sum
of the amplitudes associated with a specific instrument in the past.
In regular intervals, we sort the
instruments in the dictionary by the ratio of the amplitude sum versus the
number of iterations since its initialization (minus a small head
start that benefits new instrument entries); then, we prune the
dictionary by reinitializing the entries for those supposed
instruments where the ratio is lowest, leaving the $N_{\mathrm{ins}}$
instruments with the highest ratios intact.

{Concerning the parameters for moment estimation and
parameter update in Adam, the default values
(cf.\ Section \ref{sec:pseudocode}) {have turned out to be}
a good choice for the
majority of applications.  In our case, a step-size of
$\kappa=\num{e-3}$ means that if the gradient is constant, the
dominant component will go from $0$ to $1$ in the dictionary within 
less than $1000$ iterations, which is fast enough
if $N_{\mathrm{trn}}\geq10000$.  While lowering $\kappa$ is a common
way to improve training accuracy, this did not appear to have any
effect in our applications.}

\section{Separation and Resynthesis}
\label{sec:resynthesis}

After the dictionary has been trained by alternating between
identification and dictionary update, we represent the entire recording by
running the identification/pursuit algorithm on each time frame
$U[\cdot,t]$ for $t=0,\dotsc,n-1$ (where $n$ is the number of time
frames in the spectrogram) with those $N_{\mathrm{ins}}$
instruments in the dictionary that were left intact after the latest
{pruning}.  This time, however, we need a linear-frequency
spectrogram, {since this is much easier to convert back into
a time-domain signal,} so we apply the reverse transformation
$f(\alpha)=f_0\,2^{\alpha/\alpha_0}$ on the means of the Gaussians
and reconstruct the spectrogram for the $\eta$th instrument via:
\begin{equation}\label{eq:umodelinst}
z_\eta[f,t]\coloneqq\sum_{\substack{j,h\\\eta_{j,t}=\eta}}
a_{j,h,t}\cdot\exp\Biggl(-\frac{\bigl(f-f_{j,h,t}\bigr)^2}{2F^2\sigma_{j,t}^2}\Biggr),
\end{equation}
which is the model from \eqref{eq:fmodel} limited to one instrument.

For the generation of the time-domain signal, we use the classical
algorithm by Griffin and Lim \cite{GriffinLim84}, which iteratively
approximates the signal whose corresponding STFT magnitude spectrogram
is (in the $\ell_2$ sense) closest to the given one.  As initial
value, we give the phase of the STFT of the original signal.

While more sophisticated phase retrieval methods have been developed
recently (e.g., \cite{PfanderSalanevich19}), the
algorithm by Griffin and Lim is well-established, robust, and simple.

\subsection{Spectral Masking}
\label{sec:masking}

As an optional post-processing step, we can mask
the spectrograms from the dictionary representation with the
spectrogram from the original recording.  This
method was proposed in \cite{Jaiswal11a,Jaiswal11b}:
\begin{equation}\label{eq:mask}
\tilde{z}_{\eta}[f,t]\coloneqq
\frac{z_{\eta}[f,t]}{z[f,t]}\cdot Z[f,t].
\end{equation}
In practice, a tiny value is added to the denominator in order to
avoid division by zero.

With this procedure, we make sure that the output spectrograms do
not have any artifacts at frequencies that are not present in the
original recording.  Another benefit is mentioned in
\cite{Jaiswal11a}:  In cases where the sound of an instrument is not
perfectly invariant with respect to pitch and volume, the masking can correct
this.

A potential drawback with masking is that destructive interference in
the original spectrogram may alter the spectrograms of the isolated
instruments.

From a statistical perspective, spectral masking can also be regarded
as a (trivial) \emph{Wiener filter} (cf.\ \cite{Vincent18}).  In this
case, one would regard
the \emph{squared} magnitude spectrograms in the fraction in
\eqref{eq:mask} and treat
them as power spectra that give priors for the frequency distribution
of the signals.  However, we consider this perspective problematic, as
the masks are in fact generated from the data itself, which is already
subject to interference, and squaring the spectra could exacerbate
the error.

\section{Experimental Results and Discussion}
\label{sec:evaluation}

We generate the log-frequency spectrogram as specified in Section
\ref{sec:spectrogram}.
For the dictionary, we use $N_{\mathrm{har}}=25$ harmonics.

\subsection{Performance Measures}
\label{sec:performance}

Vincent et al.\ \cite{Vincent06} define the signal-to-distortion ratio
(SDR), the signal-to-interference ratio (SIR), and the
signal-to-artifacts ratio (SAR).  These {$\ell_2$-based} measures
have become the de facto standard for the performance evaluation of
blind audio source separation.\footnote{{In the meantime, version 3.0
of the \emph{BSS Eval} software package has become available, which
employs a slightly different definition that includes time shifts.
However, for comparability
{with \cite{Jaiswal11a,Jaiswal11b,Jaiswal13,Duan08}},
we are using the original measures as
implemented in version 2.1 \cite{Fevotte05}.}}

The SDR is an “overall” performance measure that incorporates all
kinds of errors in the reconstructed signal; {it yields}
a value of $-\infty$ if the original signal
and the reconstructed signal are uncorrelated.  The SIR is similar,
but it ignores any artifacts that are uncorrelated with the original
signals.  The SAR only measures the artifacts and ignores
interference; it is constant with respect to permutations
of the original signals.
{Those} measures are independent of the scale of
the reconstruction, but they are very sensitive to phase mismatch,
{as the} projection
of a sinusoid on its $\SI{90}{\degree}$-shifted copy will be zero,
even though the signals are otherwise identical.
In order to find the right mapping between the synthesized and the
original signals, the synthesized signals {are permuted} such that
the mean SIR over all instruments is {maximized}.

{Another method for the performance evaluation of audio source
separation is given by the PEASS \cite{Emiya11,Vincent12},
which define the overall perceptual score (OPS), the
target-related perceptual score (TPS), the
interference-related perceptual score (IPS), and the
artifacts-related perceptual score (APS), which are computed using
psychoacoustically motived measures and {were trained via
empirical listening experiments}.
The OPS and IPS correspond conceptually to the SDR and SIR, but the
artifacts as measured via the SAR are subdivided into the
TPS, which accounts for the misrepresentation of the original signal
itself, and the APS, which only comprises the remaining error that
does not originate from misrepresentation or interference.
The values of the scores range from $0$ (worst) to $100$ (best).}

\subsection{Separation of Recorder and Violin Sounds}
\label{sec:separation}

In order to generate a realistic separation
  scenario, we chose the 8th piece from the 12 Basset Horn Duos by
Wolfgang A.\ Mozart (K.\ 487) in an arrangement by Alberto Gomez Gomez
for two recorders\footnote{\RaggedRight%
\url{https://imslp.org/wiki/12_Horn_Duos,_K.487/496a_(Mozart,_Wolfgang_Amadeus)}}.
The upper part was played on a soprano recorder, and the lower part
was played on a violin.  These instruments are easily distinguishable,
as the recorder has an almost sinusoidal sound, while the sound of the
violin is sawtooth-like, with strong harmonics
\cite{FletcherRossing12}.

The instrument tracks were recorded separately {in an apartment
room ($\mathit{RT}_{60}\approx\SI{0.4}{s}$) with an audio recorder at a distance of
approximately \SI{1}{m} to the instrument}, while a
metronome/“play-along” track was provided via headphones.  Evenness of
the tone was favored over musical expression.  We combined the tracks
by adding the two digital signals with no post-processing other than
adjustment of volume and overall timing and let the algorithm run with
$N_{\mathrm{trn}}=100000$ training iterations\footnote{We already achieve
  similarly good performance with $N_{\mathrm{trn}}=10000$ iterations,
but more iterations make the result more consistent with respect to initial
values.}, with $N_{\mathrm{ins}}=2$ and
$N_{\mathrm{spr}}=1$.

This procedure was performed with random seeds $0,\dotsc,9$.  For
comparison, we further applied the algorithm developed in
\cite{Duan08} on our data.  We found that their method
is sensitive with respect to hyperparameters, and we searched for
those values that optimize separation performance for this piece,
but we could only achieve marginal improvement over the defaults
provided in the code.
For application of this algorithm, we downsampled the
audio data to $\SI{22050}{Hz}$, as this is the sampling frequency that
the algorithm was designed to operate on.  The best-case results for
both algorithms are presented in Table~\ref{tab:real}, and the
distribution over all 10 runs of our algorithm is displayed in Figure
\ref{fig:boxplot}.

\begin{table}[t]
  \centering
  \caption{Performance measures for the best-case run of the separation
    of recorder and violin.  Best numbers are marked.}
    \label{tab:real}
    \tabcolsep=4.5pt
    \renewcommand{\arraystretch}{1.1}
    \heavyrulewidth=.5pt
    \lightrulewidth=.4pt
    \cmidrulewidth=.3pt
    \aboverulesep=0.0ex
    \belowrulesep=0.0ex

  \newcolumntype Y{S[
      table-format=2.1,
      table-auto-round,
      table-text-alignment=center,
      table-number-alignment=center,
      table-space-text-post={*}]}
    \begin{tabular}{cccYYY}
    \toprule
    {\bfseries Method} & {\bfseries Mask} & {\bfseries Instrument} & {\bfseries SDR} & {\bfseries SIR} & {\bfseries SAR} \\
    \midrule
    \multirow{4}{*}{Ours} & \multirow{2}{*}{No} & Recorder & 12.8947812 & 32.48386129* & 12.94523547 \\
    & & Violin & 7.09621775 & 24.10629685* & 7.20038094\\
    \cmidrule{2-6}
    & \multirow{2}{*}{Yes} & Recorder & 15.13270565* & 32.38531432 & 15.21774855* \\
    & & Violin & 11.86338139* & 23.7800286 & 12.1702504* \\
    \cmidrule{1-6}
    \multirow{2}{*}{\cite{Duan08}} & \multirow{2}{*}{---} & Recorder & 10.6024 & 21.4279 & 11.0084 \\
    & & Violin & 5.8343 & 18.4353 & 6.1416\\    
    \bottomrule
  \end{tabular}
\end{table}

\begin{figure}[t]
  \tikzsetnextfilename{Figure_\the\numexpr(\thefigure+1)\relax}
  \footnotesize
  \begin{tikzpicture}[mark size=1.6pt,mark=+]
    \begin{groupplot}[group style={group size=1 by 3,
          vertical sep=12mm},
        ytick={1,2,3,4},yticklabels={Recorder\\[-1](unmasked),Recorder\\[-1](masked),Violin\\[-1](unmasked),Violin\\[-1](masked)},yticklabel style={align=right},height=2.8cm,width=5.5cm,
        scale only axis,
        ymin=0.5,ymax=4.5,title style={yshift=-1.5mm},y dir=reverse,ytick style={draw=none}]
      \nextgroupplot[title={SDR},xmin=0]
      \addplot[boxplot={box extend=0.4},mark=none,color=black!50] table[y index={0}] {mozart-nomask-measures.dat};
      \addplot[only marks] table[x index={0},y expr=1] {mozart-nomask-measures.dat};
      \draw[color=black!25] (axis cs:\pgfkeysvalueof{/pgfplots/xmin},1.5) -- (axis cs:\pgfkeysvalueof{/pgfplots/xmax},1.5);
      \addplot[boxplot={box extend=0.4},mark=none,color=black!50] table[y index={0}] {mozart-mask-measures.dat};
      \addplot[only marks] table[x index={0},y expr=2] {mozart-mask-measures.dat};
      \draw[color=black!25] (axis cs:\pgfkeysvalueof{/pgfplots/xmin},2.5) -- (axis cs:\pgfkeysvalueof{/pgfplots/xmax},2.5);
      \addplot[boxplot={box extend=0.4},mark=none,color=black!50] table[y index={1}] {mozart-nomask-measures.dat};
      \addplot[only marks] table[x index={1},y expr=3] {mozart-nomask-measures.dat};
      \draw[color=black!25] (axis cs:\pgfkeysvalueof{/pgfplots/xmin},3.5) -- (axis cs:\pgfkeysvalueof{/pgfplots/xmax},3.5);
      \addplot[boxplot={box extend=0.4},mark=none,color=black!50] table[y index={1}] {mozart-mask-measures.dat};
      \addplot[only marks] table[x index={1},y expr=4] {mozart-mask-measures.dat};
      \nextgroupplot[title={SIR},xmin=20]
      \addplot[boxplot={box extend=0.4},mark=none,color=black!50] table[y index={2}] {mozart-nomask-measures.dat};
      \addplot[only marks] table[x index={2},y expr=1] {mozart-nomask-measures.dat};
      \draw[color=black!25] (axis cs:\pgfkeysvalueof{/pgfplots/xmin},1.5) -- (axis cs:\pgfkeysvalueof{/pgfplots/xmax},1.5);
      \addplot[boxplot={box extend=0.4},mark=none,color=black!50] table[y index={2}] {mozart-mask-measures.dat};
      \addplot[only marks] table[x index={2},y expr=2] {mozart-mask-measures.dat};
      \draw[color=black!25] (axis cs:\pgfkeysvalueof{/pgfplots/xmin},2.5) -- (axis cs:\pgfkeysvalueof{/pgfplots/xmax},2.5);
      \addplot[boxplot={box extend=0.4},mark=none,color=black!50] table[y index={3}] {mozart-nomask-measures.dat};
      \addplot[only marks] table[x index={3},y expr=3] {mozart-nomask-measures.dat};
      \draw[color=black!25] (axis cs:\pgfkeysvalueof{/pgfplots/xmin},3.5) -- (axis cs:\pgfkeysvalueof{/pgfplots/xmax},3.5);
      \addplot[boxplot={box extend=0.4},mark=none,color=black!50] table[y index={3}] {mozart-mask-measures.dat};
      \addplot[only marks] table[x index={3},y expr=4] {mozart-mask-measures.dat};
      \nextgroupplot[title={SAR},xmin=0]
      \addplot[boxplot={box extend=0.4},mark=none,color=black!50] table[y index={4}] {mozart-nomask-measures.dat};
      \addplot[only marks] table[x index={4},y expr=1] {mozart-nomask-measures.dat};
      \draw[color=black!25] (axis cs:\pgfkeysvalueof{/pgfplots/xmin},1.5) -- (axis cs:\pgfkeysvalueof{/pgfplots/xmax},1.5);
      \addplot[boxplot={box extend=0.4},mark=none,color=black!50] table[y index={4}] {mozart-mask-measures.dat};
      \addplot[only marks] table[x index={4},y expr=2] {mozart-mask-measures.dat};
      \draw[color=black!25] (axis cs:\pgfkeysvalueof{/pgfplots/xmin},2.5) -- (axis cs:\pgfkeysvalueof{/pgfplots/xmax},2.5);
      \addplot[boxplot={box extend=0.4},mark=none,color=black!50] table[y index={5}] {mozart-nomask-measures.dat};
      \addplot[only marks] table[x index={5},y expr=3] {mozart-nomask-measures.dat};
      \draw[color=black!25] (axis cs:\pgfkeysvalueof{/pgfplots/xmin},3.5) -- (axis cs:\pgfkeysvalueof{/pgfplots/xmax},3.5);
      \addplot[boxplot={box extend=0.4},mark=none,color=black!50] table[y index={5}] {mozart-mask-measures.dat};
      \addplot[only marks] table[x index={5},y expr=4] {mozart-mask-measures.dat};
    \end{groupplot}
  \end{tikzpicture}
  \caption{Distribution of the performance measures of the separation
    of violin and piano over 10 runs, without and with spectral masking}
  \label{fig:boxplot}
\end{figure}

Our criterion for the best run {in our algorithm}
was the mean SDR over both
instruments.  This was achieved by a random seed of
$7$ for this sample.  When the algorithm is used in a real-world
scenario in which the original tracks are not available, the
performance measures are unknown to the user.  In this case, the user
can select the \enquote{best-sounding} result from all $10$
candidates, perhaps guided by the value of the loss function as a
proxy measure.  {The notion of ensemble learning does not apply to
the algorithm in \cite{Duan08}, as it is a clustering method and does
not have an initial dictionary.  Instead, we there consider the result
that we achieve with the hand-optimized parameters as best-case.}

With our algorithm, the recorder is universally better represented than
the violin, and spectral masking leads to considerable improvements in
SDR and SAR especially for the violin.  This complies with the
explanation in \cite{Jaiswal11a} that spectral masking helps represent
instruments with more diverse spectra, such as the
violin, which has 4 different strings and a sound that is very
sensitive to technique.
{When we compare the outcomes in pairs without and with spectral
masking over the random seeds $0,\dotsc,9$ respectively, the}
improvement in SDR achieved by spectral masking is
statistically significant at
$p_{\mathrm{Recorder}}=p_{\mathrm{Violin}}=\num{9.8e-4}$ in a
one-sided Wilcoxon signed-rank test \cite{R}\footnote{%
  Briefly speaking, the Wilcoxon signed-rank test has the null
  hypothesis that the differences in the pairs are symmetrically
  distributed around $0$.  For this, the sum of the signed ranks of
  the differences is computed.  In the one-sided test, the acceptance
  region for this sum is asymmetric.}, as for each dictionary,
spectral masking leads to a consistent improvement of the separation
result.

The algorithm from \cite{Duan08} reacts in a similar way, yielding
better performance for the recorder than for the violin.  However, the
working principle is different:  Rather than trying to represent
both instruments, it clusters the peaks from the spectrum in order to
make out a \enquote{dominant} instrument, while the second
\enquote{instrument} is just the collection of residual peaks.  In our
example, the violin was identified as the dominant instrument, but
nonetheless the representation of the recorder is better.
However, our algorithm provides superior performance for both
instruments, even without spectral masking.

For phase reconstruction, we used merely one iteration (i.e., only one
magnitude adjustment and one projection) of the
Griffin-Lim algorithm in order to preserve the phase of the
original spectrogram as much as possible.

The aural impression of the results with different random seeds is
largely very similar.  While some artifacts and interference are
audible, the generated audio data provides a good aural representation
of the actually played tracks.  The only tone\footnote{which occurs 4
  times in total,
  due to repetitions of the passage} that is
misidentified over a long period of time is a recorder tone that
interferes with the
even-numbered harmonics of the violin tone that is played at the same
time and is one octave lower.  In this case, the third harmonic of the
violin tone is erroneously identified as the recorder tone.

\begin{table}[t]
  \centering
  \caption{{PEASS scores for the best-case run of the separation
    of recorder and violin.  Best numbers are marked.}}
    \label{tab:realpeass}
    \tabcolsep=4.5pt
    \renewcommand{\arraystretch}{1.1}
    \heavyrulewidth=.5pt
    \lightrulewidth=.4pt
    \cmidrulewidth=.3pt
    \aboverulesep=0.0ex
    \belowrulesep=0.0ex

  \newcolumntype Y{S[
      table-format=2,
      table-auto-round,
      table-text-alignment=center,
      table-number-alignment=center,
      table-space-text-post={*}]}
  \newcolumntype Z{S[
      table-format=3,
      table-auto-round,
      table-text-alignment=center,
      table-number-alignment=center,
      table-space-text-post={*}]}
    \begin{tabular}{cccYZYY}
    \toprule
    {\bfseries Method} & {\bfseries Mask} & {\bfseries Inst.} & {\bfseries OPS} & {\bfseries TPS} & {\bfseries IPS} & {\bfseries APS}\\
    \midrule
    \multirow{4}{*}{Ours} & \multirow{2}{*}{No} & Rec. & 34.1606* & 30.8666 & 69.6187* & 38.2803* \\
    & & Vln. & 34.3471* & 27.0487 & 70.7032* & 36.9768* \\
    \cmidrule{2-7}
    & \multirow{2}{*}{Yes} & Rec. & 24.6135 & 63.7181 & 38.5417 & 38.2138 \\
    & & Vln. & 13.4361 & 99.5885* & 32.5808 & 52.4661 \\
    \cmidrule{1-7}
    \multirow{2}{*}{\cite{Duan08}} & \multirow{2}{*}{---} & Rec. & 27.9334 & 84.2784* & 26.0979 & 35.2104 \\
    & & Vln. & 31.7427 & 19.0793 & 70.6994 & 30.3333 \\    
    \bottomrule
  \end{tabular}
\end{table}

{The PEASS scores for the same runs and parameters are given in
Table~\ref{tab:realpeass}.  Surprisingly, the results without spectral masking
are now mostly preferred.  Our only explanation is that as discussed in
Section~\ref{sec:masking}, spectral masking can cause interference
in overlapping tones, as can be seen in the drop in both SIR and IPS.
While the SDR still increases overall with spectral masking, this
interference might have a large negative impact on the OPS.
However, we did not find this discrepancy in
most of the other samples, so it does not appear to be a general pattern.}

\begin{figure}[htbp]
  \tikzsetnextfilename{Figure_\the\numexpr(\thefigure+1)\relax}
  \centering
  \footnotesize
  \begin{tikzpicture}
  \begin{groupplot}[group style={group size=1 by 3,
                                 vertical sep=20mm},
                    enlargelimits=false,
                    axis on top,
                    scale only axis,
                    width=6.6cm,
                    height=5.1cm,
                    ymode=log,
                    xlabel={Time [$\si{s}$]},
                    ylabel={Frequency [$\si{kHz}$]},
                    log ticks with fixed point]
    \nextgroupplot
    \addplot
    graphics[xmin=0,xmax=8.1920,ymin=0.02,ymax=20.48] {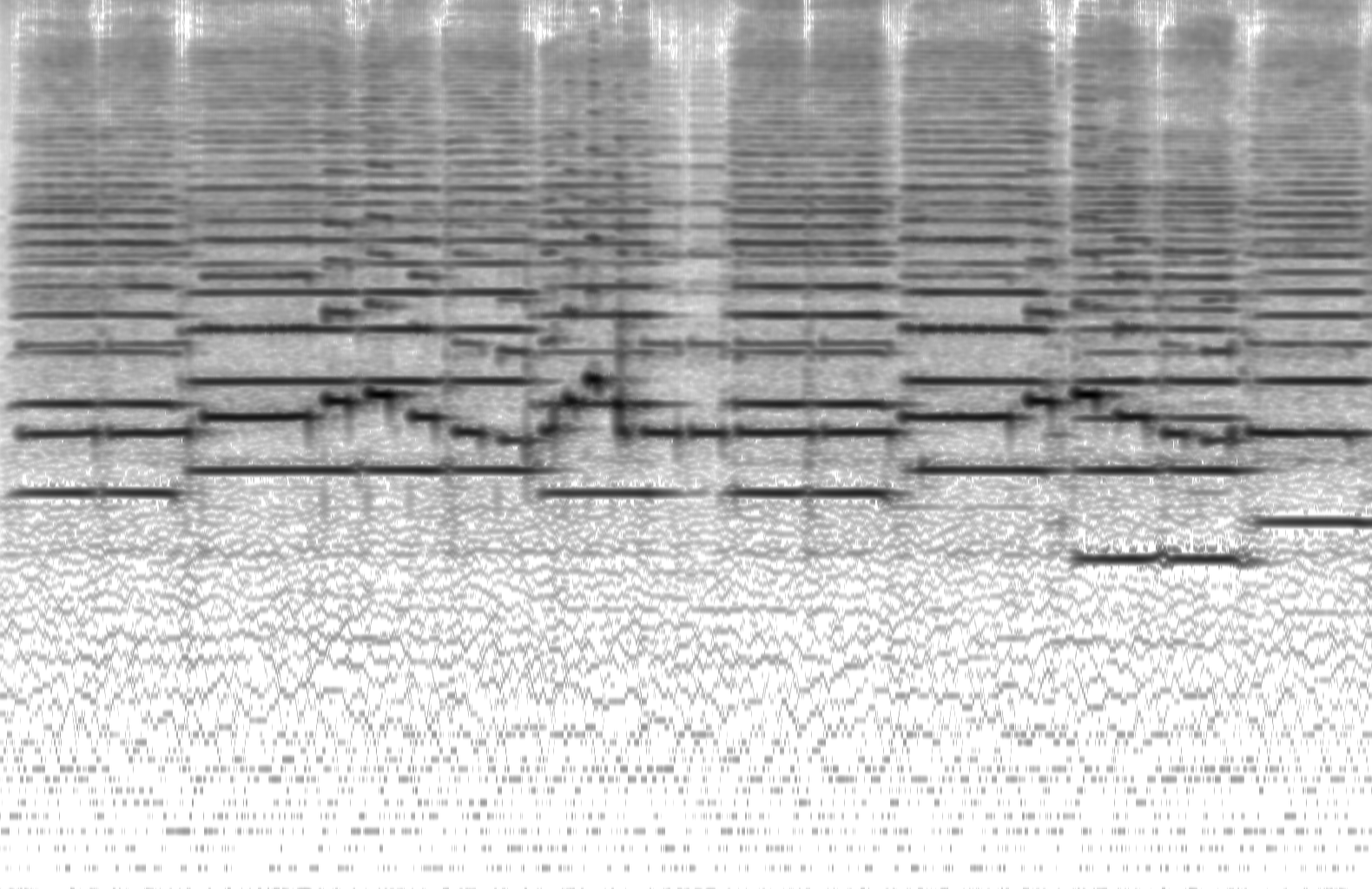};
    \nextgroupplot
    \addplot
    graphics[xmin=0,xmax=8.1920,ymin=0.02,ymax=20.48] {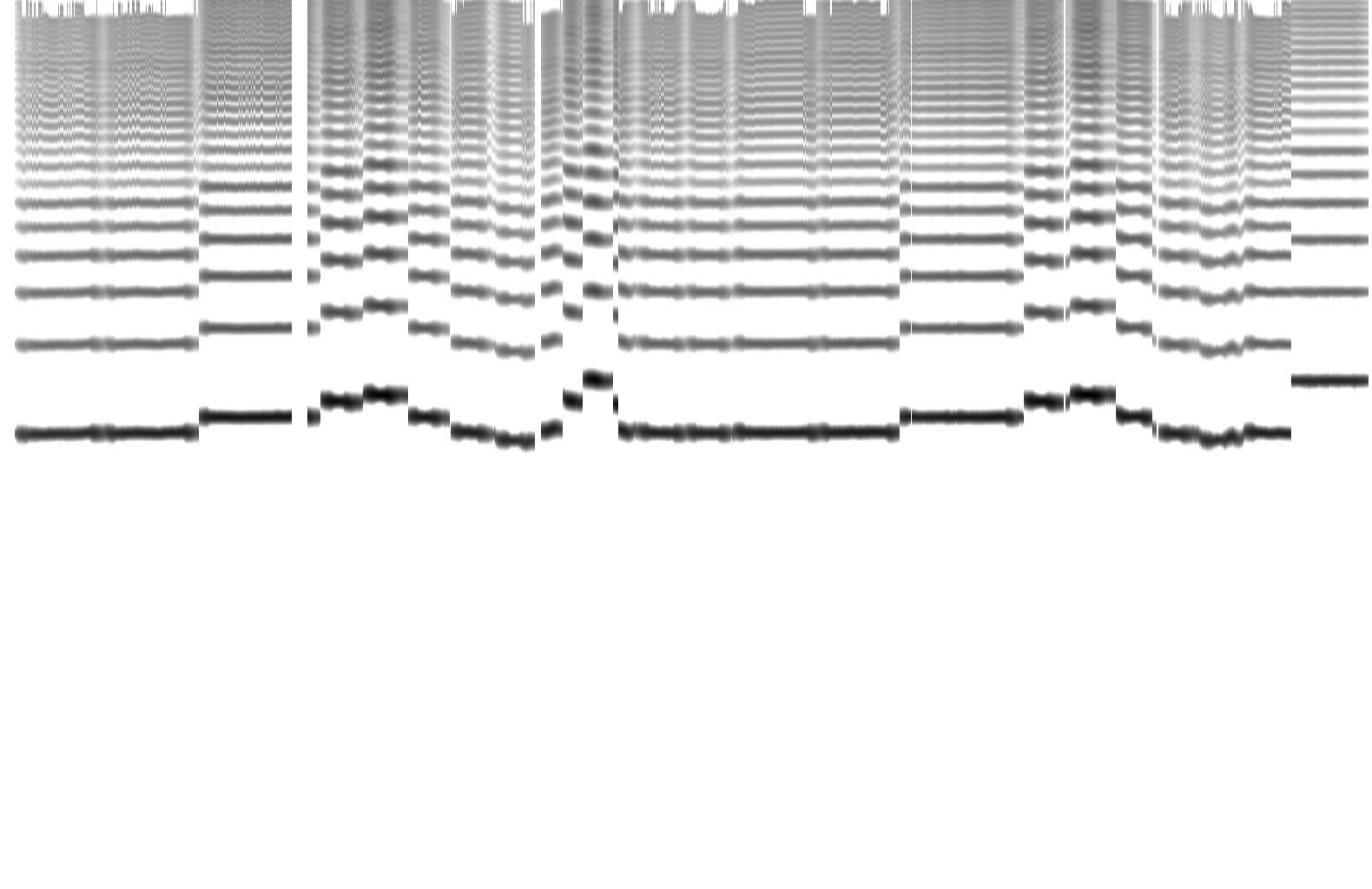};
    \nextgroupplot
    \addplot
    graphics[xmin=0,xmax=8.1920,ymin=0.02,ymax=20.48] {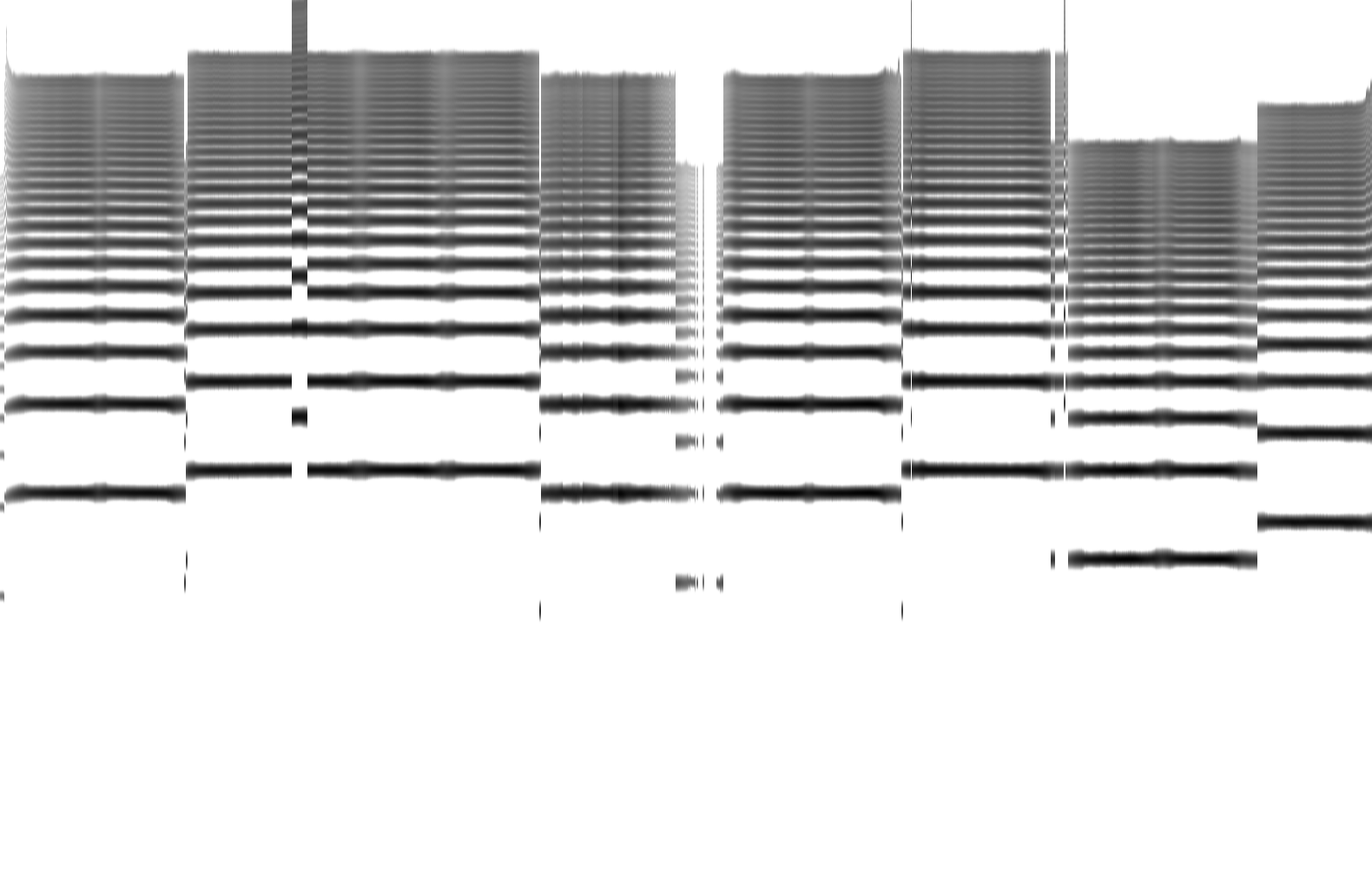};
  \end{groupplot}
  \node[below] at (group c1r1.outer south) {\sffamily (a) Original, generated via the sparse pursuit method};
  \node[below] at (group c1r2.outer south) {\sffamily (b) Synthesized recorder track};
  \node[below] at (group c1r3.outer south) {\sffamily (c) Synthesized violin track};
  \end{tikzpicture}
  \caption{Log-frequency spectrograms for beginning of the recorded piece and the
    synthesized tracks.
    The grayscale axis is logarithmic and normalized to a dynamic
    range of $\SI{100}{dB}$ for each plot.}
  \label{fig:spect}
\end{figure}

Spectrograms of the original recording and the synthesized
representations (with the random seed of $7$ that maximizes the SDR)
are displayed in Figure~\ref{fig:spect}.  The original
spectrogram contains broad-spectrum components (“noise”) that do not
fit the dictionary model and thus cannot be represented, so they are
not found in the output spectrograms.  The choice of
$N_{\mathrm{har}}=25$ must be regarded as a compromise:  Although the
sound of the violin could be represented more accurately with an even
higher numbers of harmonics, this would increase both the computation
time of the algorithm and also the number of variables to be trained.
The incorrectly identified recorder tone corresponds to the rightmost
set of horizontal lines
in Figure~\ref{fig:spect}b.  It is not audible when the
synthesized audio files are mixed back together.

Since spectral masking is only applied on the linear-frequency
spectrograms, its effects cannot be seen in Figure~\ref{fig:spect}.

\subsection{Separation of Clarinet and Piano Sounds}
\label{sec:clarinet}

We recorded the same piece on clarinet and piano {using the same
set-up} as for recorder and violin, {except that the instruments
were played in a rehearsal hall ($\mathit{RT}_{60}\approx\SI{1.4}{s}$)}.
{The algorithm was also run} under the same
conditions.  The
distribution of the results over random seeds $0,\dotsc,9$ is displayed
in Figure \ref{fig:boxplot-cl}.
The best-case results of our algorithm with a random seed value of $6$
as well as those for the algorithm from \cite{Duan08} (with again, the
data downsampled to $\SI{22050}{Hz}$) are presented in
Table \ref{tab:cl-piano}.

The separation quality with our algorithm is much worse than the for
recorder and violin, and representation of the piano is especially
problematic.  We have several explanations for this:

\begin{enumerate}
\item {Piano} tones
  exhibit non-negligible inharmonicity, which makes it harder to
  identify them in the spectrum.  Even though our model incorporates
  this inharmonicity, cross-correlation does not.
\item Compared to the rather steady tone of recorder, violin, and
  clarinet, the piano tone has a very characteristic onset
  (\emph{attack}), which exhibits different spectral characteristics
  than the rest of the tone.
\end{enumerate}

\begin{table}[h]
  \centering
  \caption{Performance measures for the best-case run of the separation
    of clarinet and piano, with spectral masking.  Best numbers are marked.}
  \label{tab:cl-piano}
  \tabcolsep=4.5pt
  \renewcommand{\arraystretch}{1.1}
  \heavyrulewidth=.5pt
  \lightrulewidth=.4pt
  \cmidrulewidth=.3pt
  \aboverulesep=0.0ex
  \belowrulesep=0.0ex

  \newcolumntype Y{S[
      table-format=2.1,
      table-auto-round,
      table-text-alignment=center,
      table-number-alignment=center,
      table-space-text-post={*}]}
    \begin{tabular}{ccYYY}
      \toprule
    {\bfseries Method} & {\bfseries Instrument} & {\bfseries SDR} & {\bfseries SIR} & {\bfseries SAR} \\\midrule
    \multirow{2}{*}{Ours} & Clarinet & 4.05190929 & 24.30443735* & 4.10916925 \\
    & Piano & 2.11202768 & 9.28523716 & 3.52042693 \\
    \cmidrule{1-5}
    \multirow{2}{*}{\cite{Duan08}} & Clarinet & 6.7415* & 21.3302 & 6.9270* \\
    & Piano & 5.4587* & 16.3777* & 5.9240* \\
    \bottomrule
  \end{tabular}
\end{table}

\begin{figure}[p]
  \tikzsetnextfilename{Figure_\the\numexpr(\thefigure+1)\relax}
  \footnotesize
  \begin{tikzpicture}[mark size=1.6pt,mark=+]
    \begin{groupplot}[group style={group size=1 by 3,
          vertical sep=12mm},
        ytick={1,2},yticklabels={Clarinet,Piano},height=1.4cm,width=5.5cm,scale only axis,
        ymin=0.5,ymax=2.5,title style={yshift=-1.5mm},y dir=reverse,ytick style={draw=none}]
      \nextgroupplot[title={SDR}]
      \addplot[boxplot={box extend=0.4},mark=none,color=black!50] table[y index={0}] {mozart-cl-measures.dat};
      \addplot[only marks] table[x index={0},y expr=1] {mozart-cl-measures.dat};
      \draw[color=black!25] (axis cs:\pgfkeysvalueof{/pgfplots/xmin},1.5) -- (axis cs:\pgfkeysvalueof{/pgfplots/xmax},1.5);
      \addplot[boxplot={box extend=0.4},mark=none,color=black!50] table[y index={1}] {mozart-cl-measures.dat};
      \addplot[only marks] table[x index={1},y expr=2] {mozart-cl-measures.dat};
      \nextgroupplot[title={SIR},xmin=0]
      \addplot[boxplot={box extend=0.4},mark=none,color=black!50] table[y index={2}] {mozart-cl-measures.dat};
      \addplot[only marks] table[x index={2},y expr=1] {mozart-cl-measures.dat};
      \draw[color=black!25] (axis cs:\pgfkeysvalueof{/pgfplots/xmin},1.5) -- (axis cs:\pgfkeysvalueof{/pgfplots/xmax},1.5);
      \addplot[boxplot={box extend=0.4},mark=none,color=black!50] table[y index={3}] {mozart-cl-measures.dat};
      \addplot[only marks] table[x index={3},y expr=2] {mozart-cl-measures.dat};
      \nextgroupplot[title={SAR},xmin=0]
      \addplot[boxplot={box extend=0.4},mark=none,color=black!50] table[y index={4}] {mozart-cl-measures.dat};
      \addplot[only marks] table[x index={4},y expr=1] {mozart-cl-measures.dat};
      \draw[color=black!25] (axis cs:\pgfkeysvalueof{/pgfplots/xmin},1.5) -- (axis cs:\pgfkeysvalueof{/pgfplots/xmax},1.5);
      \addplot[boxplot={box extend=0.4},mark=none,color=black!50] table[y index={5}] {mozart-cl-measures.dat};
      \addplot[only marks] table[x index={5},y expr=2] {mozart-cl-measures.dat};
    \end{groupplot}
  \end{tikzpicture}
  \caption{Distribution of the performance measures of the separation
    of clarinet and piano over 10 runs, with spectral masking}
  \label{fig:boxplot-cl}
\end{figure}

{This raises the question whether our algorithm can represent
piano tones at all.  In order to test this, we ran it on the original
piano track.
The result was very stable, with a maximum
SDR of $\SI{8.7}{dB}$ for a random seed of $9$ -- without spectral
masking, as this would not make sense for a single instrument.
In Figure~\ref{fig:piano}, we show a time frame from the spectrogram
within the first tone of the piano ($t=100$).  The fundamental
frequency
was identified by the algorithm as $f_1^\circ=\SI{441.8}{Hz}$ and the
inharmonicity as $b=\num{5.3e-4}$.  In Figure~\ref{fig:piano}a,
the original spectrum is displayed with the predicted frequencies of the
harmonics when inharmonicity is neglected, and the deviation upward
from the 5th harmonic becomes clearly recognizable.
In Figure~\ref{fig:piano}b,
the computed inharmonicity is incorporated, and so the predicted
frequencies of the harmonics match those from the original spectrum
almost perfectly.  Figure~\ref{fig:piano}c represents the {reconstructed}
spectrogram time frame as returned by the separation algorithm
with all the other parameters considered, but without spectral
masking.}

\begin{figure}[p]
  \tikzsetnextfilename{Figure_\the\numexpr(\thefigure+1)\relax}
  \footnotesize
  \centering
  \begin{tikzpicture}\relax
    \begin{groupplot}[group style={group size=1 by 3,vertical sep=20mm},
        height=3cm,width=8.2cm,ymode=log,enlargelimits=false,restrict x to domain=0:10,xmin=0,xmax=10,ymax=1e-1,ymin=1e-6,xlabel={$f$ [$\si{kHz}$]}]
    \nextgroupplot
    \addplot[ycomb,line width=0.35pt,color=gray] table[x index=0,y expr=1e-8]
            {mozart-100-lincomb.dat};
    \addplot[line width=0.5pt] table
            {mozart-mock-orig-100.dat};
    \nextgroupplot
    \addplot[ycomb,line width=0.35pt,color=gray] table[x index=0,y expr=1e-8]
            {mozart-100-spreadcomb.dat};
    \addplot[line width=0.5pt] table
            {mozart-mock-orig-100.dat};
    \nextgroupplot
    \addplot[line width=0.5pt] table
            {mozart-mock-9-synth-lin-nomask-100.dat};
    \end{groupplot}
    \node[below,align=left] at (group c1r1.outer south) {\sffamily (a)
      Original spectrum and predicted harmonics\\\phantom{(a) }without inharmonicity};
    \node[below,align=left] at (group c1r2.outer south) {\sffamily (b)
      Original spectrum and predicted harmonics\\\phantom{(b) }with inharmonicity};
    \node[below] at (group c1r3.outer south) {\sffamily (c)
      Predicted spectrum};
  \end{tikzpicture}
  \caption{{Model representation of a piano tone (a') with
      the parameters identified by the separation algorithm when
      run on the pure piano track}}
  \label{fig:piano}
\end{figure}

{Thus, our algorithm does not have any issue \emph{representing}
the piano tones; the difficulty in this case is to identify them in
the presence of the clarinet tones.}

The algorithm from \cite{Duan08} performs comparatively well.  This is,
again, due to the different approach:  Rather than trying to represent
both instruments, {this} algorithm only finds the clarinet tones as the
dominant cluster and assigns the remaining parts of the spectrum to
the piano.  Thus, even though {their} model cannot represent the
piano, as
it does not include inharmonicity at all, it can still separate it
under the assumption that the clarinet is modeled correctly.
{However, for this recording, it is essential to hand-tune the
hyperparameters:  Those that were used for the separation of
recorder and violin still work reasonably well for clarinet and piano,
but with the default values, the algorithm fails.}

{In terms of the PEASS (Table~\ref{tab:cl-pianopeass}), the results of
both algorithms achieve very similar overall scores.  While our reconstruction
of the piano sound is inferior in terms of TPS, the interference and artifacts
are evaluated as perceptually less severe.}

\begin{table}[htbp]
  \centering
  \caption{{PEASS scores for the best-case run of the separation
    of clarinet and piano, with spectral masking.  Best numbers are marked.}}
  \label{tab:cl-pianopeass}
  \tabcolsep=4.5pt
  \renewcommand{\arraystretch}{1.1}
  \heavyrulewidth=.5pt
  \lightrulewidth=.4pt
  \cmidrulewidth=.3pt
  \aboverulesep=0.0ex
  \belowrulesep=0.0ex

  \newcolumntype Y{S[
      table-format=2,
      table-auto-round,
      table-text-alignment=center,
      table-number-alignment=center,
      table-space-text-post={*}]}
    \begin{tabular}{ccYYYY}
      \toprule
    {\bfseries Method} & {\bfseries Instrument} & {\bfseries OPS} & {\bfseries TPS} & {\bfseries IPS} & {\bfseries APS} \\\midrule
    \multirow{2}{*}{Ours} & Clarinet & 39.4241* & 55.8897* & 68.2744* & 45.8562* \\
    & Piano & 25.4503 & 37.0007 & 58.6843* & 30.8109* \\
    \cmidrule{1-6}
    \multirow{2}{*}{\cite{Duan08}} & Clarinet & 38.5320 & 45.7046 & 62.1917 & 41.3190 \\
    & Piano & 25.6517* & 86.6133* & 29.9177 & 21.9864 \\
    \bottomrule
  \end{tabular}
\end{table}

\subsection{Generalization Experiment}

Usually, we train our dictionary on the same audio recording that we
aim to separate.  In this experiment, however, our goal is to
ascertain whether a dictionary that was trained on one recording can
be used for the separation of another recording without additional
training.

{Under the recording conditions specified in Section~\ref{sec:separation}},
we recorded the traditional tune \enquote{Frère Jacques} with
B$\flat$ tin whistle and viola in the key of E$\flat$ major as well
as with C tin whistle and violin in the key of F major.  The violin
and viola were offset by two bars compared to the tin whistles in
order to create a musically realistic \emph{canon}.  The lowest
frequency of the B$\flat$ tin whistle was measured as $\SI{463}{Hz}$,
and the lowest frequency of the C tin whistle was measured as
$\SI{534}{Hz}$. Thus, they do \emph{not} fit in the same
equal-temperament tuning, and the intervals on these instruments are
not very consistent, either. Their tuning was mimicked by ear when
recording the viola/violin tracks.

First, the separation was performed with random seeds $0,\dotsc,9$ on
the recording with B$\flat$ tin whistle and viola.  Then the
dictionaries obtained from this separation were used on the
recording with C tin whistle and violin without any further training.
The experiment was repeated vice versa with the recordings permuted.

For the viola and B$\flat$ whistle combination, the dictionary from
the run with seed $8$ was optimal, but that from a seed of
$0$ was best when applying on the violin and C whistle recording.
Vice versa, when training on the violin and C whistle recording, the
seed of $0$ was also ideal for separation of that recording, but the
dictionary from a random seed of $2$ was better when applying on the
B$\flat$ whistle and viola recording.  All the best-case number are
presented in Table \ref{tab:gen}.

\begin{table}[htbp]
  \centering
  \caption{Performance measures for the best-case run of the separation
    of B$\flat$/C tin whistle and viola/violin, with spectral
    masking.  Results indicated as \enquote{Orig.} were generated from
    the dictionary that was trained on that recording, while
    \enquote{Gen.} means that the dictionary was trained on the other
    recording.  Best numbers are marked.}
  \label{tab:gen}
  \tabcolsep=4.5pt
  \renewcommand{\arraystretch}{1.1}
  \heavyrulewidth=.5pt
  \lightrulewidth=.4pt
  \cmidrulewidth=.3pt
  \aboverulesep=0.0ex
  \belowrulesep=0.0ex

  \newcolumntype Y{S[
      table-format=2.1,
      table-auto-round,
      table-text-alignment=center,
      table-number-alignment=center,
      table-space-text-post={*}]}
    \begin{tabular}{ccYYY}
      \toprule
    {\bfseries Mode} & {\bfseries Instrument} & {\bfseries SDR} & {\bfseries SIR} & {\bfseries SAR} \\\midrule
    \multirow{4}{*}{Orig.} & Tin whistle B$\flat$ & 14.97735648 & 29.25793458 & 15.14768632 \\
    & Viola & 10.47730705 & 26.8753361 & 10.58691021 \\
    \cmidrule{2-5}
    & Tin whistle C & 17.12817229* & 27.02304012 & 17.60619814* \\
    & Violin & 12.07540245* & 36.42649312* & 12.09236741* \\
    \cmidrule{1-5}
    \multirow{4}{*}{Gen.} & Tin whistle B$\flat$ & 15.94490565* & 30.02940108* & 16.12218056* \\
    & Viola & 11.17222528* & 28.37305596* & 11.26207277* \\
    \cmidrule{2-5}
    & Tin whistle C & 16.70059649 & 27.25414223* & 17.10897029 \\
    & Violin & 11.59176913 & 34.52949045 & 11.61543635 \\
    \bottomrule
  \end{tabular}
\end{table}

Overall, the performance figures are similar to those from recorder
and violin, as could be expected because those are similar
instruments.  To our surprise, the performance in the generalization
even sometimes exceeds that from direct training and separation.

{For a better analysis, we} gathered the data from seeds $0,\dotsc,29$ and displayed
the distribution in Figure~\ref{fig:boxplot-gen}.
{This reveals a paradox:  Maximum SDR performance for each
  instrument is achieved on a dictionary that was trained on the recording with
  C tin whistle and violin.  At the same time, when comparing the
  performance of each instrument over all random seeds pairwise
  between the recordings that the dictionaries were trained on, the
  Wilcoxon signed rank sums for each instrument indicate a better
  performance when training on the
  recording with B$\flat$ tin whistle and viola.  Thus, while the
  former recording yields a better-performing best-case dictionary with a
  sufficient number of runs, the training is also more likely to fail
  than with the latter recording.}

\begin{figure}[htbp]
  \tikzsetnextfilename{Figure_\the\numexpr(\thefigure+1)\relax}
  \footnotesize
  \begin{tikzpicture}[mark size=1.6pt,mark=+]
    \begin{groupplot}[group style={group size=1 by 3,
          vertical sep=12mm},
        ytick={1,2,3,4,5,6,7,8},yticklabels={Tin whistle\\[-1](B$\flat$),Tin whistle\\[-1]{(B$\flat$, gen.)},Tin whistle\\[-1]{(C)},Tin whistle\\[-1]{(C, gen.)},Viola,Viola\\[-1](gen.),Violin,Violin\\[-1](gen.)},yticklabel style={align=right},height=5.6cm,width=5.5cm,
        scale only axis,
        ymin=0.5,ymax=8.5,title style={yshift=-1.5mm},y dir=reverse,ytick style={draw=none}]
      \nextgroupplot[title={SDR}]
      \addplot[boxplot={box extend=0.4},mark=none,color=black!50] table[y index={0}] {bb-mask-measures.dat};
      \addplot[only marks] table[x index={0},y expr=1] {bb-mask-measures.dat};
      \draw[color=black!25] (axis cs:\pgfkeysvalueof{/pgfplots/xmin},1.5) -- (axis cs:\pgfkeysvalueof{/pgfplots/xmax},1.5);
      \addplot[boxplot={box extend=0.4},mark=none,color=black!50] table[y index={0}] {bb-gen-mask-measures.dat};
      \addplot[only marks] table[x index={0},y expr=2] {bb-gen-mask-measures.dat};
      \draw[color=black!25] (axis cs:\pgfkeysvalueof{/pgfplots/xmin},2.5) -- (axis cs:\pgfkeysvalueof{/pgfplots/xmax},2.5);
      \addplot[boxplot={box extend=0.4},mark=none,color=black!50] table[y index={0}] {c-mask-measures.dat};
      \addplot[only marks] table[x index={0},y expr=3] {c-mask-measures.dat};
      \draw[color=black!25] (axis cs:\pgfkeysvalueof{/pgfplots/xmin},3.5) -- (axis cs:\pgfkeysvalueof{/pgfplots/xmax},3.5);
      \addplot[boxplot={box extend=0.4},mark=none,color=black!50] table[y index={0}] {c-gen-mask-measures.dat};
      \addplot[only marks] table[x index={0},y expr=4] {c-gen-mask-measures.dat};
      \draw[color=black!25] (axis cs:\pgfkeysvalueof{/pgfplots/xmin},4.5) -- (axis cs:\pgfkeysvalueof{/pgfplots/xmax},4.5);
      \addplot[boxplot={box extend=0.4},mark=none,color=black!50] table[y index={1}] {bb-mask-measures.dat};
      \addplot[only marks] table[x index={1},y expr=5] {bb-mask-measures.dat};
      \draw[color=black!25] (axis cs:\pgfkeysvalueof{/pgfplots/xmin},5.5) -- (axis cs:\pgfkeysvalueof{/pgfplots/xmax},5.5);
      \addplot[boxplot={box extend=0.4},mark=none,color=black!50] table[y index={1}] {bb-gen-mask-measures.dat};
      \addplot[only marks] table[x index={1},y expr=6] {bb-gen-mask-measures.dat};
      \draw[color=black!25] (axis cs:\pgfkeysvalueof{/pgfplots/xmin},6.5) -- (axis cs:\pgfkeysvalueof{/pgfplots/xmax},6.5);
      \addplot[boxplot={box extend=0.4},mark=none,color=black!50] table[y index={1}] {c-mask-measures.dat};
      \addplot[only marks] table[x index={1},y expr=7] {c-mask-measures.dat};
      \draw[color=black!25] (axis cs:\pgfkeysvalueof{/pgfplots/xmin},7.5) -- (axis cs:\pgfkeysvalueof{/pgfplots/xmax},7.5);
      \addplot[boxplot={box extend=0.4},mark=none,color=black!50] table[y index={1}] {c-gen-mask-measures.dat};
      \addplot[only marks] table[x index={1},y expr=8] {c-gen-mask-measures.dat};
      \nextgroupplot[title={SIR}]
      \addplot[boxplot={box extend=0.4},mark=none,color=black!50] table[y index={2}] {bb-mask-measures.dat};
      \addplot[only marks] table[x index={2},y expr=1] {bb-mask-measures.dat};
      \draw[color=black!25] (axis cs:\pgfkeysvalueof{/pgfplots/xmin},1.5) -- (axis cs:\pgfkeysvalueof{/pgfplots/xmax},1.5);
      \addplot[boxplot={box extend=0.4},mark=none,color=black!50] table[y index={2}] {bb-gen-mask-measures.dat};
      \addplot[only marks] table[x index={2},y expr=2] {bb-gen-mask-measures.dat};
      \draw[color=black!25] (axis cs:\pgfkeysvalueof{/pgfplots/xmin},2.5) -- (axis cs:\pgfkeysvalueof{/pgfplots/xmax},2.5);
      \addplot[boxplot={box extend=0.4},mark=none,color=black!50] table[y index={2}] {c-mask-measures.dat};
      \addplot[only marks] table[x index={2},y expr=3] {c-mask-measures.dat};
      \draw[color=black!25] (axis cs:\pgfkeysvalueof{/pgfplots/xmin},3.5) -- (axis cs:\pgfkeysvalueof{/pgfplots/xmax},3.5);
      \addplot[boxplot={box extend=0.4},mark=none,color=black!50] table[y index={2}] {c-gen-mask-measures.dat};
      \addplot[only marks] table[x index={2},y expr=4] {c-gen-mask-measures.dat};
      \draw[color=black!25] (axis cs:\pgfkeysvalueof{/pgfplots/xmin},4.5) -- (axis cs:\pgfkeysvalueof{/pgfplots/xmax},4.5);
      \addplot[boxplot={box extend=0.4},mark=none,color=black!50] table[y index={3}] {bb-mask-measures.dat};
      \addplot[only marks] table[x index={3},y expr=5] {bb-mask-measures.dat};
      \draw[color=black!25] (axis cs:\pgfkeysvalueof{/pgfplots/xmin},5.5) -- (axis cs:\pgfkeysvalueof{/pgfplots/xmax},5.5);
      \addplot[boxplot={box extend=0.4},mark=none,color=black!50] table[y index={3}] {bb-gen-mask-measures.dat};
      \addplot[only marks] table[x index={3},y expr=6] {bb-gen-mask-measures.dat};
      \draw[color=black!25] (axis cs:\pgfkeysvalueof{/pgfplots/xmin},6.5) -- (axis cs:\pgfkeysvalueof{/pgfplots/xmax},6.5);
      \addplot[boxplot={box extend=0.4},mark=none,color=black!50] table[y index={3}] {c-mask-measures.dat};
      \addplot[only marks] table[x index={3},y expr=7] {c-mask-measures.dat};
      \draw[color=black!25] (axis cs:\pgfkeysvalueof{/pgfplots/xmin},7.5) -- (axis cs:\pgfkeysvalueof{/pgfplots/xmax},7.5);
      \addplot[boxplot={box extend=0.4},mark=none,color=black!50] table[y index={3}] {c-gen-mask-measures.dat};
      \addplot[only marks] table[x index={3},y expr=8] {c-gen-mask-measures.dat};
      \nextgroupplot[title={SAR}]
      \addplot[boxplot={box extend=0.4},mark=none,color=black!50] table[y index={4}] {bb-mask-measures.dat};
      \addplot[only marks] table[x index={4},y expr=1] {bb-mask-measures.dat};
      \draw[color=black!25] (axis cs:\pgfkeysvalueof{/pgfplots/xmin},1.5) -- (axis cs:\pgfkeysvalueof{/pgfplots/xmax},1.5);
      \addplot[boxplot={box extend=0.4},mark=none,color=black!50] table[y index={4}] {bb-gen-mask-measures.dat};
      \addplot[only marks] table[x index={4},y expr=2] {bb-gen-mask-measures.dat};
      \draw[color=black!25] (axis cs:\pgfkeysvalueof{/pgfplots/xmin},2.5) -- (axis cs:\pgfkeysvalueof{/pgfplots/xmax},2.5);
      \addplot[boxplot={box extend=0.4},mark=none,color=black!50] table[y index={4}] {c-mask-measures.dat};
      \addplot[only marks] table[x index={4},y expr=3] {c-mask-measures.dat};
      \draw[color=black!25] (axis cs:\pgfkeysvalueof{/pgfplots/xmin},3.5) -- (axis cs:\pgfkeysvalueof{/pgfplots/xmax},3.5);
      \addplot[boxplot={box extend=0.4},mark=none,color=black!50] table[y index={4}] {c-gen-mask-measures.dat};
      \addplot[only marks] table[x index={4},y expr=4] {c-gen-mask-measures.dat};
      \draw[color=black!25] (axis cs:\pgfkeysvalueof{/pgfplots/xmin},4.5) -- (axis cs:\pgfkeysvalueof{/pgfplots/xmax},4.5);
      \addplot[boxplot={box extend=0.4},mark=none,color=black!50] table[y index={5}] {bb-mask-measures.dat};
      \addplot[only marks] table[x index={5},y expr=5] {bb-mask-measures.dat};
      \draw[color=black!25] (axis cs:\pgfkeysvalueof{/pgfplots/xmin},5.5) -- (axis cs:\pgfkeysvalueof{/pgfplots/xmax},5.5);
      \addplot[boxplot={box extend=0.4},mark=none,color=black!50] table[y index={5}] {bb-gen-mask-measures.dat};
      \addplot[only marks] table[x index={5},y expr=6] {bb-gen-mask-measures.dat};
      \draw[color=black!25] (axis cs:\pgfkeysvalueof{/pgfplots/xmin},6.5) -- (axis cs:\pgfkeysvalueof{/pgfplots/xmax},6.5);
      \addplot[boxplot={box extend=0.4},mark=none,color=black!50] table[y index={5}] {c-mask-measures.dat};
      \addplot[only marks] table[x index={5},y expr=7] {c-mask-measures.dat};
      \draw[color=black!25] (axis cs:\pgfkeysvalueof{/pgfplots/xmin},7.5) -- (axis cs:\pgfkeysvalueof{/pgfplots/xmax},7.5);
      \addplot[boxplot={box extend=0.4},mark=none,color=black!50] table[y index={5}] {c-gen-mask-measures.dat};
      \addplot[only marks] table[x index={5},y expr=8] {c-gen-mask-measures.dat};
    \end{groupplot}
  \end{tikzpicture}
  \caption{Separation of tin whistle (B$\flat$/C) and viola/violin
    with spectral masking over 30 runs.  Results labeled as
    \enquote{gen.} were obtained by applying the the dictionaries
    trained on the other instrument combination.}
  \label{fig:boxplot-gen}
\end{figure}

  {We conclude that as intended, the model does not overfit to
  the specific recording, but it instead provides a dictionary that
  can be applied to a different recording even if slightly different
  instruments are used and the key is changed ({confirming} pitch-invariance).
  For a practical scenario, this means that if a dictionary for a
  specific combination of instruments is already available, it can be
  applied to other similar recordings, which saves computation time.\footnote{%
  {For the sample with B$\flat$ tin whistle and viola which has
    a duration of $\SI{24}{s}$, the computation of the log-frequency spectrogram
    lasted $\SI{137}{min}$.  Training took $\SI{212}{min}$ for each of the
    10 dictionaries (with $N_{\mathrm{trn}}=100000$ iterations), while separation
    and resynthesis with a given dictionary were performed within $\SI{7}{min}$.
    All computations were conducted on an Intel i5-4460 microprocessor using 2 cores
    for multiprocessing.  Note that there is still significant potential for saving computation
    time by reducing redundancy in the sampling of the STFT and decreasing the number of
    training iterations.
  }}
  In fact, re-using a well-trained dictionary can lead to superior
  separation results than training on the recording itself.
}

\subsection{Comparison on other data}

To our knowledge, there exists no standard benchmark
database with the kind of samples that our algorithm is designed for.
While the \emph{BASS-dB} set \cite{Vincent05} was created with blind
source separation in mind, it contains instruments which violate the
structural assumptions that we make about the sounds, and the
polyphony levels are not sufficiently controlled.  A similar issue
occurs with the databases that are used for supervised learning,
such as in the SiSEC 2018 \cite{Stoeter18}.

For score-informed separation, the \emph{Bach10} \cite{Duan11} and
\emph{URMP} {\cite{Li18}} databases are popular, which contain
recordings of melodic acoustic instruments.  {In} terms
of polyphony and similarity of the instruments in these samples, one
cannot expect to obtain reasonable performance from blind
separation {on most of the samples.  However, a subset of the
two-instrument recordings in URMP appeared to be usable, so
we are incorporating it in our evaluation.}

{Also, we} were able to obtain the data used by Jaiswal et
al.\ \cite{Jaiswal11a,Jaiswal11b,Jaiswal13}. As it does not contain
any samples with acoustic instruments, it is not ideal for
evaluation of our method, but being able to perform the separation
provides a proof of concept.

Further, we used the publicly available data from Duan et
al.\ \cite{Duan08}, which does contain a sample with acoustic
instruments.

\subsubsection{URMP}
\label{sec:urmp}

{The URMP dataset \cite{Li18} contains a total number of
44 audio samples arranged from classical music that were recorded
using acoustic musical
instruments.  In many of these samples, the instruments are very
similar, so we selected suitable samples based on the following
criteria:
\begin{itemize}
\item No instrument should be duplicated.
\item No two bowed string instruments should appear in one recording.
\item No two brass instruments should appear in one recording.
\item If two woodwinds appear together, one should be a reed
  instrument and the other one should not.
\end{itemize}}

{The samples with three or more instruments quickly turned out to
be too difficult for our blind separation algorithm.  From the total
number of 11 duets, this therefore left us with 4 samples:
\begin{enumerate}
\item \emph{Dance of the Sugar Plum Fairy} 
  by P.\ Tchaikovsky with
  flute and clarinet,
\item \emph{Jesus bleibet meine Freude} 
  by J.\ S.\ Bach with
  trumpet and violin,
\item \emph{March from Occasional Oratorio} 
  by G.\ F.\ Handel with
  trumpet and saxophone,
\item \emph{Ave Maria} 
  by F.\ Schubert with oboe and cello.
\end{enumerate}}

{Considering the combination of trumpet and saxophone, we were
doubtful whether a separation would be possible.  Even though the
sound production principle is very different, their sound
appears somewhat similar, which is supported by the roles
of these instruments in jazz ensembles.  We decided to include the
sample anyway in order to see how the algorithm reacts.}

{Again, we are taking the best-case number from 10
runs with $N_{\mathrm{trn}}=100000$ training iterations, and for
comparison, we are using the algorithm from \cite{Duan08} with
hand-optimized hyperparameters on the data {(as downsampled to
$\SI{22050}{Hz}$)}.  The results with the classical
measures are shown in Table~\ref{tab:urmp}.}

\begin{table}[htbp]
  \centering
  \caption{{Performance measures for the best-case runs over a
    selection of samples from the URMP \cite{Li18} dataset.
    Best numbers are marked.}}
  \label{tab:urmp}
  \tabcolsep=4.5pt
  \renewcommand{\arraystretch}{1.1}
  \heavyrulewidth=.5pt
  \lightrulewidth=.4pt
  \cmidrulewidth=.3pt
  \aboverulesep=0.0ex
  \belowrulesep=0.0ex

  \newcolumntype Y{S[
      table-format=2.1,
      table-auto-round,
      table-text-alignment=center,
      table-number-alignment=center,
      table-space-text-post={*}]}
  \newcolumntype Z{S[
      table-format=-1.1,
      table-auto-round,
      table-text-alignment=center,
      table-number-alignment=center,
      table-space-text-post={*}]}
    \begin{tabular}{ccZYZ}
      \toprule
    {\bfseries Method} & {\bfseries Instrument} & {\bfseries SDR} & {\bfseries SIR} & {\bfseries SAR} \\\midrule
    \multirow{7}{*}{Ours} & Flute & 2.44635917 & 9.46831118 & 3.87247707* \\
    & Clarinet & 6.22754594* & 25.27585464* & 6.29482429* \\
    \cmidrule{2-5}
    & Trumpet & 5.27036917* & 16.55001473* & 5.70157274* \\
    & Violin & 7.70812429* & 25.11747363* & 7.80105425* \\
    \cmidrule{2-5}
    & Trumpet & -2.40914079 & 1.07264129 & 2.68289795* \\
    & Saxophone & 0.12786161 & 22.48902291* & 0.17756585 \\
    \cmidrule{2-5}
    & Oboe & 6.31074976* & 17.04997418* & 6.7782757* \\
    & Cello & 4.19790633* & 17.10544068* & 4.50990864 \\
    \cmidrule{1-5}
    \multirow{7}{*}{\cite{Duan08}} & Flute & 3.4085* & 19.5506* & 3.5633 \\
    & Clarinet & 2.0911 & 5.9174 & 5.4046 \\
    \cmidrule{2-5}
    & Trumpet & {---} & {---} & {---} \\
    & Violin & {---} & {---} & {---} \\
    \cmidrule{2-5}
    & Trumpet & 1.1638* & 9.4487* & 2.3286 \\
    & Saxophone & 6.9220* & 17.2436 & 7.4265* \\
    \cmidrule{2-5}
    & Oboe & -0.7868 & 13.1102 & -0.3989 \\
    & Cello & 3.3892 & 6.4313 & 7.2580* \\
    \bottomrule
  \end{tabular}
\end{table}

{The piece for flute and clarinet was challenging for both
algorithms (perhaps because both instruments are woodwinds).  The
algorithm from \cite{Duan08}}
{isolated the clarinet as the dominant instrument
but only achieved inferior performance on it, whereas the residual has good
resemblence with the flute track.}
{On the piece with trumpet and violin,
our algorithm performed quite well, but the algorithm from
\cite{Duan08} got stuck in an apparently endless loop, so we could not
get a comparison result.  With the piece for trumpet and saxophone,
which we had already considered problematic beforehand, our algorithm
failed to give an acceptable result in terms of SDR and SIR (in
contrast to the PEASS evaluation, as we will discuss later).  The compared
algorithm gives better figures when separating the trumpet as the
dominant instrument, but the result cannot be considered good, either;
however, the residual signal gives a decent separation of the saxophone
track.}  {By contrast, in the piece with
oboe and cello, the algorithm from \cite{Duan08} separated the cello as the
dominant instrument comparatively well, whereas it failed on the oboe.
For both instruments, the results from our algorithm are better.}

{As before, it turned out that adjustment of the hyperparameters for
every sample was crucial in application of the algorithm from
\cite{Duan08}, as the clustering depends on the amount of variation in
the sound of the dominant instrument as well as on the similarity of
the sounds of both instruments.}

{The corresponding PEASS scores are given in
Table~\ref{tab:urmppeass}.  The main difference is that our separation
of the trumpet in the third piece that received very bad SDR/SIR/SAR
values was given very good perceptual scores, mostly exceeding those
of the compared method.  Listening to the separated trumpet tracks
ourselves, we find that while ours certainly has issues, large parts
are much more usable than the SDR suggests, and we can understand why
one would perceive the errors as less disruptive than in the track
that was isolated by the algorithm from \cite{Duan08}.}

\begin{table}[htbp]
  \centering
  \caption{{PEASS scores for the best-case runs over a
    selection of samples from the URMP \cite{Li18} dataset.
    Best numbers are marked.}}
  \label{tab:urmppeass}
  \tabcolsep=4.5pt
  \renewcommand{\arraystretch}{1.1}
  \heavyrulewidth=.5pt
  \lightrulewidth=.4pt
  \cmidrulewidth=.3pt
  \aboverulesep=0.0ex
  \belowrulesep=0.0ex

  \newcolumntype Y{S[
      table-format=2,
      table-auto-round,
      table-text-alignment=center,
      table-number-alignment=center,
      table-space-text-post={*}]}
  \newcolumntype Z{S[
      table-format=-1.1,
      table-auto-round,
      table-text-alignment=center,
      table-number-alignment=center,
      table-space-text-post={*}]}
    \begin{tabular}{ccYYYY}
      \toprule
    {\bfseries Method} & {\bfseries Instrument} & {\bfseries OPS} &
    {\bfseries TPS} & {\bfseries IPS} & {\bfseries APS} \\\midrule
    \multirow{7}{*}{Ours} & Flute & 28.1734 & 45.7278 & 66.4962* & 28.7159 \\
    & Clarinet & 36.0336* & 58.2186* & 70.8365 & 38.5229* \\
    \cmidrule{2-6}
    & Trumpet & 29.5444* & 66.6389* & 47.2786* & 36.1293* \\
    & Violin & 30.9201* & 33.1396* & 68.9686* & 35.8290* \\
    \cmidrule{2-6}
    & Trumpet & 47.1746* & 69.1253* & 62.7971 & 53.7664* \\
    & Saxophone & 24.1759 & 22.8240 & 70.0301* & 15.3862 \\
    \cmidrule{2-6}
    & Oboe & 17.9278* & 6.8533 & 60.1928* & 7.0506 \\
    & Cello & 30.1588* & 42.2368* & 58.1360 & 41.5621* \\
    \cmidrule{1-6}
    \multirow{7}{*}{\cite{Duan08}} & Flute & 34.9979* & 74.6492* & 38.0635 & 45.8107* \\
    & Clarinet & 26.6379 & 27.9426 & 75.7467* & 24.5313 \\
    \cmidrule{2-6}
    & Trumpet & {---} & {---} & {---} & {---} \\
    & Violin & {---} & {---} & {---} & {---} \\
    \cmidrule{2-6}
    & Trumpet & 41.7824 & 52.4926 & 63.5646* & 45.9306 \\
    & Saxophone & 26.3325* & 71.9387* & 21.9816 & 58.7136* \\
    \cmidrule{2-6}
    & Oboe & 15.4774 & 53.6512* & 28.0907 & 18.8949* \\
    & Cello & 19.5795 & 15.5406 & 66.9432* & 21.9385 \\
    \bottomrule
  \end{tabular}
\end{table}

{We believe that one key challenge with this dataset is that the
instruments were played with the mindset of a musical performance, and
thus there is more variation in playing technique than with our own
samples.}

\subsubsection{Jaiswal et al.}
\label{sec:jaiswal}

We ran our algorithm on the data that was used in
\cite{Jaiswal11a,Jaiswal11b,Jaiswal13}, which consists of
computer-synthesized samples with two instruments, each playing one
tone at a time.
Due to the large number of samples, and since we are only interested
in best-case numbers, we set $N_{\mathrm{trn}}=10000$ and selected the
best result (in terms of mean SDR) out of 10 runs (with random seeds
$0,\dotsc,9$) for each sample.  No further adjustments to our
algorithm were conducted.  The performance measures are displayed in
Figure~\ref{fig:jaiswal} and Table~\ref{table:jaiswal}.

\begin{figure}[t]
  \tikzsetnextfilename{Figure_\the\numexpr(\thefigure+1)\relax}
  \footnotesize
  \centering
  \begin{tikzpicture}
    \hypersetup{hidelinks}
    \begin{groupplot}[group style={group size=2 by 1, 
        horizontal sep=0mm},
        scale only axis,
        height=6cm,
        enlargelimits=false,
        ymin=-2.5,ymax=52,
        xmajorgrids,
        grid style={color=gray!20!white}]
    \nextgroupplot[xmin=0,xmax=26,
        xlabel={Sample},ylabel={[\si{dB}]},
        extra x ticks={1,...,25},
        extra x tick labels=\empty,
        scale only axis,
        width=5.40cm,legend pos=north west]
    \addplot[only marks,mark=o] table[y index=1] {jaiswal.dat};
    \addlegendentry{SDR}
      \addplot[only marks,mark=x] table[y index=2] {jaiswal.dat};
    \addlegendentry{SIR}
      \addplot[only marks,mark=+] table[y index=3] {jaiswal.dat};
    \addlegendentry{SAR}
    \nextgroupplot[xmin=0,xmax=3,
        xlabel={\strut},
        xtick={1,2},
        xticklabels={Mean, \cite{Jaiswal13}},
        yticklabel=\empty,
        width=1.35cm,
        tick label style={rotate=90}
    ]
    \addplot[only marks,mark=o] coordinates {(1, 7.5486003) (2, 11.11)};
    \addplot[only marks,mark=x] coordinates {(1, 23.05845728) (2, 32.13)};
    \addplot[only marks,mark=+] coordinates {(1, 8.37824845) (2, 11.47)};
    \end{groupplot}
    \pgfresetboundingbox
    \useasboundingbox (group c1r1.outer north west) rectangle (group c1r1.outer south east -| group c2r1.outer south east);
  \end{tikzpicture}
  \caption{Performance of our algorithm applied on the audio samples
    from \cite{Jaiswal11a,Jaiswal11b,Jaiswal13} (best-case run out of
    10 for each sample).  The means over the samples with our algorithm
    are compared to the mean values given in \cite{Jaiswal13}.}
  \label{fig:jaiswal}
\end{figure}

It can be seen that for certain samples, our algorithm performs very
well, while for others, it fails to produce acceptable results.  When
comparing the means, our algorithm is inferior to
\cite{Jaiswal11a,Jaiswal11b,Jaiswal13}.\footnote{We could not compare
  the performance on the individual samples, as those numbers are not
  available to us.}

Our explanation for this is that our algorithm assumes much looser
constraints on the data that it gets, as it accepts arbitrary tones
in the audible range.  By contrast, in
\cite{Jaiswal11a,Jaiswal11b,Jaiswal13}, the expected fundamental
frequencies for the instruments are hardcoded in the algorithm due to
prior knowledge.  In
\cite{Jaiswal11a}, 7 values are allowed per sample, while in
\cite{Jaiswal11b}, this number was invidually adjusted to 4--9 values
for each sample in order to achieve maximum performance figures; in
\cite{Jaiswal13}, those were 5--12 values.  Further, the algorithms
can exploit the fact that the tone ranges for the respective
instruments in the samples were chosen to have little or no overlap.
In the case of no overlap, such distinctive information would even
make it possible to separate instruments with identical frequency
spectra, but this would violate our notion of
blind separation.

\begin{table}[t]
  \centering
  \caption{Comparison of our algorithm to
    \cite{Jaiswal11a,Jaiswal11b,Jaiswal13} on the data used therein
    (means over all instruments and all samples in the best cases).
    Best numbers are marked.}
  \label{table:jaiswal}
  \tabulinesep=2pt
  \tabcolsep=4.5pt

  \newcolumntype Y{S[
      table-format=2.1,
      table-auto-round,
      table-text-alignment=center,
      table-number-alignment=center,
      table-space-text-post={*}]}
  \tabucolumn{Y}
  \begin{tabu}{cYYY}
    \tabucline[.5pt]{-}
    \rowfont{\bfseries}
    {Method} & {SDR} & {SIR} & {SAR} \\\tabucline[.4pt]{-}
    \cite{Jaiswal11a} & 8.94 & 23.69 & 9.72 \\
    \cite{Jaiswal11b} & 10.88 & 25.44 & 11.47 \\
    \cite{Jaiswal13} & 11.11* & 32.13* & 11.47* \\
    Ours & 7.5486003 & 23.05845728 & 8.37824845 \\
    \tabucline[.5pt]{-}
  \end{tabu}
\end{table}

As can be seen in
Table~\ref{table:jaiswal}, the individual adjustments that were
conducted in \cite{Jaiswal11b} had a much greater effect on the
performance than the algorithmic improvements in \cite{Jaiswal13}.

Applying the algorithms in \cite{Jaiswal11a,Jaiswal11b,Jaiswal13} to
our data would not be meaningful, as those algorithms require, due to
their data representation, perfectly consistent equal temperament
tuning, which wind instruments and string instruments without frets do
not satisfy.

We conclude that the out-of-the-box performance of
our algorithm is on average inferior to the figures in
\cite{Jaiswal11a,Jaiswal11b,Jaiswal13} on the samples used therein,
but this is compensated by its vastly greater flexibility, which
enables it to operate on real-world acoustic signals and eliminates
the need for prior specification of the tuning or range of the instruments.

\subsubsection{Duan et al.}
\label{sec:duan}

From the data used in \cite{Duan08}, we selected the samples that we
deemed suitable for our algorithm, skipping the ones that contain
human voice components, as those cannot be represented by our model.

The three samples that we therefore consider are composed as follows:
\begin{enumerate}
\item Acoustic oboe and acoustic euphonium,
\item Synthesized piccolo and synthesized organ,
\item Synthesized piccolo, synthesized organ, and synthesized
  oboe.
\end{enumerate}

The original samples are sampled at $f_{\mathrm{s}}=\SI{22050}{Hz}$.
We upsampled them to $f_{\mathrm{s}}=\SI{44100}{Hz}$ in order to apply
them to our algorithm.  We again ran the algorithm with
$N_{\mathrm{trn}}=10000$ iterations and picked the best-case runs from
random seeds $0,\dotsc,9$, respectively.  The results are displayed in
Table \ref{tab:duan}.

\begin{table}[htbp]
  \centering
  \caption{Performance measures for the best-case runs of different
    instrument combinations, with spectral masking.  Instruments
    labeled as \enquote{s.} are synthetic, those labeled as
    \enquote{a.} are acoustic.  Best numbers are marked.}
  \label{tab:duan}
  \tabcolsep=4.5pt
  \renewcommand{\arraystretch}{1.1}
  \heavyrulewidth=.5pt
  \lightrulewidth=.4pt
  \cmidrulewidth=.3pt
  \aboverulesep=0.0ex
  \belowrulesep=0.0ex

  \newcolumntype Y{S[
      table-format=2.1,
      table-auto-round,
      table-text-alignment=center,
      table-number-alignment=center,
      table-space-text-post={*}]}
    \begin{tabular}{ccYYY}
      \toprule
    {\bfseries Method} & {\bfseries Instrument} & {\bfseries SDR} & {\bfseries SIR} & {\bfseries SAR} \\\midrule
    \multirow{7}{*}{\cite{Duan08}} & Oboe (a.) & 8.7 & 25.8 & 8.8 \\
    & Euphonium (a.) & 4.6 & 14.5 & 5.3 \\
    \cmidrule{2-5}
    & Piccolo (s.) & 14.2* & 27.9* & 14.4* \\
    & Organ (s.) & 11.8* & 25.1* & 12.1* \\
    \cmidrule{2-5}
    & Piccolo (s.) & 6.5* & 20.0 & 6.7* \\
    & Organ (s.) & 6.6* & 17.3 & 7.1* \\
    & Oboe (s.) & 9.0* & 21.9* & 9.2* \\
    \cmidrule{1-5}
    \multirow{7}{*}{Ours} & Oboe (a.) & 18.63132098* & 33.5532616* & 18.77536405* \\
    & Euphonium (a.) & 14.65085084* & 31.46362964* & 14.74537616* \\
    \cmidrule{2-5}
    & Piccolo (s.) & 11.17925459 & 25.93718057 & 11.33799989 \\
    & Organ (s.) & 10.10624241 & 20.73031387 & 10.5362565 \\
    \cmidrule{2-5}
    & Piccolo (s.) & 4.22638963 & 24.79923549* & 4.27897973 \\
    & Organ (s.) & 6.04851133 & 19.96833418* & 6.27182761 \\
    & Oboe (s.) & 5.25789918 & 12.43157896 & 6.42346892 \\
    \bottomrule
  \end{tabular}
\end{table}

The main goal of our algorithm was to provide good performance for
acoustic instruments, and in fact, on the combination of two acoustic
instruments, it exceeds the original performance of the compared
method by roughly \SI{10}{dB} in SDR.
For the synthetic instruments, the performance achieved by the
algorithm in \cite{Duan08} is mostly superior, while our algorithm
still attains acceptable performance for piccolo and organ, and we
demonstrate that it can at least in principle also be applied to
combinations of more than two instruments.

{The corresponding PEASS scores for the separated tracks are given
in Table~\ref{tab:duanpeass}.  Here, in the example with two
acoustic instruments, the separation of the oboe track by the
algorithm in \cite{Duan08} receives a higher OPS and IPS, suggesting
that the overall quality of our separation is perceptually worse and
this is at least partly caused by interference.  However, according to
our own listening opinion, the result from our algorithm matches the
original signal very well and contains no audible interference while
the result from the compared algorithm contains very obvious
interference and also other representation errors, so we cannot
explain the outcome of this evaluation.  On the other hand, with the
synthetic instruments, it is now often our algorithm that is
preferred.}

\begin{table}[tbp]
  \centering
  \caption{{PEASS scores for the best-case runs of different
    instrument combinations, with spectral masking.  Instruments
    labeled as \enquote{s.} are synthetic, those labeled as
    \enquote{a.} are acoustic.  Best numbers are marked.
    The APS in the fourth row for each method was a perfect tie.}}
  \label{tab:duanpeass}
  \tabcolsep=4.5pt
  \renewcommand{\arraystretch}{1.1}
  \heavyrulewidth=.5pt
  \lightrulewidth=.4pt
  \cmidrulewidth=.3pt
  \aboverulesep=0.0ex
  \belowrulesep=0.0ex

  \newcolumntype Y{S[
      table-format=2,
      table-auto-round,
      table-text-alignment=center,
      table-number-alignment=center,
      table-space-text-post={*}]}
    \begin{tabular}{ccYYYY}
      \toprule
    {\bfseries Method} & {\bfseries Instrument} & {\bfseries OPS} & {\bfseries TPS} & {\bfseries IPS} & {\bfseries APS}\\\midrule
    \multirow{7}{*}{\cite{Duan08}} & Oboe (a.) & 23.7930* & 32.7538 & 82.0716* & 9.3787 \\
    & Euphonium (a.) & 23.7911 & 65.5141 & 43.1052* & 5.1622 \\
    \cmidrule{2-6}
    & Piccolo (s.) & 48.3636* & 74.0279 & 59.2677 & 54.1870 \\
    & Organ (s.) &  40.5255 & 85.5256 & 73.0786 & 86.7384 \\
    \cmidrule{2-6}
    & Piccolo (s.) & 22.0234 & 66.9663* & 34.9453 & 35.3938*  \\
    & Organ (s.) & 27.7719 & 62.9366* & 59.6174* & 57.8997 \\
    & Oboe (s.) &  43.9763* & 69.7916* & 58.2238 & 57.2847 \\
    \cmidrule{1-6}
    \multirow{7}{*}{Ours} & Oboe (a.) & 18.7314 & 99.3543* & 44.4212 & 65.9758* \\
    & Euphonium (a.) & 33.8985* & 69.8733* & 38.1936 & 59.5047* \\
    \cmidrule{2-6}
    & Piccolo (s.) & 24.3047 & 82.6787* & 68.7563* & 76.8533* \\
    & Organ (s.) & 78.6313* & 92.7679* & 86.3911* & 86.7384 \\
    \cmidrule{2-6}
    & Piccolo (s.) & 27.0385* & 28.5166 & 56.2023* & 32.3926 \\
    & Organ (s.) & 37.7011* & 52.5770 & 50.0225 & 51.6044 \\
    & Oboe (s.) & 20.1115 & 61.0290 & 67.7760* & 80.1698* \\
    \bottomrule
  \end{tabular}
\end{table}

\section{Conclusion and Future Work}

We developed a novel algorithm to represent discrete mixture spectra as
sparse shifted linear combinations of analytically given non-negative
continuous patterns.  We applied this algorithm to spectrograms of
audio recordings, first to convert an STFT magnitude spectrogram into
a log-frequency spectrogram, then to identify patterns of peaks
related to the sounds of musical instruments in the context of a
dictionary learning algorithm based on Adam, a method that originates
from the field of deep learning.

This led us to an algorithm to perform blind source separation
on polyphonic music recordings of wind and string instruments, making
only minimal structural assumptions about the data.  In its model,
the spectral properties of the musical instruments are fixed and
pitch-invariant.  Thus, instruments that
satisfy this assumption can be represented irrespectively of their
tuning.  The only parameters that have to be known a-priori are the
number of instruments and an upper bound for the sparsity level.

When applied to recordings of appropriate acoustic instruments, the
performance of our algorithm surpasses that of comparable literature.
Further, we show that once a dictionary has been trained on a certain
combination of instruments, it can be applied to combinations of
\enquote{related} instruments, even if those have a different tuning.

We note, however, that blind source separation always needs
favorable data:  Representing other kinds of instruments would require
a different model, and instruments with a pronounced attack sound are
also problematic.  The sound of the instruments must be sufficiently
pitch- and volume-invariant with only little overall variation in
the harmonic structure, and the sparsity level must be rather strict.

While the pitch-invariant spectrogram substantially facilitates the
identification of the instrument sounds, it has a lower resolution in
the high frequencies, and therefore some information from the STFT
spectrogram is lost.  Also, any phase information is lost
completely.  Despite an inharmonicity parameter being included in our
model, instruments with strong inharmonicity are problematic to
identify.

Overall, while our algorithm appears to work well in certain
settings, the framework that was created in order to bring the
computational complexity under control is not very flexible.
Instead of the hand-crafted pursuit algorithm, one may also consider
the application of a neural network for identification, while still
doing blind separation via a parametric model.

In our application, the frequency shift of the spectrum is caused by a
change in pitch.  Another common source for frequency shifts in
various areas (such as communication technology or astronomy) is the
Doppler effect.  We believe that this could open new applications for
our pursuit algorithm and potentially also the dictionary learning
algorithm.  {Specifically, the pursuit algorithm could be used as
an alternative to continuous basis pursuit \cite{Ekanadham11}, which
is advertised as a method for radar and geological seismic data and
has been used for the analysis of neural spike trains \cite{Ekanadham14}.}

\section{Appendix}
\label{sec:appendix}

We provide pseudo-code with a description of the implementation details as well as some additional figures with commentary.

\subsection{Pseudo-Code}
\label{sec:pseudocode}

We will now present the mentioned algorithms in more detail via
pseudo-code.  First, Algorithm~\ref{algo:pursuit} describes the sparse
pursuit/identification algorithm.  It takes as arguments the
dictionary, the sample vector, a selector function, the sparsity levels,
and the sum of the
previous amplitudes for each pattern (which will become important for
dictionary {pruning}).  In the non-linear optimization step, it calls
the L-BFGS-B minimizer to minimize the loss $L_D$ with respect to the given
parameters.

\begin{algoenv}[!h]
\caption{Sparse identification algorithm}
\label{algo:pursuit}
\flushleft\small\hspace{-0.92em}%
\begin{varwidth}{\textwidth}
\begin{algorithmic}
  \Function{pursuit}{$D,Y,\textrm{select},N_{\mathrm{pre}},N_{\mathrm{spr}},A$}
    \State $\mathcal{J}\leftarrow\emptyset$
    \State $r\leftarrow Y^q$
    \Loop{$N_{\mathrm{itr}}$ times}
      \State $a_{j}\leftarrow0$, $\theta_{j}\leftarrow\theta_{\mathrm{nil}}$
      \State\hspace{\algorithmicindent}for $j\in\{1,\dotsc,N_{\mathrm{spr}}+N_{\mathrm{pre}}\}\setminus\mathcal{J}$
      \State $\mathcal{J}_{\mathrm{new}}\leftarrow\operatorname{sort}(\{1,\dotsc,N_{\mathrm{spr}}+N_{\mathrm{pre}}\}\setminus\mathcal{J})[1,\dotsc,N_{\mathrm{pre}}]$
      \State $a_{\mathcal{J}_{\mathrm{new}}},\mu_{\mathcal{J}_{\mathrm{new}}},\eta_{\mathcal{J}_{\mathrm{new}}}$
      \State\hspace{\algorithmicindent}
      $\leftarrow$\;\Call{select}{$r,y_0,\dotsc,y_{N_{\mathrm{pat}}-1},\mathcal{J}_{\mathrm{new}},N_{\mathrm{pre}}$}\;
      \State $\mathcal{J}\leftarrow\mathcal{J}\cup\mathcal{J}_{\mathrm{new}}$
      \State $a_{\mathcal{J}},\mu_{\mathcal{J}},\theta_{\mathcal{J}}\leftarrow\operatorname{bfgs}(L_D,Y,a_{\mathcal{J}},\mu_{\mathcal{J}},\eta_{\mathcal{J}},\theta_{\mathcal{J}})$
      \State\hspace{\algorithmicindent}$a_{\mathcal{J}}\geq0$, $\theta_{\mathcal{J}}\in\Omega_\theta$
      \For{$\eta=0,\dotsc,N_{\mathrm{pat}}-1$}  
        \State $\mathcal{J}_\eta\leftarrow\bigl(\operatorname{arg\,sort}_{j\in\{1,\dotsc,N_{\mathrm{spr}}+N_{\mathrm{pre}}\},\,\eta_j=\eta}a_{j}\bigr)$
        \State\hspace{\algorithmicindent}$[1,\dotsc,N_{\mathrm{spr}}]$
      \EndFor
      \State $\mathcal{J}\leftarrow\bigcup_{\eta\in\{0,\dotsc,N_{\mathrm{pat}-1}\}}\mathcal{J}_\eta$
      \State $a_{\mathcal{J}},\mu_{\mathcal{J}},\theta_{\mathcal{J}}\leftarrow\operatorname{bfgs}(L_D,Y,a_{\mathcal{J}},\mu_{\mathcal{J}},\eta_{\mathcal{J}},\theta_{\mathcal{J}})$
      \State\hspace{\algorithmicindent}$a_{\mathcal{J}}\geq0$, $\theta_{\mathcal{J}}\in\Omega_\theta$
      \State $r\leftarrow Y^q-\bigl(\sum_{j\in\mathcal{J}}a_jy_{\eta_{j},\theta_{j}}(\cdot-\mu_j)\bigr)^q$
      \State $\theta\leftarrow\snorm{r}_2$
      \If{$\snorm{r}_2\geq\lambda\theta$}
        \State restore values from previous iteration
        \State \textbf{break}
      \EndIf
    \EndLoop
    \For{$\eta=0,\dotsc,N_{\mathrm{pat}}-1$}  
      \State $A[\eta]\leftarrow A[\eta]+\sum_{\eta_{j}=\eta}a_{j}$
    \EndFor
    \State \Return{$\mathcal{J},a_{\mathcal{J}},\mu_{\mathcal{J}},\eta_{\mathcal{J}},\theta_{\mathcal{J}},A$}
    \EndFunction
\end{algorithmic}
\end{varwidth}
\end{algoenv}

The $\textrm{sel\_xcorr}$ function in Algorithm~\ref{algo:xcorr} is
used in the separation.  It selects up to $N_{\mathrm{pre}}$ patterns
based on cross-correlation, and it computes their discrete amplitudes
and shifts.  In the implementation, this is accelerated via the
FFT convolution theorem.
The $\textrm{sel\_peaks}$ function in Algorithm~\ref{algo:peaks} ignores
the patterns, and it simply returns the $N_{\mathrm{pre}}$ largest
local maxima with dominance $N_{\mathrm{dom}}$ (typically,
$N_{\mathrm{dom}}=3$).

The dictionary learning algorithm (Algorithm~\ref{algo:adam}) is largely
identical to the original formulation of Adam (with values $\beta_1=0.9$,
$\beta_2=0.999$, $\eps=\num{e-8}$, and a step-size of
$\kappa=\num{e-3}$), except that $v_2$ is
averaged over all the harmonics for one instrument.  It counts the
number of training iterations $\tau[\eta]$ for each instrument $\eta$
individually.  The dictionary is initialized by the function in
Algorithm~\ref{algo:init}, which creates a new dictionary column with
random values.  The function in Algorithm~\ref{algo:prune} removes
seldom-used instruments in the dictionary by comparing their average
amplitude but with a head start which is half the length of
the pruning interval: $\tau_0=N_{\mathrm{prn}}/2$.

The $\text{logspect}$ function in Algorithm~\ref{algo:logspect} takes an
STFT magnitude spectrogram and applies the sparse pursuit algorithm in
order to convert it into a log-frequency spectrogram with a height of
$m=1024$.
Finally, the $\text{separate}$ function in Algorithm~\ref{algo:separate}
performs the overall separation procedure.  It {prunes} the dictionary
every $N_{\mathrm{prn}}=500$ steps.

\begin{algoenv}[!h]
\caption{Selector function based on cross-correlation}
\label{algo:xcorr}
\flushleft\small\hspace{-0.92em}%
\begin{varwidth}{\textwidth}
\begin{algorithmic}
  \Function{sel\_xcorr}{$r,y_0,\dotsc,y_{N_{\mathrm{pat}-1}},\mathcal{J},N_{\mathrm{pre}}$}
     \State $\rho[\mu,\eta]\leftarrow\sum_{i=0}^{m-1} r[i]\,\bigl(y_{\eta,\theta_{\mathrm{nil}}}[i-\mu]\bigr)^q/\snorm{y_{\eta,\theta_{\mathrm{nil}}}[\cdot]^q}_2$
     \State\hspace{\algorithmicindent}for $\mu=0,\dotsc,m-1$ and $\eta=0,\dotsc,N_{\mathrm{pat}}-1$
     \State $(\mu_{\mathcal{J}},\eta_{\mathcal{J}})\leftarrow\operatorname{arg\,sort}_{(\mu,\eta)}(-\rho[\mu,\eta])[:N_{\mathrm{pre}}]$
     \State $a_{\mathcal{J}}\leftarrow(\rho[\mu_{\mathcal{J}},\eta_{\mathcal{J}}]/\snorm{y_{\eta_{\mathcal{J}},\theta_{\mathrm{nil}}}[\cdot]^q}_2)^{1/q}$
     \State $\mathcal{J}\leftarrow\{j\in\mathcal{J}:a_j>0\}$
     \State \Return{$a_{\mathcal{J}},\mu_{\mathcal{J}},\eta_{\mathcal{J}}$}
  \EndFunction
\end{algorithmic}
\end{varwidth}
\end{algoenv}

\begin{algoenv}[!h]
\caption{Selector function based on peaks}
\label{algo:peaks}
\flushleft\small\hspace{-0.92em}%
\begin{varwidth}{\textwidth}
\begin{algorithmic}
  \Function{sel\_peaks}{$r,\_,\mathcal{J},N_{\mathrm{pre}}$}
  \State $\mu_{\mathcal{J}}\leftarrow\operatorname{arg\,sort}_{\mu}$
  \State\hspace{\algorithmicindent}$\{a_\mu:r[\mu]\geq r[\mu+k],\:\sabs{k}\leq N_{\mathrm{dom}}\}[:N_{\mathrm{pre}}]$
  \State $\mathcal{J}\leftarrow\{j\in\mathcal{J}:r[j]>0\}$
  \State \Return{$r_{\mathcal{J}},\mu_{\mathcal{J}},0$}
  \EndFunction
\end{algorithmic}
\end{varwidth}
\end{algoenv}

\begin{algoenv}[!h]
\caption{Dictionary learning function}
\label{algo:adam}
\flushleft\small\hspace{-0.92em}%
\begin{varwidth}{\textwidth}
\begin{algorithmic}\small
  \Function{adam}{$D,\tau,v_1,v_2,g$}
    \For{$\eta=0,\dotsc,N_{\mathrm{pat}}-1$}
      \State $\tau[\eta]\leftarrow\tau[\eta]+1$
      \State $v_1[\cdot,\eta]\leftarrow\beta_1\cdot v_1[\cdot,\eta]+(1-\beta_1)\cdot g[\cdot,\eta]$
      \State $v_2[\eta]\leftarrow\beta_2\cdot v_2[\eta]+(1-\beta_2)\cdot \operatorname{mean}(g[\cdot,\eta]^2)$
      \State $\hat{v}_1[\cdot,\eta]\leftarrow v_1[\cdot,\eta]/(1-\beta_1^{\tau[\eta]})$
      \State $\hat{v}_2[\eta]\leftarrow v_2[\eta]/(1-\beta_2^{\tau[\eta]})$
      \State $D[\cdot,\eta]\leftarrow D[\cdot,\eta]-\kappa\cdot\hat{v}_1[\cdot,\eta]/(\sqrt{\hat{v}_2[\eta]+\eps})$
      \State $D[\cdot,\eta]\leftarrow \max\bigl(0,\min\bigl(1,D[\cdot,\eta]\bigr)\bigr)$
    \EndFor
    \State \Return{$D,\tau,v_1,v_2$}
  \EndFunction
\end{algorithmic}
\end{varwidth}
\end{algoenv}

\begin{algoenv}[!h]
\caption{Dictionary initialization function}
\label{algo:init}
\flushleft\small\hspace{-0.92em}%
\begin{varwidth}{\textwidth}
\begin{algorithmic}\small
  \Function{init}{\hspace{0pt}}
    \State $e\leftarrow\mathrm{Par}(1,0.5)$
    \For{$h=1,\dotsc,N_{\mathrm{har}}$}
      \State $d[h]\leftarrow\mathcal{U}[0,1)$
      \State $d[h]\leftarrow d[h]/h^e$
    \EndFor
  \State\Return $d[\cdot]$
  \EndFunction
\end{algorithmic}
\end{varwidth}
\end{algoenv}

\begin{algoenv}[!h]
\caption{Dictionary pruning function}
\label{algo:prune}
\flushleft\small\hspace{-0.92em}%
\begin{varwidth}{\textwidth}
\begin{algorithmic}
\Function{prune}{$A,D,\tau$}
    \State $\mathcal{I}\leftarrow\bigl(\operatorname{arg\,sort}_{\eta\in\{0,\dotsc,N_{\mathrm{pat}}-1\}} A[\eta]/(\tau[\eta]-\tau_0)\bigr)$
    \State\hspace{\algorithmicindent} $[0,\dotsc,N_{\mathrm{ins}}]$
    \State $\tau[\mathcal{I}^\complement]=0$,
      $v_1[\cdot,\mathcal{I}^\complement]=0$,
      $v_2[\mathcal{I}^\complement]=0$,
      $A[\mathcal{I}^\complement]=0$
    \For{$\eta\in\mathcal{I}^\complement$}
      \State $D[\cdot,\eta]\leftarrow$\;\Call{init}{\hspace{0pt}}
    \EndFor
    \State \Return{$\mathcal{I},\tau,v_1,v_2,A,D$}
  \EndFunction
\end{algorithmic}
\end{varwidth}
\end{algoenv}

\begin{algoenv}[!h]
  \caption{Log-spectrogram generation function}
  \label{algo:logspect}
  \flushleft\small\hspace{-0.92em}%
  \begin{varwidth}{\textwidth}
  \begin{algorithmic}
  \Function{logspect}{$Z,N_{\mathrm{pre}},N_{\mathrm{spr}}$}
  \For{$t=0,\dotsc,n-1$}
    \State $\mathcal{J}_t,a_{\mathcal{J}_t,t},\mu_{\mathcal{J}_t,t},\eta_{\mathcal{J}_t,t},\theta_{\mathcal{J}_t,t},\_$
    \State\hspace{\algorithmicindent}$\leftarrow$\;\Call{pursuit}{$[1],Z[\cdot,t],\textrm{sel\_peaks},1,N_{\mathrm{spr}},\_$}
  \EndFor
  \State$U[\alpha,t]\leftarrow\sum_{j,h}a_{j,h,t}\operatorname{exp}(-(\alpha-\alpha(\mu_{\mathcal{J}_t,t}))^2/(2F^2\sigma_{j,t}^2))$
  \State\hspace{\algorithmicindent}for $\alpha=0,\dotsc,m-1$, $t=0,\dotsc,n-1$
  \State\Return$U$
  \EndFunction
  \end{algorithmic}
  \end{varwidth}
\end{algoenv}

\begin{algoenv}[!h]
  \caption{Dictionary learning and separation function}
  \label{algo:separate}
  \flushleft\small\hspace{-0.92em}%
  \begin{varwidth}{\textwidth}
  \begin{algorithmic}
  \Function{separate}{$U,N_{\mathrm{pre}},N_{\mathrm{spr}}$}
  \For{$\eta=0,\dotsc,N_{\mathrm{pat}}-1$}
    \State $D[\cdot,\eta]\leftarrow$\;\Call{init}{\hspace{0pt}}
    \State $\tau[\eta]\leftarrow0$, $v_1[\cdot,\eta]\leftarrow0$, $v_2[\eta]\leftarrow0$, $A[\eta]\leftarrow0$
  \EndFor
  \Loop{a multiple of $N_{\mathrm{prn}}$ times}
    \State $t\leftarrow\operatorname{random}(\{0,\dotsc,n-1\})$
    \State $\mathcal{J},a_{\mathcal{J}},\mu_{\mathcal{J}},\eta_{\mathcal{J}},\theta_{\mathcal{J}},A$
    \State\hspace{\algorithmicindent}$\leftarrow$\;\Call{pursuit}{$D,U[\cdot,t],\textrm{sel\_xcorr},1,N_{\mathrm{spr}},A$}
    \State $g\leftarrow \nabla_D L_D(Y,a_{\mathcal{J}},\mu_{\mathcal{J}},\eta_{\mathcal{J}},\theta_{\mathcal{J}})$
    \State $D,\tau,v_1,v_2\leftarrow$\;\Call{adam}{$D,\tau,v_1,v_2,g$}
    \If{$\min(\tau)\!\!\mod N_{\mathrm{prn}}=0$}
      \State $\mathcal{I},\tau,v_1,v_2,A,D\leftarrow$\;\Call{prune}{$A,D,\tau$}
    \EndIf
  \EndLoop
  \For{$t=0,\dotsc,n-1$}
    \State $\mathcal{J}_t,a_{\mathcal{J}_t,t},\mu_{\mathcal{J}_t,t},\eta_{\mathcal{J}_t,t},\theta_{\mathcal{J}_t,t},\_$
    \State\hspace{\algorithmicindent}$\leftarrow$\;\Call{pursuit}{$D[\cdot,\mathcal{I}],U[\cdot,t],\textrm{sel\_xcorr},1,N_{\mathrm{spr}},\_$}
  \EndFor
  \State\Return $\{\mathcal{J}_t,a_{\mathcal{J}_t,t},\mu_{\mathcal{J}_t,t},\eta_{\mathcal{J}_t,t},\theta_{\mathcal{J}_t,t}$
  \State\hspace{\algorithmicindent}$:t=0,\dotsc,n-1\}$
  \EndFunction
  \end{algorithmic}
  \end{varwidth}
\end{algoenv}

\subsection{Benefits Over the Mel Spectrogram}
\label{sec:mel}

In Figure~\ref{fig:brahms}, we compared our log-spectrogram that was
computed via the sparse pursuit method to the mel spectrogram and the
constant-Q transform.  We concluded in Section~\ref{sec:spectrogram}
that as the CQT uses windows of
different length for different frequencies, it is not a good choice
for our dictionary representation.

The mel spectrogram does not have this particular problem, but the
Heisenberg uncertainty principle constrains the time-log-frequency
resolution according to the lowest frequency to be represented.  In
Figure~\ref{fig:brahms}a, we cut the spectrogram at $\SI{530}{Hz}$
(which corresponds to $\SI{577}{Hz}$ when compensating for the
different sampling frequency),
but for our sample with recorder and violin, this is not sufficient,
as it contains notes as low as c'.  Thus, we chose the lowest
frequency as $\SI{200}{Hz}$, sacrificing some resolution.

\begin{figure}[p]
  \tikzsetnextfilename{Figure_\the\numexpr(\thefigure+1)\relax}
  \centering
  \footnotesize
  \begin{tikzpicture}
  \begin{groupplot}[group style={group size=1 by 3,
                                 vertical sep=20mm},
                    enlargelimits=false,
                    axis on top,
                    scale only axis,
                    width=6.6cm,
                    height=5.1cm,
                    ymode=log,
                    xlabel={Time [$\si{s}$]},
                    ylabel={Frequency [$\si{kHz}$]},
                    log ticks with fixed point]
    \nextgroupplot
    \addplot graphics[xmin=0,xmax=8.1920,ymin=0.2,ymax=20.48] {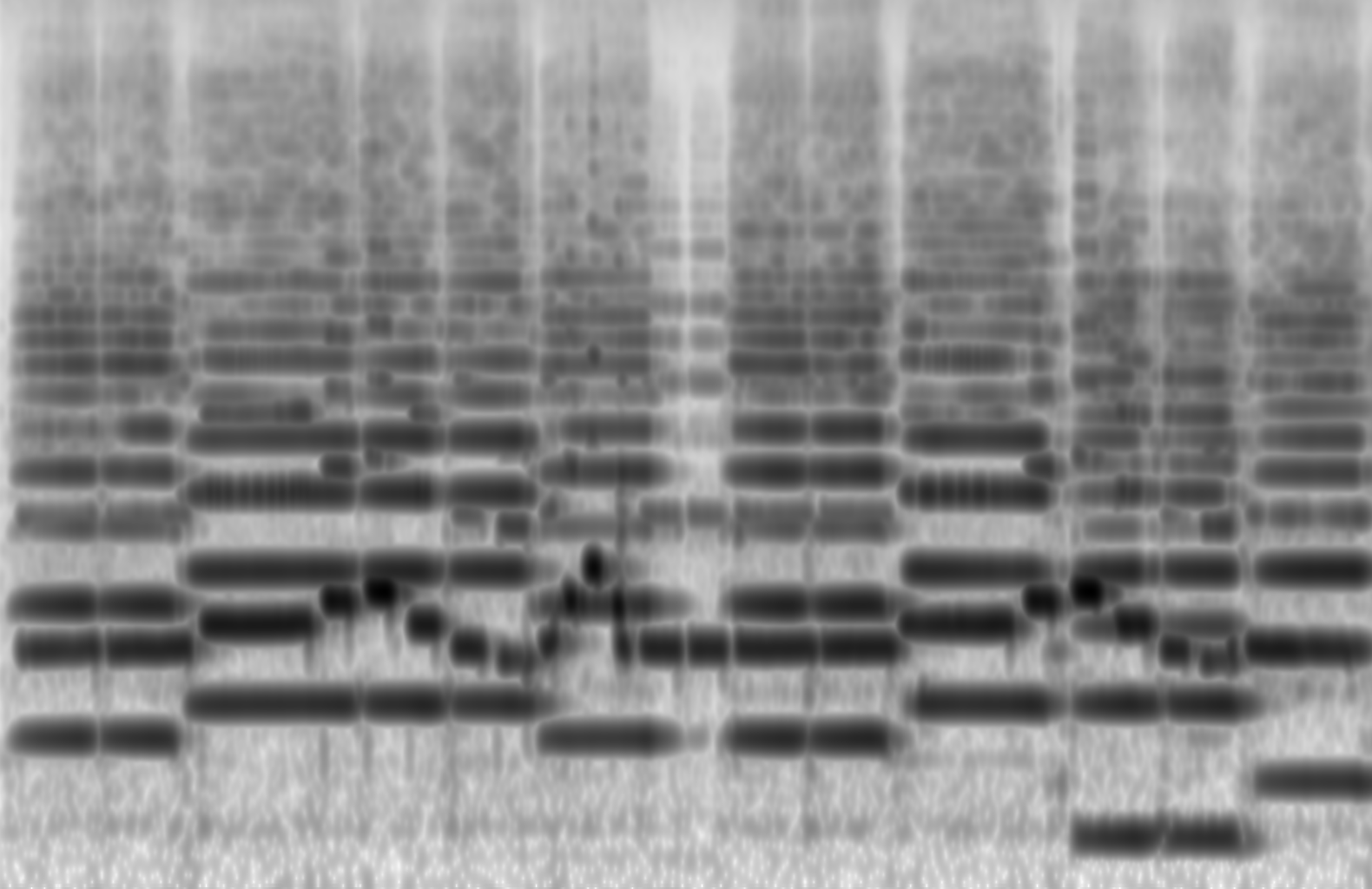};
    \nextgroupplot
    \addplot graphics[xmin=0,xmax=8.1920,ymin=0.2,ymax=20.48] {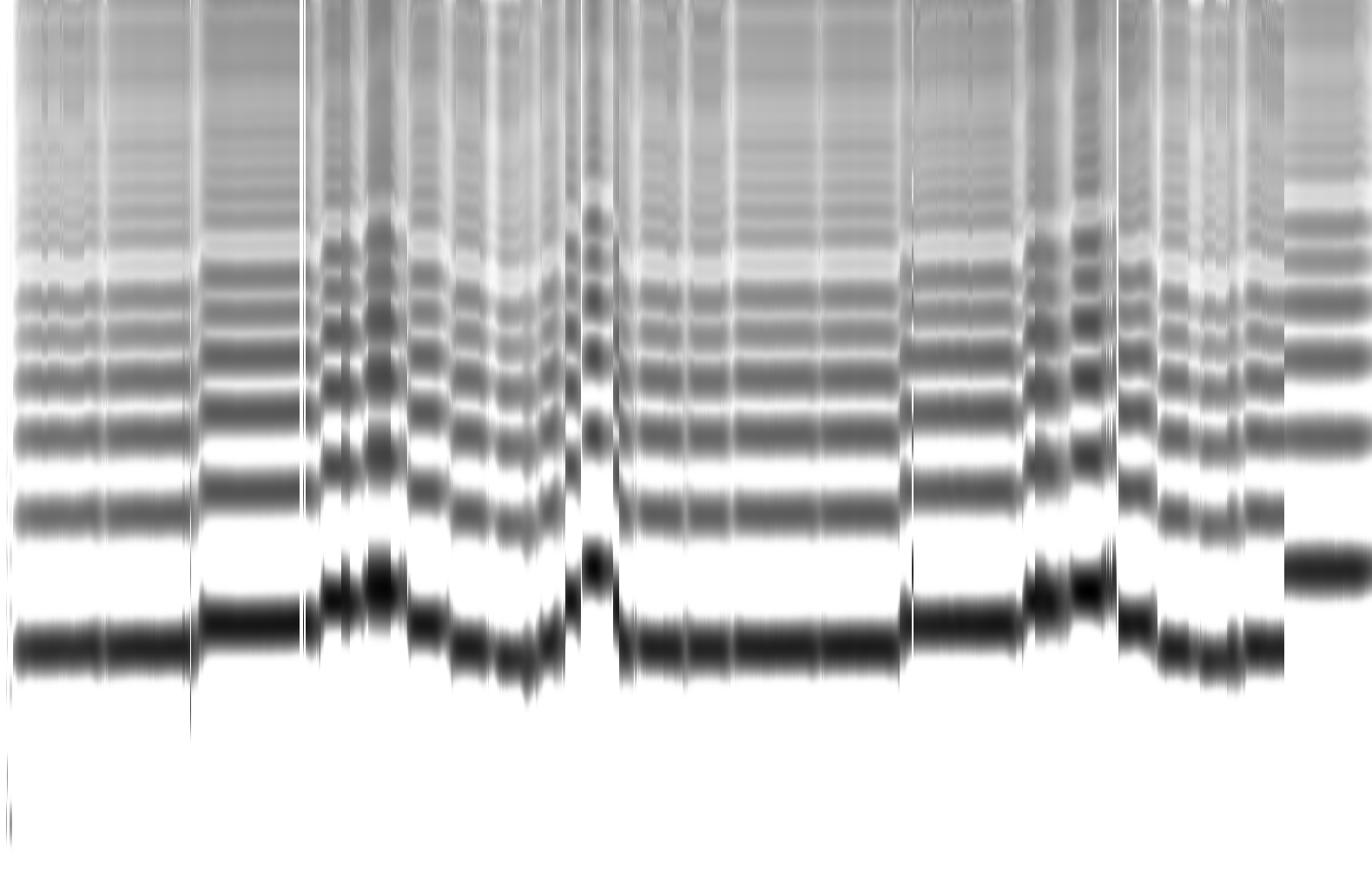};
    \nextgroupplot
    \addplot graphics[xmin=0,xmax=8.1920,ymin=0.2,ymax=20.48] {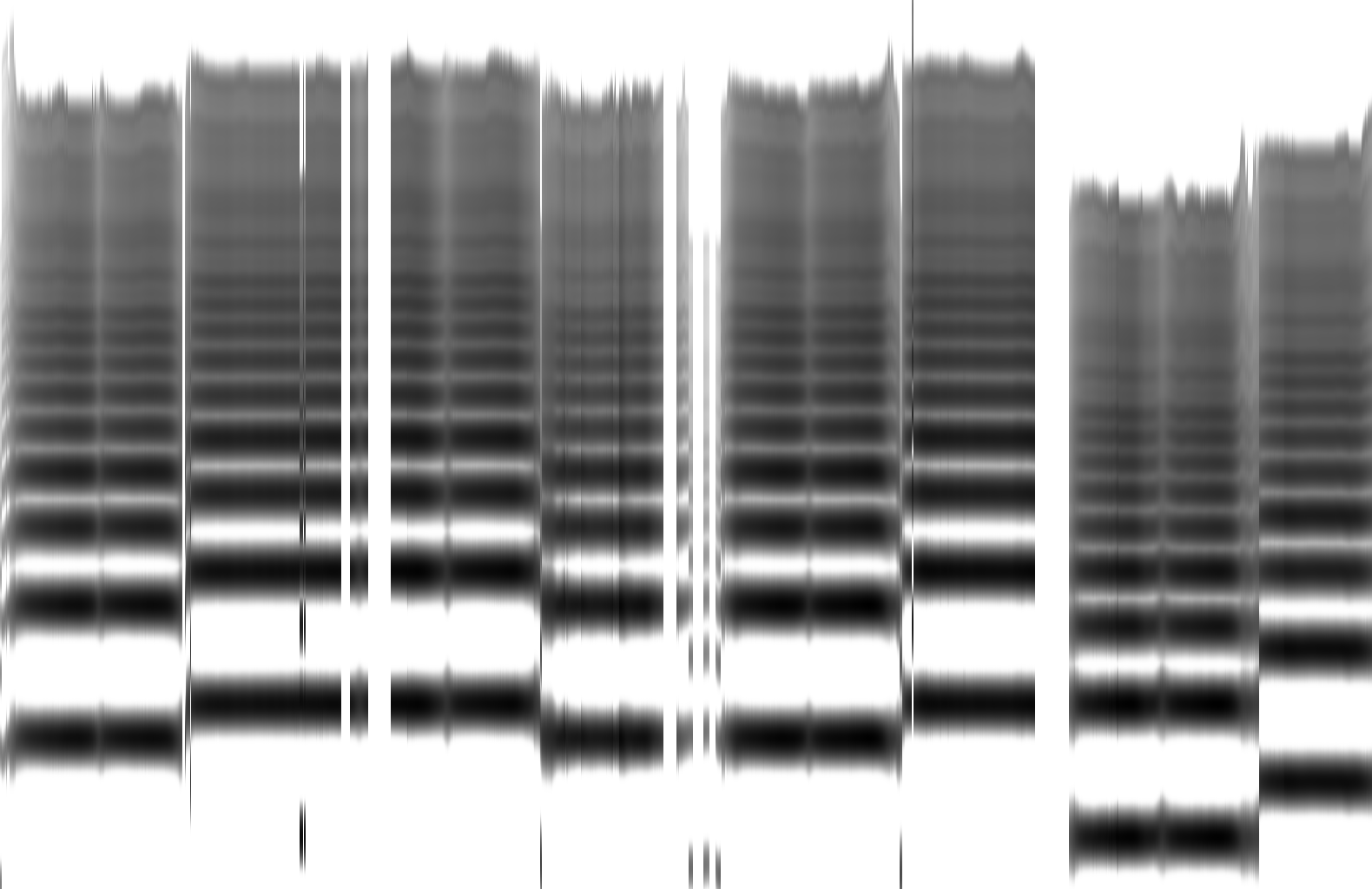};
  \end{groupplot}
  \node[below] at (group c1r1.outer south) {\sffamily (a) Mel spectrogram for the original sample};
  \node[below] at (group c1r2.outer south) {\sffamily (b) Synthesized recorder track};
  \node[below] at (group c1r3.outer south) {\sffamily (c) Synthesized violin track};
  \end{tikzpicture}
\caption{Mel spectrogram for the recorded piece and log-frequency
    spectrograms for the synthesized tracks that were generated based
    on the mel spectrogram.}
  \label{fig:mozart-mel}
\end{figure}

We computed the mel spectrogram on this sample and ran the separation
algorithm $10$ times with $N_{\mathrm{trn}}=100000$ training
iterations in order to obtain a fair comparison.  The performance
figures are given in Table~\ref{tab:mozart-mel} {and
Figure~\ref{fig:boxplot-mel}}, and the results from
the best-case run with a random seed of $7$ are displayed in
Figure~\ref{fig:mozart-mel}.
It can be seen that the performance does not reach what we
achieved with a spectrogram generated via the sparse pursuit
method (cf.\ Figure~\ref{fig:spect} and Table~\ref{tab:real}).

Using again a one-sided Wilcoxon signed-rank test, we find that
without spectral masking, the SDR when using the mel spectrogram is
worse at $p_{\mathrm{Recorder}}=\num{9.8e-4}$ and
$p_{\mathrm{Violin}}=\num{2.0e-3}$.  With spectral masking applied, we
achieve $p_{\mathrm{Recorder}}=p_{\mathrm{Violin}}=\num{9.8e-4}$, as
for each random seed $0,\dotsc,9$, the results from our representation
are consistently better.

We thus conclude that our use of the sparse pursuit algorithm for
generating a log-frequency spectrogram provides a notable benefit for
the subsequent processing.

\begin{table}[p!]
  \centering
  \caption{Performance measures for the best-case run of the separation
    of recorder and violin using the mel spectrogram.  Best numbers are marked.}
    \label{tab:mozart-mel}
    \tabcolsep=4.5pt
    \renewcommand{\arraystretch}{1.1}
    \heavyrulewidth=.5pt
    \lightrulewidth=.4pt
    \cmidrulewidth=.3pt
    \aboverulesep=0.0ex
    \belowrulesep=0.0ex

  \newcolumntype Y{S[
      table-format=2.1,
      table-auto-round,
      table-text-alignment=center,
      table-number-alignment=center,
      table-space-text-post={*}]}
    \begin{tabular}{ccYYY}
    \toprule
    {\bfseries Mask} & {\bfseries Instrument} & {\bfseries SDR} & {\bfseries SIR} & {\bfseries SAR} \\
    \midrule
    \multirow{2}{*}{No} & Recorder &  10.58883825& 31.86674483* &10.62414297  \\
    & Violin & 5.79540404 & 22.47776298* & 5.91412612\\
    \cmidrule{1-5}
    \multirow{2}{*}{Yes} & Recorder & 13.43457382* & 31.46845024 & 13.50650863* \\
    & Violin & 9.25377752* & 21.00994404& 9.58803162* \\
    \bottomrule
  \end{tabular}
\end{table}

\begin{figure}[p!]
  \tikzsetnextfilename{Figure_\the\numexpr(\thefigure+1)\relax}
  \footnotesize
  \begin{tikzpicture}[mark size=1.6pt,mark=+]
    \begin{groupplot}[group style={group size=1 by 3,
          vertical sep=12mm},
        ytick={1,2,3,4},yticklabels={Recorder\\[-1](unmasked),Recorder\\[-1](masked),Violin\\[-1](unmasked),Violin\\[-1](masked)},yticklabel style={align=right},height=2.8cm,width=5.5cm,
        scale only axis,
        ymin=0.5,ymax=4.5,title style={yshift=-1.5mm},y dir=reverse,ytick style={draw=none}]
      \nextgroupplot[title={SDR}]
      \addplot[boxplot={box extend=0.4},mark=none,color=black!50] table[y index={0}] {mozart-mel-nomask-measures.dat};
      \addplot[only marks] table[x index={0},y expr=1] {mozart-mel-nomask-measures.dat};
      \draw[color=black!25] (axis cs:\pgfkeysvalueof{/pgfplots/xmin},1.5) -- (axis cs:\pgfkeysvalueof{/pgfplots/xmax},1.5);
      \addplot[boxplot={box extend=0.4},mark=none,color=black!50] table[y index={0}] {mozart-mel-mask-measures.dat};
      \addplot[only marks] table[x index={0},y expr=2] {mozart-mel-mask-measures.dat};
      \draw[color=black!25] (axis cs:\pgfkeysvalueof{/pgfplots/xmin},2.5) -- (axis cs:\pgfkeysvalueof{/pgfplots/xmax},2.5);
      \addplot[boxplot={box extend=0.4},mark=none,color=black!50] table[y index={1}] {mozart-mel-nomask-measures.dat};
      \addplot[only marks] table[x index={1},y expr=3] {mozart-mel-nomask-measures.dat};
      \draw[color=black!25] (axis cs:\pgfkeysvalueof{/pgfplots/xmin},3.5) -- (axis cs:\pgfkeysvalueof{/pgfplots/xmax},3.5);
      \addplot[boxplot={box extend=0.4},mark=none,color=black!50] table[y index={1}] {mozart-mel-mask-measures.dat};
      \addplot[only marks] table[x index={1},y expr=4] {mozart-mel-mask-measures.dat};
      \nextgroupplot[title={SIR}]
      \addplot[boxplot={box extend=0.4},mark=none,color=black!50] table[y index={2}] {mozart-mel-nomask-measures.dat};
      \addplot[only marks] table[x index={2},y expr=1] {mozart-mel-nomask-measures.dat};
      \draw[color=black!25] (axis cs:\pgfkeysvalueof{/pgfplots/xmin},1.5) -- (axis cs:\pgfkeysvalueof{/pgfplots/xmax},1.5);
      \addplot[boxplot={box extend=0.4},mark=none,color=black!50] table[y index={2}] {mozart-mel-mask-measures.dat};
      \addplot[only marks] table[x index={2},y expr=2] {mozart-mel-mask-measures.dat};
      \draw[color=black!25] (axis cs:\pgfkeysvalueof{/pgfplots/xmin},2.5) -- (axis cs:\pgfkeysvalueof{/pgfplots/xmax},2.5);
      \addplot[boxplot={box extend=0.4},mark=none,color=black!50] table[y index={3}] {mozart-mel-nomask-measures.dat};
      \addplot[only marks] table[x index={3},y expr=3] {mozart-mel-nomask-measures.dat};
      \draw[color=black!25] (axis cs:\pgfkeysvalueof{/pgfplots/xmin},3.5) -- (axis cs:\pgfkeysvalueof{/pgfplots/xmax},3.5);
      \addplot[boxplot={box extend=0.4},mark=none,color=black!50] table[y index={3}] {mozart-mel-mask-measures.dat};
      \addplot[only marks] table[x index={3},y expr=4] {mozart-mel-mask-measures.dat};
      \nextgroupplot[title={SAR}]
      \addplot[boxplot={box extend=0.4},mark=none,color=black!50] table[y index={4}] {mozart-mel-nomask-measures.dat};
      \addplot[only marks] table[x index={4},y expr=1] {mozart-mel-nomask-measures.dat};
      \draw[color=black!25] (axis cs:\pgfkeysvalueof{/pgfplots/xmin},1.5) -- (axis cs:\pgfkeysvalueof{/pgfplots/xmax},1.5);
      \addplot[boxplot={box extend=0.4},mark=none,color=black!50] table[y index={4}] {mozart-mel-mask-measures.dat};
      \addplot[only marks] table[x index={4},y expr=2] {mozart-mel-mask-measures.dat};
      \draw[color=black!25] (axis cs:\pgfkeysvalueof{/pgfplots/xmin},2.5) -- (axis cs:\pgfkeysvalueof{/pgfplots/xmax},2.5);
      \addplot[boxplot={box extend=0.4},mark=none,color=black!50] table[y index={5}] {mozart-mel-nomask-measures.dat};
      \addplot[only marks] table[x index={5},y expr=3] {mozart-mel-nomask-measures.dat};
      \draw[color=black!25] (axis cs:\pgfkeysvalueof{/pgfplots/xmin},3.5) -- (axis cs:\pgfkeysvalueof{/pgfplots/xmax},3.5);
      \addplot[boxplot={box extend=0.4},mark=none,color=black!50] table[y index={5}] {mozart-mel-mask-measures.dat};
      \addplot[only marks] table[x index={5},y expr=4] {mozart-mel-mask-measures.dat};
    \end{groupplot}
  \end{tikzpicture}
  \caption{Distribution of the performance measures of the separation
    of violin and piano over 10 runs using the mel spectrogram, without
    and with spectral masking}
  \label{fig:boxplot-mel}
\end{figure}

\subsection{Log-Frequency Spectrograms of the Instrument Tracks}

\begin{figure}[p]
  \tikzsetnextfilename{Figure_\the\numexpr(\thefigure+1)\relax}
  \centering
  \footnotesize
  \begin{tikzpicture}
  \hypersetup{hidelinks}
  \begin{groupplot}[group style={group size=1 by 3,
                                 vertical sep=20mm},
                    enlargelimits=false,
                    axis on top,
                    scale only axis,
                    width=6.6cm,
                    height=5.1cm,
                    ymode=log,
                    xlabel={Time [$\si{s}$]},
                    ylabel={Frequency [$\si{kHz}$]},
                    log ticks with fixed point]
    \nextgroupplot
    \addplot graphics[xmin=0,xmax=8.1920,ymin=0.02,ymax=20.48] {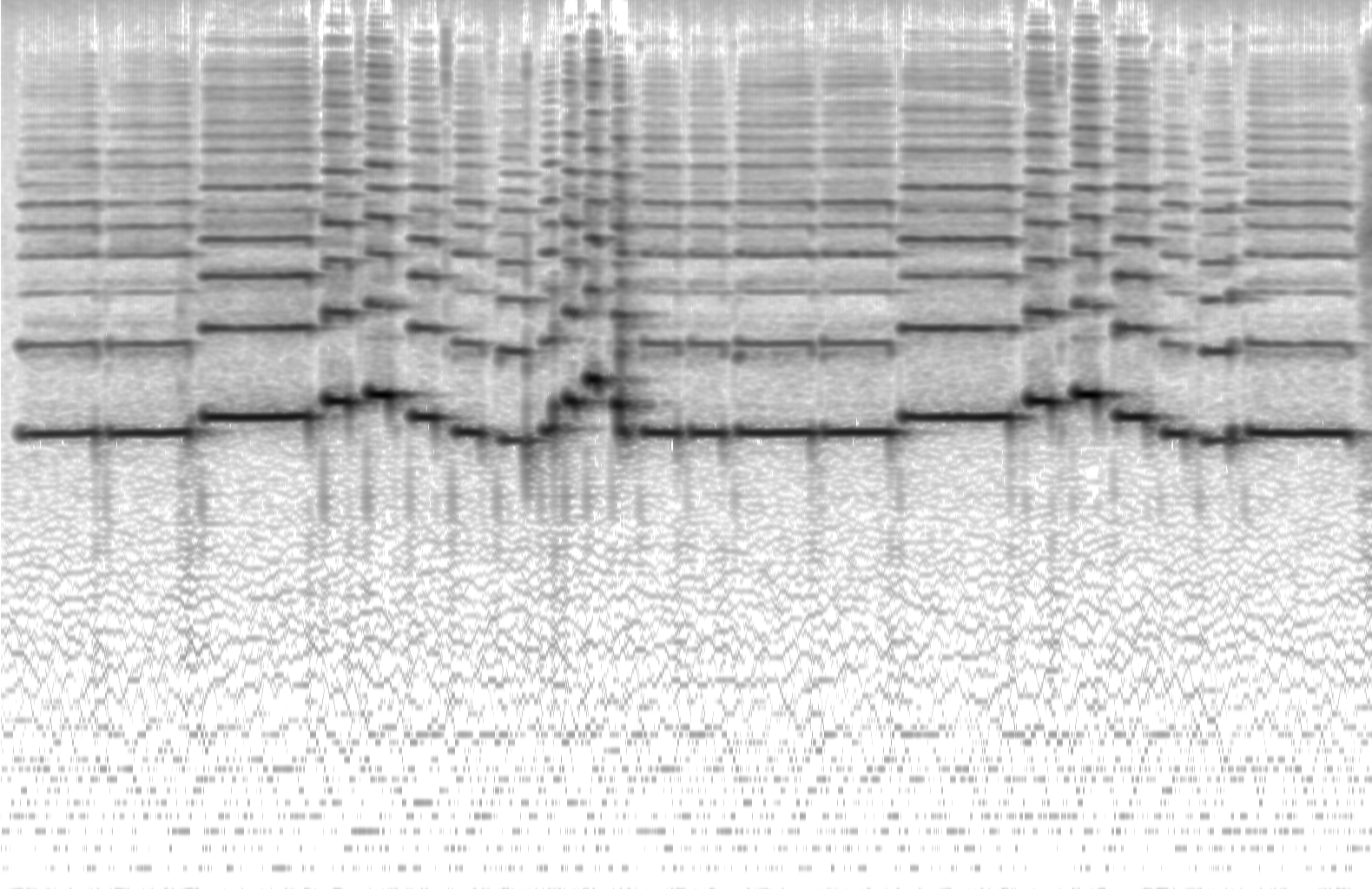};
    \nextgroupplot
    \addplot graphics[xmin=0,xmax=8.1920,ymin=0.02,ymax=20.48] {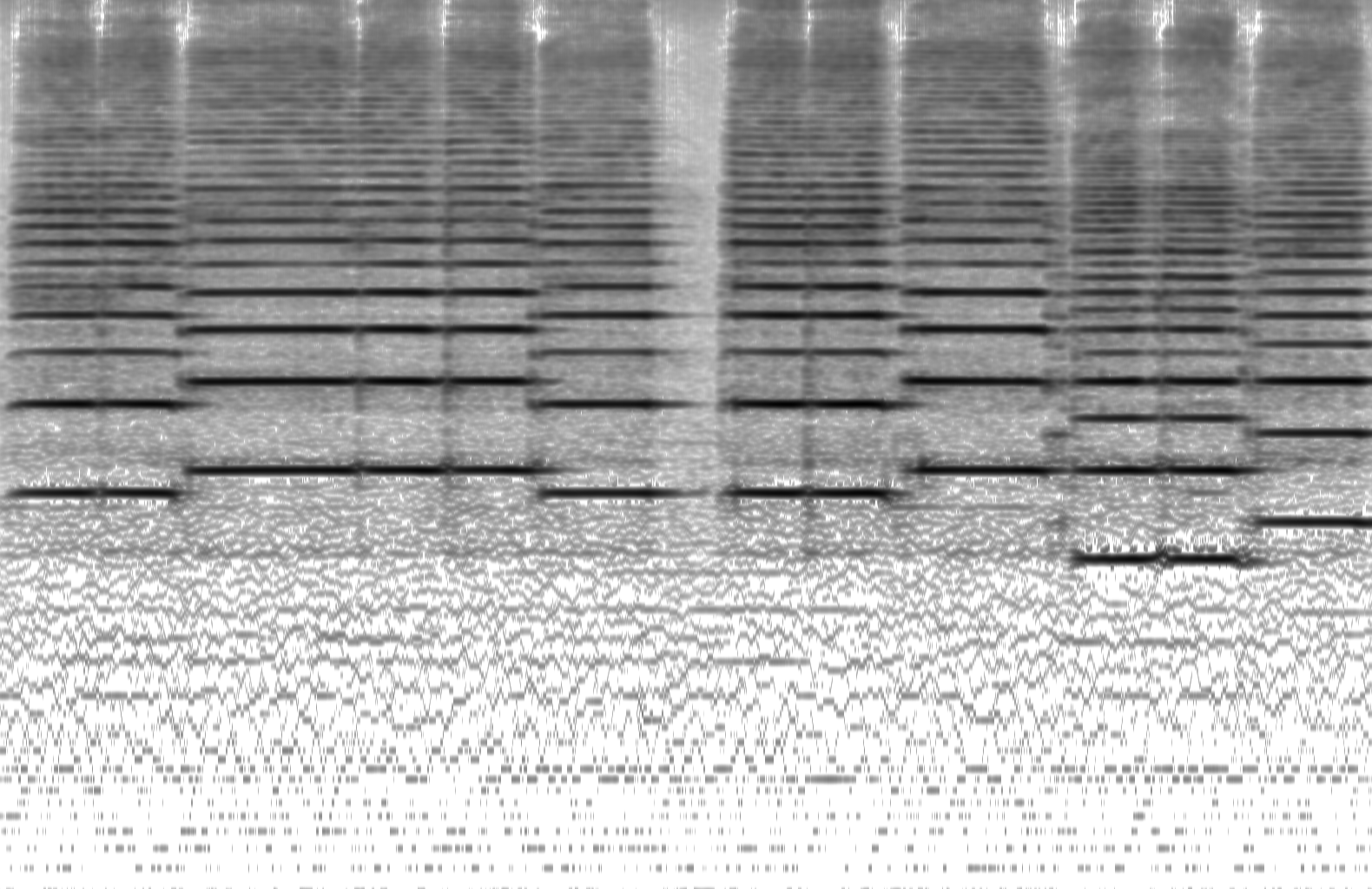};
    \nextgroupplot
    \addplot graphics[xmin=0,xmax=8.1920,ymin=0.02,ymax=20.48] {mozart-log.png};
  \end{groupplot}
  \node[below] at (group c1r1.outer south) {\sffamily (a) Recorder};
  \node[below] at (group c1r2.outer south) {\sffamily (b) Violin};
  \node[below] at (group c1r3.outer south) {\sffamily (c) Mixture (copy of Figure~\ref{fig:spect}a)};
  \end{tikzpicture}
  \caption{Sparsity-derived log-frequency spectrograms of the original instrument
    samples\strut}
  \label{fig:instorig}
\end{figure}

\begin{figure}[p]
  \tikzsetnextfilename{Figure_\the\numexpr(\thefigure+1)\relax}
  \centering
  \footnotesize
  \begin{tikzpicture}
  \begin{groupplot}[group style={group size=1 by 3,
                                 vertical sep=20mm},
                    enlargelimits=false,
                    axis on top,
                    scale only axis,
                    width=6.6cm,
                    height=5.1cm,
                    ymode=log,
                    xlabel={Time [$\si{s}$]},
                    ylabel={Frequency [$\si{kHz}$]},
                    log ticks with fixed point]
    \nextgroupplot
    \addplot graphics[xmin=0,xmax=8.1920,ymin=0.02,ymax=20.48] {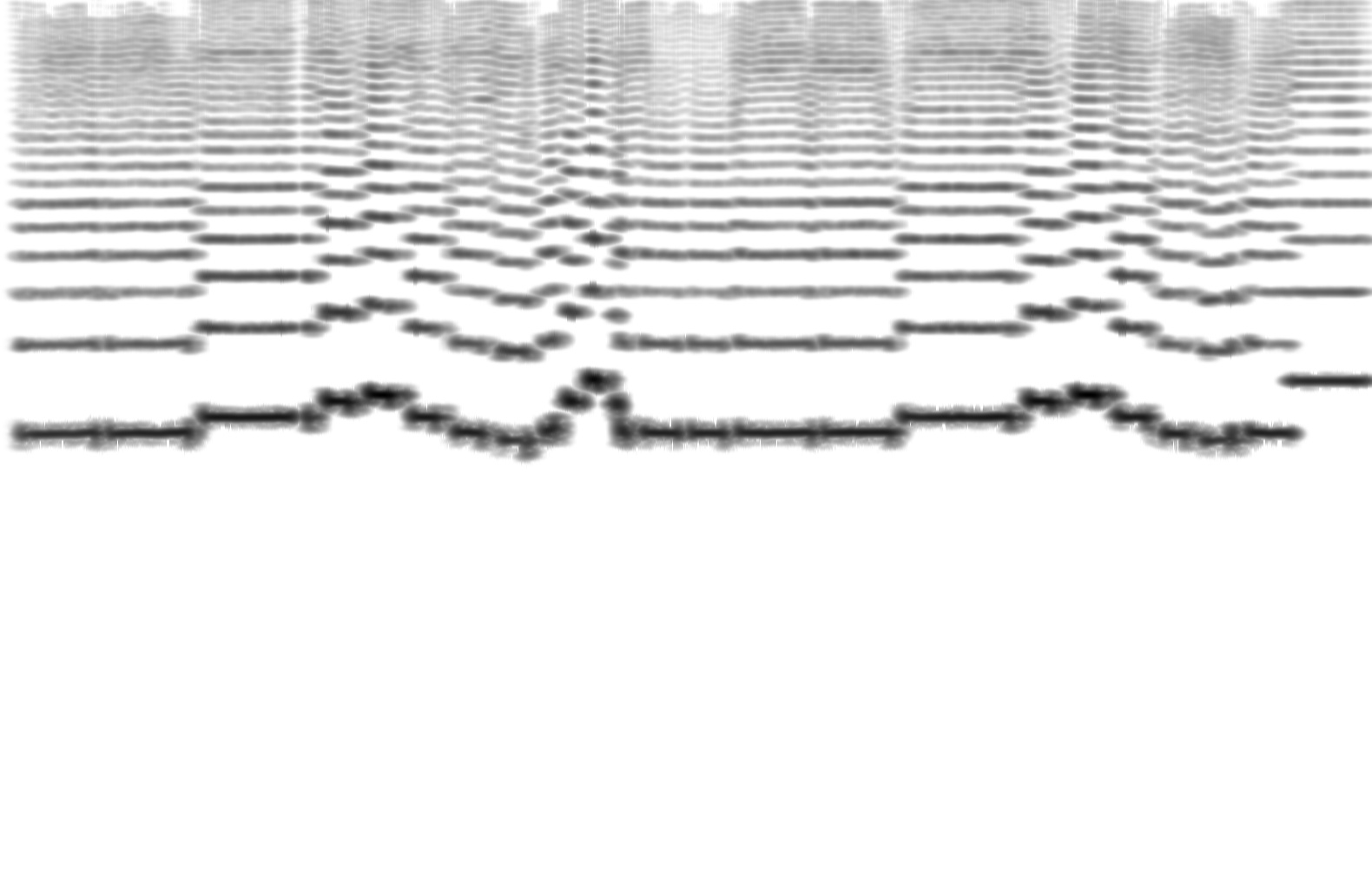};
    \nextgroupplot
    \addplot graphics[xmin=0,xmax=8.1920,ymin=0.02,ymax=20.48] {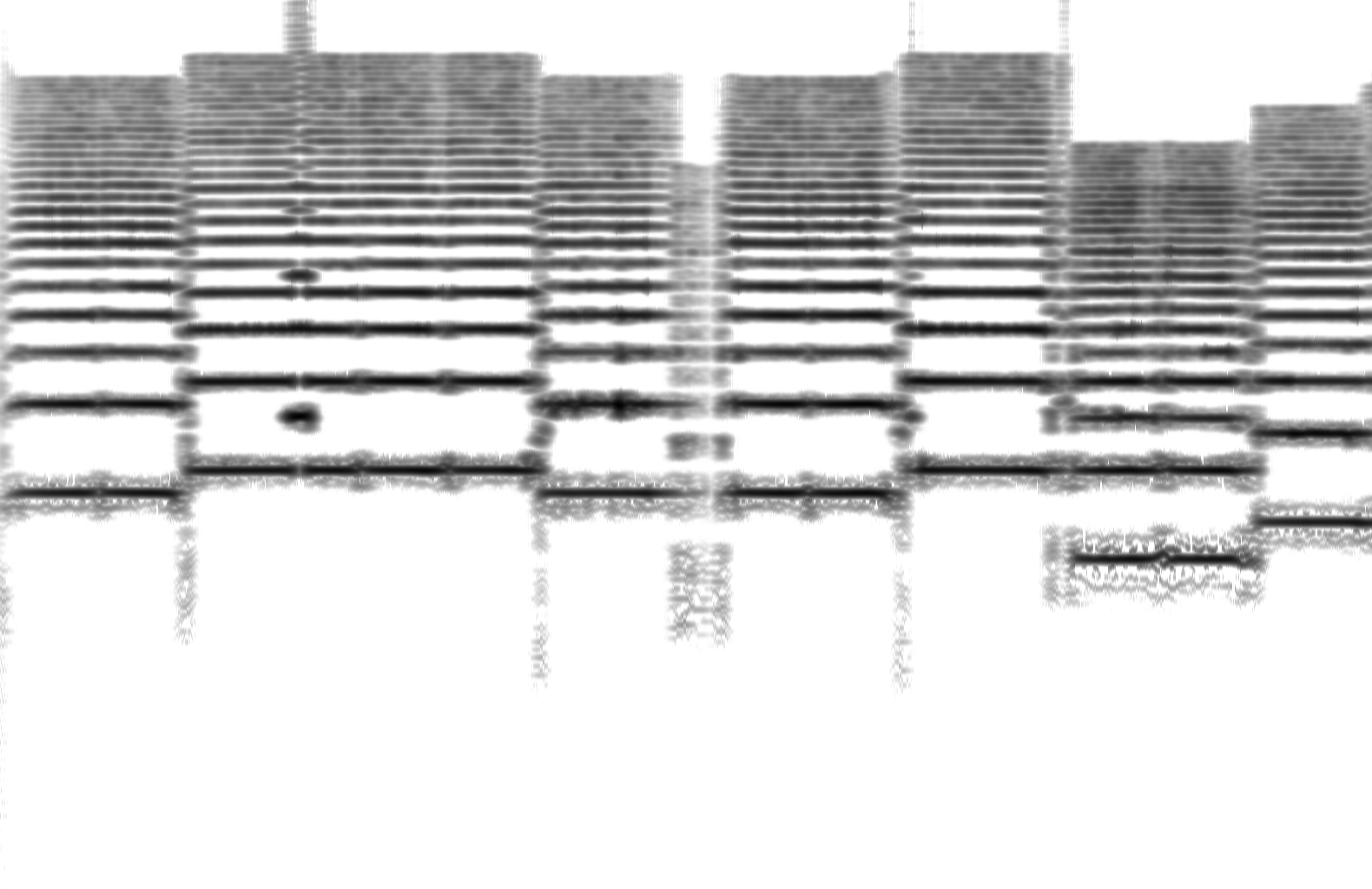};
    \nextgroupplot
    \addplot graphics[xmin=0,xmax=8.1920,ymin=0.02,ymax=20.48] {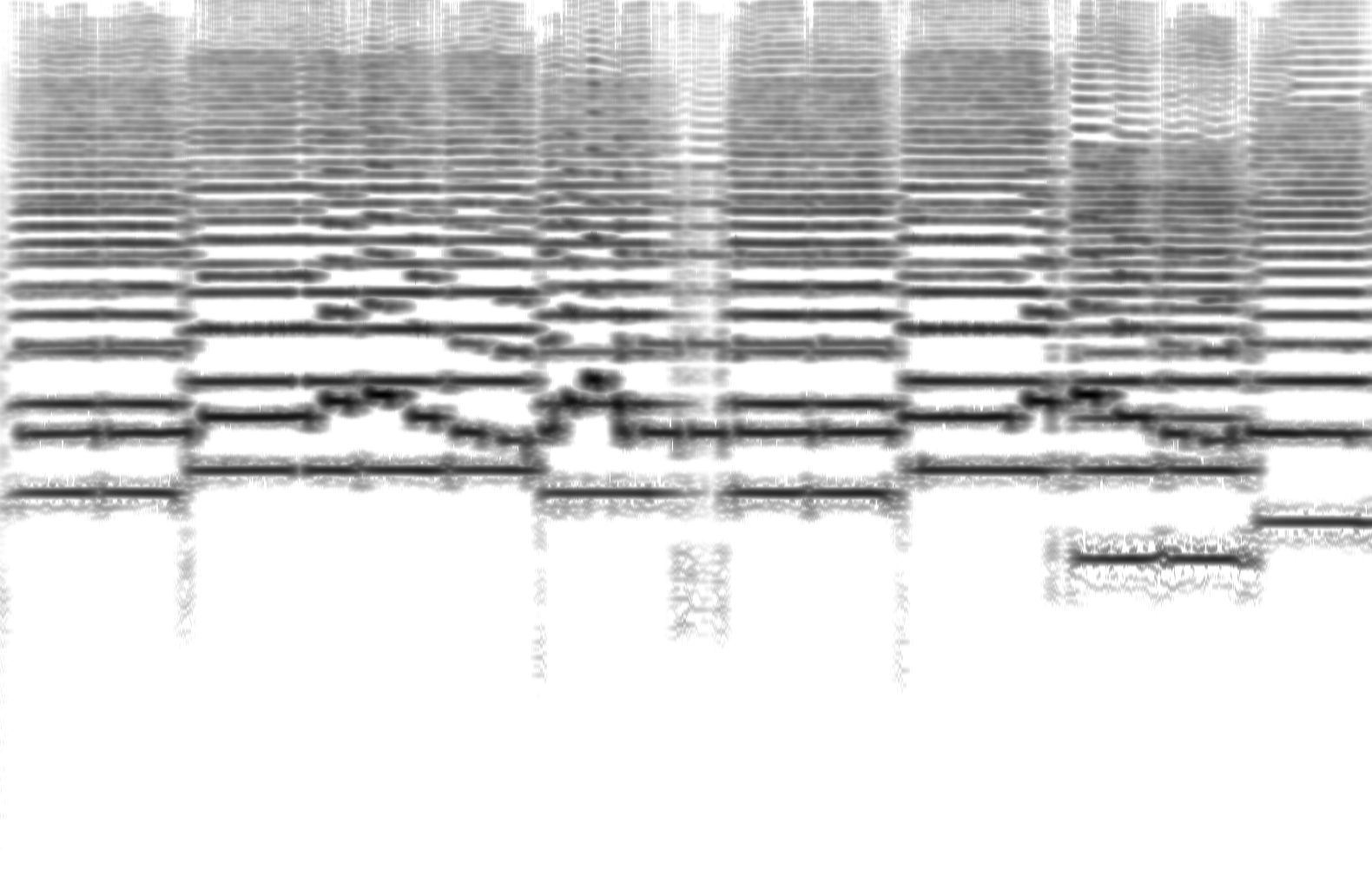};
  \end{groupplot}
  \node[below] at (group c1r1.outer south) {\sffamily (a) Recorder};
  \node[below] at (group c1r2.outer south) {\sffamily (b) Violin};
  \node[below] at (group c1r3.outer south) {\sffamily (c) Mixture};
  \end{tikzpicture}
  \caption{Sparsity-derived log-frequency spectrograms of the separated audio tracks and its mixture\strut}
  \label{fig:instcomp}
\end{figure}

For additional comparison, we computed the log-frequency spectrograms
of the original (Figure~\ref{fig:instorig}) and the computed
(Figure~\ref{fig:instcomp}) instrument tracks.

It should be noted that these spectrograms are not used
anywhere in the computation or evaluation process, and due to
artifacts from the sparse pursuit algorithm, they are not an accurate
representation of the time-domain signal.

Nevertheless, two effects can be seen when comparing
Figure~\ref{fig:instcomp} to Figure~\ref{fig:spect}:
\begin{enumerate}
\item Due to spectral masking, the harmonics now have different
  intensities.
\item The Griffin-Lim phase reconstruction algorithm
  \emph{smoothes} some of the artifacts that were introduced by the
  sparse pursuit algorithm.  This is because not every
  two-dimensional image is actually a valid spectrogram that corresponds
  to an audio signal; instead, the Griffin-Lim algorithm aims to find
  an audio signal whose spectrogram is as close as possible to the
  given image, and it uses the phase of the original mixed sample as
  the initial value.
\end{enumerate}


\clearpage

\begin{backmatter}

\section*{Abbreviations}

Adam:\ Adaptive moment estimation; %
{APS:\ Artifacts-related perceptual score;}
BASS-dB:\ Blind Audio Source Separation Evaluation Database; %
CD:\ Compact disc;
EM:\ Expectation maximization;
FFT:\ Fast Fourier transform; %
CQT:\ Constant-Q transform;
ICA:\ Independent component analysis;
{IPS:\ Interference-related perceptual score;}
K-SVD:\ K-singular value decomposition; %
{LASSO:\ Least absolute shrinkage and selection operator;} %
LSTM:\ Long short-term memory;
L-BFGS-B:\ Limited-memory Broyden-Fletcher–Goldfarb-Shanno with bounds; %
{MPE:\ Multiple pitch estimation;}
NMF:\ Non-negative matrix factorization;
NMFD:\ Non-negative matrix factor deconvolution;
NMF2D:\ Non-negative matrix factor two-dimensional deconvolution;
OMP:\ Orthogonal matching pursuit;
{OPS:\ Overall perceptual score;}
{PEASS:\ Perceptual Evaluation methods for Audio Source Separation;} %
PLCA:\ Probabilistic latent component analysis;
SiSEC:\ Signal Separation Evaluation Campaign;
SAR:\ Signal-to-artifacts ratio;
SDR:\ Signal-to-distortion ratio;
SIR:\ Signal-to-interference ratio;
STFT:\ Short-time Fourier transform;
{TPS:\ Target-related perceptual score;}
URMP:\ University of Rochester Multi-Modal Music Performance Dataset %

\section*{Availability of data and materials}

The software is presented along with audio samples and the original
input data on the institute
website\footnote{%
\url{https://www.math.colostate.edu/~king/software.html\#Musisep}}. 
The source code is available on
GitHub\footnote{\url{https://github.com/rgcda/Musisep}} under the GNU General Public License (version 3).

\section*{Competing interests}
The authors declare that they have no competing interests.

\section*{Funding}
  The first author acknowledges funding by the Deut\-sche
For\-schungs\-ge\-mein\-schaft (DFG, German Research Foundation) --
Pro\-jekt\-num\-mer 281474342/GRK2224/1.

\section*{Author's contributions}
SS devised the algorithm, wrote the source code, and drafted the manuscript.  EJK supervised the research and revised the manuscript.  All authors read and approved the final manuscript.

\section*{Acknowledgements}

The authors would like to thank Bernhard G.\ Bodmann, Gitta Kutyniok,
and Monika Dörfler for engaging discussions on the subject and Kara
Tober for playing the clarinet samples.
{Further, we thank the anonymous reviewers, whose
valuable and constructive comments helped us improve the manuscript.}


\bibliographystyle{bmc-mathphys} 
\bibliography{bmc_article}      







\end{backmatter}

\end{document}